\documentclass[pdflatex,sn-mathphys-num]{sn-jnl}

\usepackage{graphicx}%
\usepackage{multirow}%
\usepackage{amsmath,amssymb,amsfonts}%
\usepackage{amsthm}%
\usepackage{mathrsfs}%
\usepackage[title]{appendix}%
\usepackage{xcolor}%
\usepackage{textcomp}%
\usepackage{manyfoot}%
\usepackage{booktabs}%
\usepackage{algorithm}%
\usepackage{algorithmicx}%
\usepackage{algpseudocode}%
\usepackage{listings}%
\usepackage[normalem]{ulem}

\usepackage{anyfontsize}
\newcommand{\orcidID}[1]{\unskip\space(\href{https://orcid.org/#1}{ORCID: #1})}
\usepackage{hyperref}

\theoremstyle{thmstyleone}%
\theoremstyle{thmstyletwo}%

\theoremstyle{thmstylethree}%

\newcommand{\TODO}[1]{\textcolor{red}{[TODO: #1]}}

\newcommand{\supfig}[2]{\IfFileExists{#2}{\includegraphics[width=#1\textwidth]{#2}}%
  {\fbox{\parbox{\dimexpr#1\textwidth-2\fboxsep-2\fboxrule\relax}{\centering\vspace{2cm}\TODO{\detokenize{#2} not found}\vspace{2cm}}}}}
\newcommand{\secDatacollection}{\ref{Datacollection}}

\newcommand{\secGeometry}{\ref{TVED}}

\newcommand{\secMetrics}{\ref{sec:metrics}}
\newcommand{\secSensitivity}{\ref{sec:sensitivity}}

\newcommand{\tabCohort}{\ref{tab:cohort}}
\newcommand{\tabOperators}{\ref{tab:operators}}
\newcommand{\tabDensity}{\ref{tab:density}}
\newcommand{\tabLeafletStiff}{\ref{tab:leaflet_stiffness}}

\newcommand{\tabCtStiff}{\ref{tab:ctstiff}}
\newcommand{\figWorkflow}{\ref{fig1_workflow}}
\newcommand{\figInputs}{\ref{fig_exvivo_inputs}}
\newcommand{\figCohortMaps}{\ref{fig_cohort_maps}}
\newcommand{\eqWleaflet}{\ref{eq:leaflet_W}}

\begin{document}

\title[]{Patient-specific Modeling of the Tricuspid Valve from 3D Transesophageal Echocardiography}

\author[1]{\fnm{Davide} \sur{Tondi}\orcidID{0009-0009-8038-5476}}\email{davide.tondi@polimi.it}
\author[2]{\fnm{Vijay K.} \sur{Dubey}\orcidID{0000-0003-3365-4003}}\email{dvijay@utexas.edu}
\author[2]{\fnm{Collin E.} \sur{Haese}\orcidID{0009-0005-0561-2545}}\email{collinhaese@utexas.edu}
\author[3]{\fnm{Colton J.} \sur{Kostelnik}\orcidID{0000-0001-5930-8024}}\email{colton.kostelnik@heart.org}
\author[4]{\fnm{Mrudang} \sur{Mathur}\orcidID{0000-0002-9273-5586}}\email{mrudang@stanford.edu}
\author[2]{\fnm{Sumedh} \sur{Seetharam}\orcidID{0009-0000-6831-5736}}\email{sseetharam12@my.utexas.edu}
\author[5]{\fnm{Amin} \sur{Silvani}}\email{asilvani@seas.upenn.edu}
\author[5]{\fnm{Alison M.} \sur{Pouch}\orcidID{0000-0001-8854-5301}}\email{pouch@pennmedicine.upenn.edu}
\author[6, 7]{\fnm{Jakob Christoph} \sur{Voran}\orcidID{0009-0000-4547-9260}}\email{JakobChristoph.Voran@uksh.de}
\author[8]{\fnm{Felix} \sur{Kreidel}}\email{f.kreidel@asklepios.com}
\author[8]{\fnm{Marnie} \sur{Goldmann}}\email{m.goldmann@asklepios.com}
\author[9]{\fnm{Adrian} \sur{Buganza Tepole}\orcidID{0000-0001-8531-0603}}\email{ab6035@columbia.edu}
\author[3]{\fnm{William D.} \sur{Meador}\orcidID{0000-0001-5797-9553}}
\author[10]{\fnm{Marcin} \sur{Malinowski}\orcidID{0000-0002-4526-2843}}\email{mmalinowski@icloud.com}
\author[10]{\fnm{Magda} \sur{Piekarska}\orcidID{0000-0002-9889-3251}}\email{magdalucynapiekarska@gmail.com}
\author[10]{\fnm{Austin} \sur{Goodyke}\orcidID{0000-0003-3989-1026}}\email{Austin.Goodyke@corewellhealth.org}
\author[11]{\fnm{Issam D.} \sur{Moussa}\orcidID{0000-0002-5091-624X}}\email{imoussa@illinois.edu}
\author[12]{\fnm{Matthew R.} \sur{Summers}}\email{mrsumme1@sentara.com}
\author[4]{\fnm{Robyn} \sur{Fong}}\email{fongr@stanford.edu}
\author[4]{\fnm{Matthew M.} \sur{Duda}\orcidID{0000-0003-1442-7614}}\email{mattduda@stanford.edu}
\author[4]{\fnm{Kelly C.} \sur{Higa}}\email{kelhiga@stanford.edu}
\author[4]{\fnm{William} \sur{Hiesinger}\orcidID{0000-0002-3548-2578}}\email{willhies@stanford.edu}
\author[10]{\fnm{Tomasz A.} \sur{Timek}}\email{Tomasz.Timek@corewellhealth.org}
\author[13,14]{\fnm{Jan N.} \sur{Fuhg}\orcidID{0000-0002-5986-3770}}\email{jan.fuhg@utexas.edu}
\author[1]{\fnm{Emiliano} \sur{Votta}\orcidID{0000-0001-7115-0151}}\email{emiliano.votta@polimi.it}
\author*[13,2,3,14]{\fnm{Manuel K.} \sur{Rausch}\orcidID{0000-0003-1337-6472}}\email{manuel.rausch@utexas.edu}

\affil[1]{\orgdiv{Department of Electronics, Information and Bioengineering},
          \orgname{Politecnico di Milano},
          \orgaddress{\city{Milan}, \country{Italy}}}

\affil[2]{\orgdiv{Walker Department of Mechanical Engineering},
          \orgname{The University of Texas at Austin},
          \orgaddress{\street{204 E Dean Keeton St}, \city{Austin}, \state{TX}, \country{United States of America}}}

\affil[3]{\orgdiv{Department of Biomedical Engineering},
          \orgname{The University of Texas at Austin},
          \orgaddress{\street{107 W Dean Keeton St}, \city{Austin}, \state{TX}, \country{United States of America}}}
          
\affil[4]{\orgdiv{Department of Cardiothoracic Surgery},
          \orgname{Stanford University},
          \orgaddress{\street{870 Quarry Rd Extension}, \city{Palo Alto},\state{CA},\country{United States of America}}}

\affil[5]{\orgdiv{Department of Radiology and Bioengineering},
          \orgname{University of Pennsylvania},
          \orgaddress{\street{Skirkanich Hall, 210 South 33rd Street}, \city{Philadelphia}, \state{PA}, \country{United States of America}}}

\affil[6]{\orgdiv{Department of Internal Medicine III, Cardiology and Critical Care},
          \orgname{University Hospital Schleswig-Holstein},
          \orgaddress{\street{Arnold-Heller-Straße 3, Haus K3 (Innere Medizin)}, \city{Kiel},\country{Germany}}}
\affil[7]{\orgname{German Centre for Cardiovascular Research (DZHK)},
          \orgaddress{\city{Partner Site North},\country{Germany}}}

\affil[8]{\orgdiv{Department of Cardiology},
          \orgname{Asklepios Hospital},
          \orgaddress{\street{Eißendorfer Pferdeweg 52}, \city{Hamburg},\country{Germany}}}

\affil[9]{\orgdiv{Department of Mechanical Engineering},
          \orgname{Columbia University},
          \orgaddress{\street{500 W 120th St}, \city{New York}, \state{NY}, \country{United States of America}}}

\affil[10]{\orgdiv{Division of Cardiothoracic Surgery},
          \orgname{Corewell Health},
          \orgaddress{\street{100 Michigan Ave SE}, \city{Grand Rapids},\state{MI},\country{United States of America}}}

\affil[11]{\orgdiv{Carle Illinois College of Medicine},
          \orgname{University of Illinois at Urbana-Champaign},
          \orgaddress{\city{Urbana},\state{IL},\country{United States of America}}}

\affil[12]{\orgdiv{Sentara Heart Valve Center},
          \orgname{Sentara Heart Hospital},
          \orgaddress{\city{Norfolk},\state{VA},\country{United States of America}}}
          
\affil[13]{\orgdiv{Department of Aerospace Engineering $\&$ Engineering Mechanics},
          \orgname{The University of Texas at Austin},
          \orgaddress{\street{2617 Wichita St}, \city{Austin}, \state{TX}, \country{United States of America}}}

\affil[14]{\orgdiv{Oden Institute for Computational Engineering and Sciences},
          \orgname{The University of Texas at Austin},
          \orgaddress{\street{201 E 24th St}, \city{Austin}, \state{TX}, \country{United States of America}}}

\abstract{
\textbf{Purpose:} Image-based patient-specific simulations of the tricuspid valve (TV) could support repair planning. However, clinical images incompletely resolve the chordal anatomy and cannot directly determine tissue properties, which hinders the development of such models. We developed and evaluated a framework that simulates the TV from clinical images and addresses these limitations using measurements on explanted human valves.

\textbf{Methods:} We reconstructed ten functionally competent human TVs from 3D transesophageal echocardiography. Following manual segmentation, we automatically built finite element models with synthetic chordae and with leaflet thickness, chordal insertion density, and tissue properties informed by \textit{ex vivo} human measurements. Next, we used shape matching and force-based chordal-length adjustment to simulate valve closure. We also assessed inter-observer segmentation variability and sensitivity to chordal insertion density, leaflet stiffness, leaflet thickness, and chordal size and material properties.

\textbf{Results:} Per-valve median geometric errors were 0.61–1.13 mm relative to the morphed references and 0.55–0.95 mm relative to the image-derived surfaces. Coaptation covered 20–42\% of the leaflet area, and median leaflet stresses ranged from 7.6 to 16.9 kPa. Inter-observer variability in segmentation most strongly affected coaptation area. Of the population-based inputs, insertion density and leaflet stiffness most affected geometric error. Thickness primarily affected stress, whereas chordal size and material properties had limited effects.

\textbf{Conclusion:} Clinical echocardiography, supplemented with \textit{ex vivo} human tissue data, enabled patient-specific simulation of ten human TVs. This framework provides a foundation for investigating valve mechanics and developing repair-planning tools, while sensitivity analyses identify priorities for model refinement and uncertainty assessment. 
}

\maketitle

\section*{Statements and Declarations}

\bmhead{Funding}
This research was supported through funding from various sources. CJK was partially supported through the AHA via its postdoctoral fellowship 25POST1375865. CEH was supported through NIH/NHLBI predoctoral fellowship 1F31HL178280. The work was further supported by (i) NIH/NHLBI award numbers 1R01HL165251 and 1R21HL161832 to MKR and TAT, (ii) AHA award number 25IPA1454136 to WH. 

\bmhead{Competing interests}
AMP holds patents on mitral valve segmentation and computing stress distribution in heart valves using 3D TEE images. FK has a speaking and consulting agreement with Edwards LifeSciences and Abbott. MKR has a speaking and consulting agreement with Edwards LifeSciences and LumenexBio. 

\bmhead{Ethics approval}
All tissue and imaging data collection complied with the requirements of the respective local institutional review and regulatory bodies. All data were de-identified. All use of such data in the current work followed the data use agreement in place between UT Austin and the respective institution.

\bmhead{Data availability}
After acceptance of this manuscript, all the data will be made available through the Texas Data Repository in \href{https://doi.org/10.18738/T8/VBJLOX}{this Dataverse}. 

\bmhead{Code availability}
After acceptance of this manuscript, all the code will be made available through \href{https://github.com/SoftTissueBiomechanicsLab/3DTEE_to_FEM}{this GitHub repository}.

\bmhead{Declaration of generative AI and AI-assisted technologies}
During the preparation of this work, the authors used Claude to assist with data processing and editing. The authors reviewed all generated or edited content as needed and take full responsibility for the published article.

\clearpage

\section{Introduction}\label{intro}
Tricuspid regurgitation (TR) affects more than 1.6 million people in the United States and an estimated 70 million worldwide \cite{kolte2020tv}, and is associated with increased morbidity and mortality \cite{topilsky2019burden,hahn2023tr}. Surgical and transcatheter repair and replacement offer an expanding range of treatment options \cite{sorajja2023triluminate, hahn2025transcatheter, praz2025esc}. However, TR recurs in 10-30\% of patients after surgical repair \cite{rodrigueztorres2022surgical,krivickiene2024recurrent}, and selecting the most appropriate treatment for an individual patient remains unclear \cite{praz2025esc}. Further, the tricuspid valve (TV) exhibits substantial variation in leaflet number and configuration \cite{hahn2021nomenclature}, making the interaction between a proposed intervention and the patient’s anatomy an important consideration in treatment planning.
 
Patient-specific computational models provide a controlled and repeatable means to investigate these interactions. They could complement anatomical assessment tools, such as the GLIDE score for tricuspid edge-to-edge repair \cite{gercek2024glide}, by estimating how interventions alter leaflet deformation, coaptation, and tissue loading. Previous studies have built finite element (FE) models from idealized geometries \cite{stevanella2010tv,johnson2021tv}, explanted valves \cite{mathur2022trivalve}, computed tomography \cite{kong2018tv}, and echocardiography of \textit{ex vivo} porcine hearts \cite{tondi2026fwa}. These models combine valve geometry, tissue mechanical properties, and loading conditions to estimate valve mechanics in health \cite{stevanella2010tv,kong2018tv}, disease \cite{mathur2022trivalve,tondi2026fwa}, and repair \cite{mathur2022trivalve,haese_understanding_2026}. However, their predictions depend on the data used to construct the models \cite{kong2018tv,mangine2026chordae}. Clinical imaging, and echocardiography in particular, provides only part of this information. 

Echocardiography often incompletely resolves the thin TV leaflets \cite{muraru2019tv3d} and the chordal structure \cite{Hahn2016revimagingTV,haese2027chordae}, and cannot directly measure tissue mechanical properties. Models must therefore supplement image-derived geometry with assumptions about leaflet thickness, chordal architecture, and material behavior. Reconstructing the geometry also requires interpretation of indistinct tissue boundaries \cite{muraru2019tv3d}, which introduces differences between observers \cite{jassar2011mv} that propagate into model predictions \cite{voss2019match,krygier2021segmentation}. Furthermore, segmented valve configurations do not establish correspondence between material points, which complicates the assessment and validation of leaflet deformation. Together, these limitations make accurate image-based valve models challenging to build.
 
To overcome these challenges, previous studies have used hyperelastic warping, or ``shape matching'', to establish correspondence between imaged leaflet configurations \cite{rego2018noninvasive,tondi2025magnetic}. Others have constructed synthetic chordal architectures that produce functionally equivalent valve deformations without requiring the complete native chordal geometry \cite{khalighi2019chordae,liu_computational_2023}. In our prior work, we combined shape matching with synthetic chordal architectures to reproduce TV closure without knowledge of subject-specific chordal attachments \cite{Mathur2026}. However, we evaluated that framework using a single, high-fidelity model reconstructed from an explanted human valve. How these methods generalize to valves reconstructed from routine clinical images, and how their predictions depend on assumed tissue properties and observer-dependent geometry, remain unknown.
 
In this work, our objective was to develop and evaluate TV models using clinical images as the only source of patient-specific anatomical data, supplemented by population-based measurements of human tissue. To this end, we constructed FE models of 10 TVs from three-dimensional transesophageal echocardiography (3D-TEE). We prescribed leaflet thickness, leaflet and chordal material properties, chordal cross-sectional areas, and regional chordal insertion densities using \textit{ex vivo} human tissue data. We then simulated valve closure and assessed agreement with the imaged end-systolic configurations. Finally, in three valves, we investigated sensitivity to modeling assumptions over ranges informed by variability in the human data. We compared these effects with the variability arising from image segmentation by three independent observers. These analyses aimed to determine how reliably the models reproduce valve closure and which inputs most influence predicted valve mechanics. This workflow provides a foundation for building patient-specific TV models from routinely acquired clinical images, which can be used to investigate the mechanical effects of repair procedures.

\clearpage
\section{Materials and Methods}\label{M&Ms}
\begin{figure*}
  \centering
  \includegraphics[width=.9\textwidth]{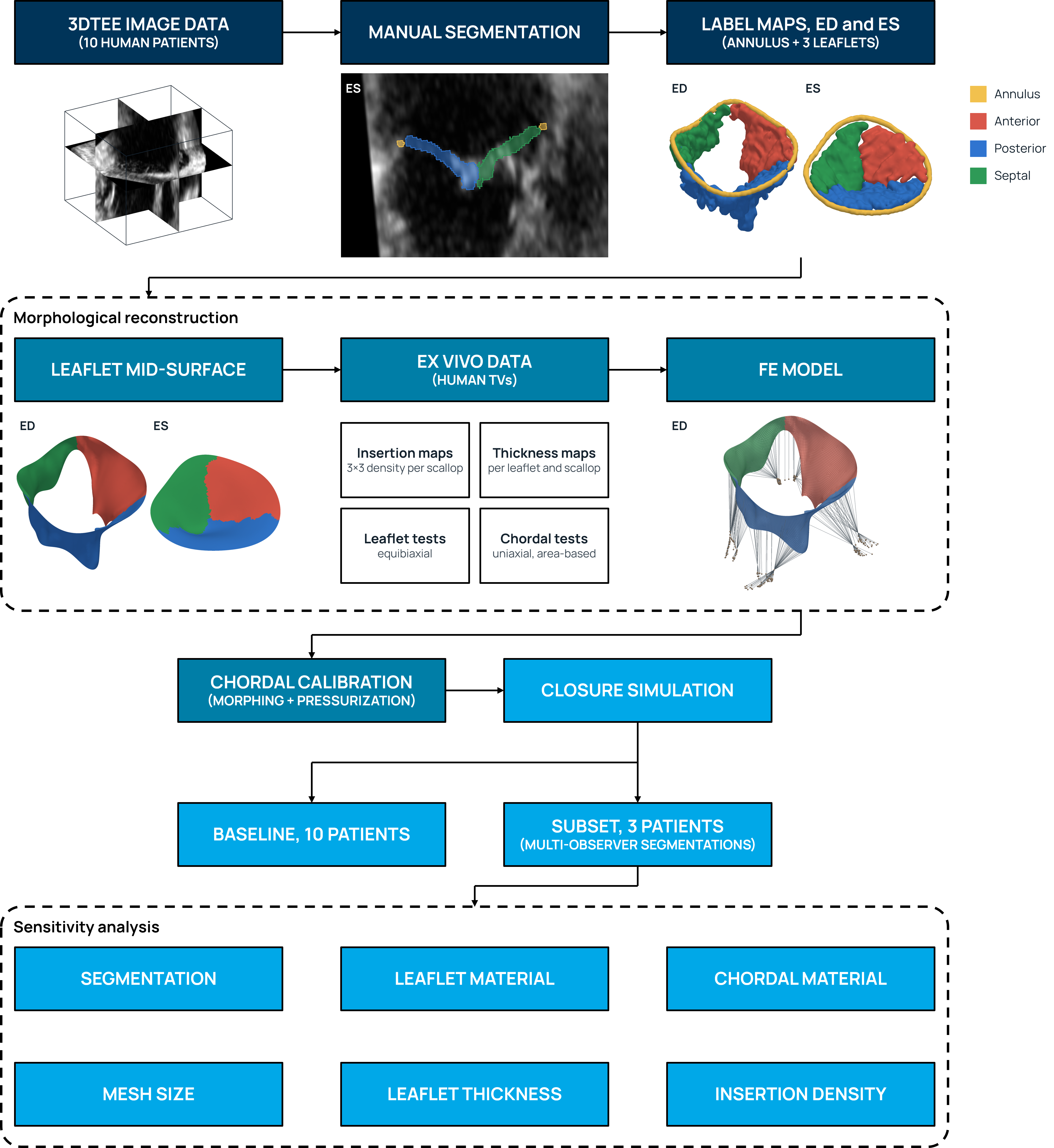}
  \caption{Patient-specific tricuspid valve (TV) modeling and simulation workflow. \textbf{Top row:} manual segmentation of the annulus and the three leaflets on the 3D-TEE images of each patient, at end-diastole (ED) and end-systole (ES). The images show the 3D-TEE volume at ES as three orthogonal slices, a long-axis slice with the manual label map, and the label maps at ED and ES (yellow: annulus; red: anterior; blue: posterior; green: septal leaflet). \textbf{Middle:} the leaflet mid-surfaces extracted at ED and ES. The ED mid-surface is combined with population data from \textit{ex vivo} human TVs (chordal insertion density maps, leaflet thickness maps, equibiaxial leaflet tests and uniaxial chordal tests) to build the FE model, shown with its shell mesh and chordae before chordal calibration. \textbf{Bottom:} the chordae are calibrated (morphing and pressurization setup simulations) before the closure simulation. Closure was simulated for all 10 patients (baseline), and for a subset of 3 patients with multi-observer segmentations. A one-at-a-time sensitivity analysis was run on six inputs.  All simulations were performed with the FE solver ABAQUS/Explicit 2023 (SIMULIA, Dassault Syst\`emes)}
  \label{fig1_workflow}
\end{figure*}

Figure \ref{fig1_workflow} summarizes the workflow used to build our image-based TV models. We manually segmented the geometry of the annulus and leaflets from three-dimensional transesophageal echocardiography (3D-TEE) at end-diastole (ED) and at end-systole (ES).  We automatically constructed FE models with prescribed leaflet thickness, chordal architecture, and tissue mechanical properties. We calibrated the stress-free length of the chordae tendineae (CT) and simulated TV closure from ED to ES. We then compared the predicted closed configuration to the segmented ES state.

\subsection{Clinical Imaging Dataset}\label{Datacollection}
We used anonymized preoperative 3D-TEE images of ten ($N = 10$) adult human TVs collected at Asklepios Hospital Harburg (Hamburg, Germany) and the University Hospital Schleswig-Holstein (Kiel, Germany). Data were used in accordance with an agreement between these institutions and the University of Texas at Austin. All patients presented with competent TVs, with no more than trace TR, but may have had other cardiac conditions. TEE images were acquired using Philips ultrasound systems and exported as three-dimensional Cartesian DICOM volumes. Voxel edge lengths ranged from 0.24-0.92 mm, with voxel diagonals of 0.57-1.13 mm. Synchronized electrocardiogram data were unavailable.

\subsection{Image segmentation}\label{Imseg}
We segmented the annulus and leaflets at ED and ES using 3D Slicer and SlicerHeart \cite{fedorov2012slicer,Lasso2022SlicerHeart} (Fig. \ref{fig1_workflow}). Frames were selected by visual inspection of the cine sequences. We identified ES as the last frame in which the TV remained closed immediately before valve opening. For ED, we chose the frame immediately before the onset of right ventricular contraction. When the leaflets could not be distinguished from the ventricular endocardium in this frame, we chose the closest subsequent frame in which the valve was clearly visible. The septum, the anterior papillary muscle, and the right-ventricular outflow tract guided valve orientation and leaflet identification following \cite{hahn2021nomenclature}. The commissures were defined as the shared boundaries between the adjacent leaflets at ED. Because coaptation obscured leaflet boundaries at ES, we used the commissures identified at ED to partition the ES reconstruction.

The baseline models used segmentations generated by Observer O1, who segmented all 10 valves. To assess inter-observer variability, O1 classified the datasets as having either good, average, or poor segmentation quality and randomly selected one valve from each group: TV3, TV6, and TV10, respectively. Two additional observers from different institutions independently segmented these three valves using the same ED and ES frames chosen by O1. Observers were blinded to one another’s segmentations and independently assigned leaflet subdivisions. These three valves also formed the sensitivity-analysis subset, see Section \ref{sec:sensitivity}.

\subsection{Geometry Reconstruction and Discretization}\label{TVED}
We processed the segmentation labels using a custom Python pipeline based on PyVista \cite{Sullivan2019PyVista}. At ED, leaflet surfaces were extracted, isolated segmentation fragments were removed, and the annulus was regularized. Radial cross-sections were used to reconstruct smooth leaflet mid-surfaces ($\Omega^\mathrm{ED}$). Leaflets were separated at the main commissures and divided into scallops at accessory commissures, when present, and each scallop was partitioned into a \(3\times3\) grid for assigning regional tissue data (Fig. \ref{fig_exvivo_inputs}b). The ED surfaces were discretized using four-node, reduced-integration shell elements with a reference size of 0.5 mm (S4R, Abaqus/Explicit 2023). Mesh sensitivity was assessed as described in Section \ref{sec:sensitivity}. We reconstructed the ES surface in the same local cylindrical coordinate system defined from the ED anatomy. Because the leaflets were in contact, their profiles were fitted and lofted into a continuous surface ($\Omega^\mathrm{ES}$) that served as the rigid geometric target for chordal calibration. Supplementary Section \ref{supp:geom} details the surface reconstruction, partitioning, and meshing procedures.

\subsection{Leaflet Thickness and Chordal Architecture}
\label{sec:thickness_chordae}
\begin{figure*}[p]
  \centering
  \includegraphics[width=\textwidth,height=0.825\textheight,keepaspectratio]{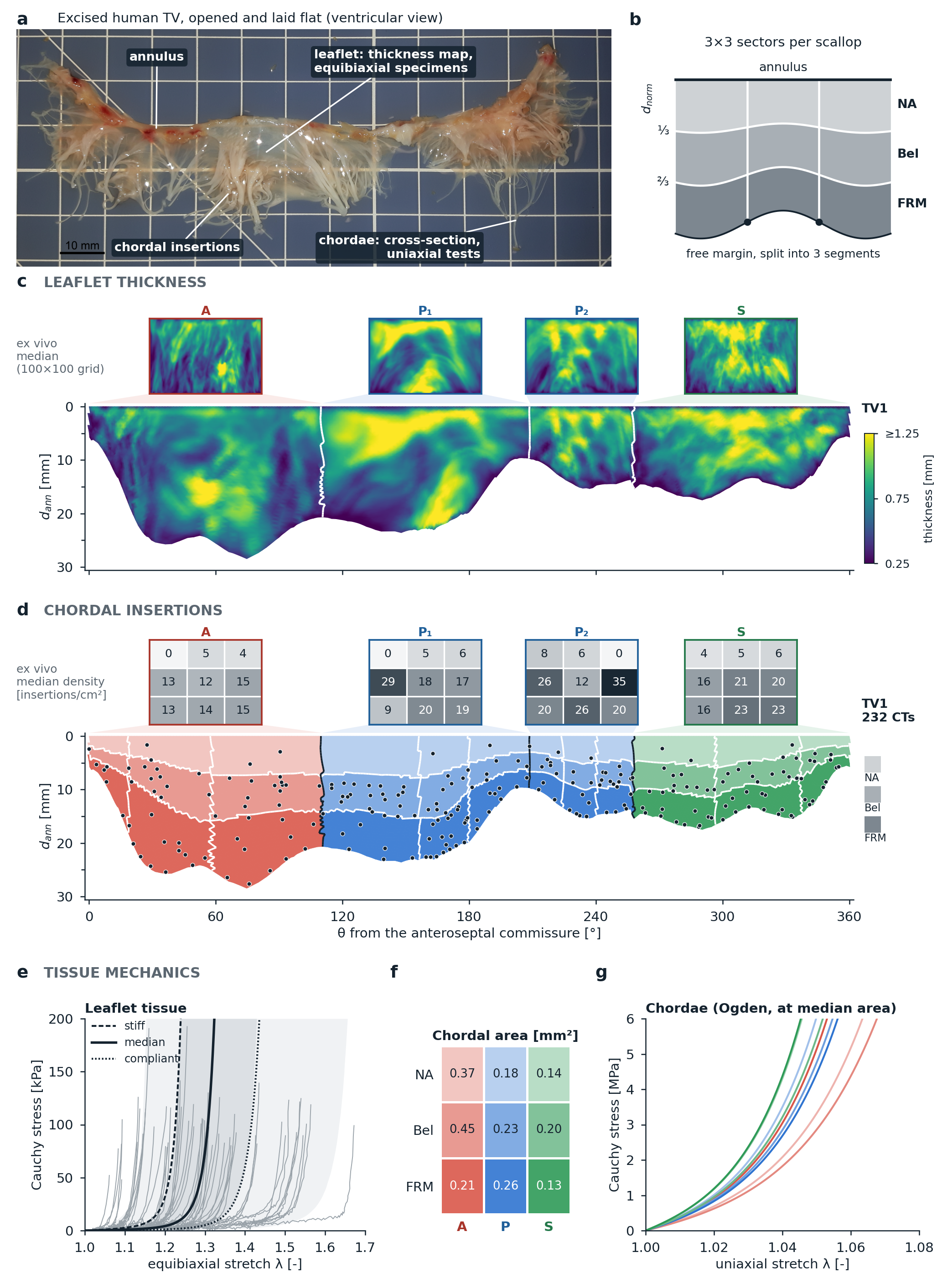}
    \caption{Population-based components of the TV models, derived from \textit{ex vivo} measurements on human TVs, and their transfer to a representative model (TV1). 
    \textbf{(a)} Excised human TV (ventricular view) \cite{haese2027chordae, haese_chordae_2026}. \textbf{(b)} Partition of each scallop into $3 \times 3$ sectors: near-annulus (NA), belly (Bel), and free-margin (FRM) bands at $d_{norm} = 1/3$ and $2/3$, and wedges bounded by the points dividing the free margin into three equal arc lengths. \textbf{(c)} Median leaflet thickness maps of the leaflet-and-scallop groups assigned to TV1, and the resulting nodal thickness of TV1. \textbf{(d)} Median chordal insertion density [insertions/cm$^2$] of the same groups, and the sectors and insertions of TV1. In (c) and (d), the leaflets of TV1 at ED are unrolled onto the angle $\theta$ about the valve axis, measured from the anteroseptal commissure, and the geodesic distance from the annulus $d_{ann}$; P$_1$ and P$_2$ are the two scallops of the posterior leaflet. \textbf{(e)} Equibiaxial response of the leaflet tissue: tested specimens (grey), 25\textsuperscript{th}-75\textsuperscript{th} and 5\textsuperscript{th}-95\textsuperscript{th} percentiles of the population-predictive response (shaded), and stiff, median, and compliant sets (Table \ref{tab:leaflet_params}). \textbf{(f)} Median chordal cross-sectional area per leaflet and band, and (g) the corresponding Ogden response (Table \ref{tab:material_params}).}
  \label{fig_exvivo_inputs}
\end{figure*}

We prescribed spatially varying leaflet thickness using maps measured from explanted human TVs \cite{kostelnik_leaflet_data_and_code_2026}. Maps from healthy valves were grouped by leaflet and scallop configuration, and cell-wise median thicknesses were mapped onto the patient-specific leaflet surfaces (Fig. \ref{fig_exvivo_inputs}c). Thicknesses below 0.25 mm were set to 0.25 mm. Section \ref{supp:thick} describes thickness measurement, map processing, and the treatment of unmatched scallop configurations.
 
We generated synthetic chordal insertions using regional insertion-density maps from 25 explanted human TVs \cite{haese2027chordae}. Maps were grouped by leaflet and scallop configuration and pooled across control and heart-failure specimens (Fig. \ref{fig_exvivo_inputs}d). Within each sector, the target number of insertions was determined by multiplying the measured insertion density by the reconstructed sector area. Insertions were sampled using a fixed random seed. Within the free edge sectors, we placed 40\% of the assigned chordal insertions at equal intervals along the free edge and sampled the remaining 60\% on the leaflet surface inward from the edge.

We could not reliably identify papillary muscles (PMs) from the images. We therefore prescribed chordal origins using three idealized papillary bodies represented by arcs beneath the commissures, with additional arcs for accessory commissures as needed. Their systolic motion was derived from annular and commissural motion while maintaining a fixed distance from the annular plane. Section \ref{supp:pm} details the geometric construction and assumptions of its location and motion. Each insertion was connected to its assigned papillary arc by a single unbranched chorda, discretized using two-node truss elements with a reference length of 2 mm (T3D2, Abaqus). Chordal cross-sectional areas were assigned by leaflet and radial band position based on human measurements \cite{haese2027chordae}. Chordal forces were distributed over a small surrounding leaflet region through coupling constraints. See Section \ref{supp:insertion} for further details on insertion sampling and coupling.

\subsection{Constitutive Modeling}
\label{sec:constitutive}
We modeled leaflet tissue as a nearly incompressible, isotropic hyperelastic material with strain-energy density given by Eq. \ref{eq:leaflet_W}, where $I_1$ is the first invariant of the right Cauchy-Green deformation tensor and $C_0$, $C_1$, and $C_2$ are material parameters. 
\begin{equation}
\label{eq:leaflet_W}
    W(I_1)=\frac{C_0}{2}(I_1-3)+\frac{C_1}{2}\left[\exp\left(C_2(I_1-3)^2\right)-1\right]
\end{equation}
Parameters were identified from the final equibiaxial unloading cycle of 28 leaflet specimens from five healthy subjects \cite{kostelnik_leaflet_data_and_code_2026}. We fitted both in-plane Cauchy stress components using a Bayesian multilevel model \cite{bol2019bayesian, pensalfini2023} and assigned the same pooled material properties to all leaflets. We selected representative stiff, median, and compliant sets from joint population-predictive draws by matching the 25th, 50th, and 75th percentiles of predicted equibiaxial stretch at fixed stress, respectively (Fig. \ref{fig_exvivo_inputs}e). Section \ref{supp:mlm} describes the tests, Bayesian fit, and selection procedure. Table \ref{tab:leaflet_params} reports the parameter sets.
 
Chordae were modeled using a nearly incompressible, one-term Ogden formulation in the Abaqus convention. Parameters were obtained from the area-dependent hierarchical Bayesian model reported by Haese et al. \cite{haese2027chordae}. Each chorda received the median cross-sectional area of its leaflet and radial band and the representative constitutive parameters predicted at that area, with valve and specimen effects set to zero (Fig. \ref{fig_exvivo_inputs}f and Fig. \ref{fig_exvivo_inputs}g). Circumferential sectors within a radial band shared the same properties. Section \ref{supp:ctmat} and Table \ref{tab:material_params} give the constitutive equations and regional parameter sets.

\subsection{Chordal Calibration and Valve Closure}
\label{sec:calibration}
We calibrated chordal stress-free lengths using shape matching and force-based length adjustment \cite{Mathur2026, tondi2026fwa}. First, the ED leaflet mesh was driven toward $\Omega^\mathrm{ES}$ using a rigid contact target, ventricular pressure, and apically directed free-margin forces. Annular nodes and chordal-origin nodes were moved from their ED to ES positions during this step. The resulting morphed leaflet configuration ($\Omega_\mathrm{M}^\mathrm{ES}$) provided estimates of the loaded chordal lengths at ES. Second, we used these lengths in a pressurization simulation with nearly inextensible chordae and a transvalvular pressure of 22 mmHg, i.e. the end-systolic transvalvular pressure measured in an \textit{ex vivo} beating human heart \cite{mathur2022trivalve}. The resulting chordal forces were divided by the assigned cross-sectional areas to calculate nominal axial stresses. We inverted the chordal constitutive relation to obtain axial stretch and calculated each stress-free length as its loaded ES length divided by that stretch.
 
We then simulated closure using the calibrated lengths and hyperelastic tissue properties. During initialization, we held the leaflets fixed and moved each chordal origin to its prescribed ED position if the move could be made without stretching the chorda. All chordal origins were moved to their ES positions during closure. Meanwhile, the transvalvular pressure was increased smoothly to 22 mmHg, and prescribed annular displacements and assumed papillary motion moved the model from ED to ES. Leaflet contact was modeled using penalty contact with a friction coefficient of 0.05. All simulations were performed in Abaqus/Explicit 2023. We assessed energy histories to evaluate inertial effects, numerical stabilization, and energy balance. Further details on calibration loads, displacement mapping, solver settings, and energy histories are provided in Section \ref{supp:numerics}.

\subsection{Output Metrics}
\label{sec:metrics}
We assessed the simulated ES configuration using geometric error, coaptation area, maximum principal stress, and major principal stretch. Because the 3D-TEE images did not resolve the individual leaflet surfaces within the coaptation region, we instead used the morphed ES configuration, which includes the coapting portion of each leaflet, as the reference for geometric error. We defined geometric error, $\varepsilon_\mathrm{D}$, as the shortest distance from each node of the simulated ES leaflet mesh to the morphed surface. Because the morphed configuration also informed chordal length calibration, this metric measures how closely the closure simulation reproduces that configuration. We also evaluated agreement over the portion of the valve visible in 3D-TEE at ES by measuring the shortest distances from the image-derived ES surface to the simulated leaflet surface, ($\varepsilon_\mathrm{ES}$; Sections \ref{supp:metrics} and \ref{supp:eps}).

Coaptation area, $A_\mathrm{C}$, was defined as the leaflet surface area in contact at ES, reported in absolute units and as a fraction of the total ES leaflet area. We calculated this area by interpolating nodal contact indicators within each element, see Section \ref{supp:metrics}. Maximum principal stress, $S_\mathrm{I}$, was defined as the in-plane maximum principal Cauchy stress at the leaflet mid-surface. Major principal stretch, $\lambda_\mathrm{I}$, described the maximum in-plane principal stretch from ED to ES. We summarized maximum principal stress, major principal stretch, and geometric error using the median and interquartile range (25\textsuperscript{th}-75\textsuperscript{th} percentiles). We calculated these summaries across all leaflets combined and for each leaflet separately. For stress and stretch, we weighted each value by its element area. For geometric error, statistics were computed over the leaflet nodes without weighting.

\subsection{Baseline Simulations and Sensitivity Analysis}
\label{sec:sensitivity}
We simulated all 10 valves using O1 segmentations, median thickness and insertion-density maps, median regional chordal areas, and representative median constitutive responses. Sensitivity analyses were performed on TV3, TV6, and TV10, for which independent segmentations were available. To this end, we first assessed mesh sensitivity using leaflet element sizes of 0.3–0.8 mm. Chordal insertions were transferred between meshes, and chordal origins and stress-free lengths were held fixed. We selected the coarsest mesh for which geometric error and coaptation area differed by no more than 5\%, and median principal stress differed by no more than 10\%, from the 0.3 mm reference model.

We assessed segmentation variability by repeating the complete modeling workflow for each additional observer’s segmentation. We then varied four population-based inputs one at a time while holding the remaining inputs at baseline values: 1) chordal insertion density, 2) leaflet material properties, 3) leaflet thickness, and 4) chordal size and material properties. Insertion-density and thickness maps were varied between their first and third quartiles. Leaflet material properties were varied between the stiff and compliant sets, i.e. the representative draws matching the first and third quartiles of the predicted equibiaxial stretch, respectively. Chordal areas were varied between their first and third quartiles together with the constitutive parameters predicted at those areas, preserving their measured size–material relationship \cite{haese2027chordae}. Table \ref{tab:sensitivity} summarizes the factors and levels used in the sensitivity study.

For each population-based input, we compared geometric error, coaptation area, and median principal stress with the corresponding baseline model. We compared the range across the tested levels with the range across the inter-observer models for the same valve. Section \ref{supp:sens} provides details on mesh transfer and the calculation of observer variability. Inter-observer variability was summarized using the coefficient of variation (CV), calculated as the sample standard deviation divided by the mean. Associations between model outputs and segmentation-derived quantities across the nine models were assessed using repeated-measures correlation \cite{bakdash2017rmcorr}, accounting for grouping by valve.

\newpage

\clearpage

\section{Results}\label{results}

All simulations reached the ES configuration with the leaflets in contact. A reference element size of 0.5 mm was the coarsest that met all the tolerances specified in Section \ref{sec:sensitivity}: relative to the 0.3 mm models, the median of $S_\mathrm{I}$ differed by $-7$ to $+8$\% (at most 1.0 kPa), the median of $\varepsilon_\mathrm{D}$ by $-3$ to $+1$\% (at most 0.04 mm), and $A_\mathrm{C}$ by $-1$ to $+4$\%, while the closure simulations took 31--37 instead of 118--140 min on 20 CPU cores (Section \ref{supp:res_mesh}, Fig. \ref{fig_mesh}). Leaflet-wise medians of $S_\mathrm{I}$ were more sensitive, changing by up to 32\%, so that 0.5 mm elements are adequate for whole-valve outputs.

\subsection{The models reproduced the imaged closed geometry with millimeter-scale error}\label{res:baseline}
\begin{figure*}
  \centering
  \includegraphics[width=\textwidth]{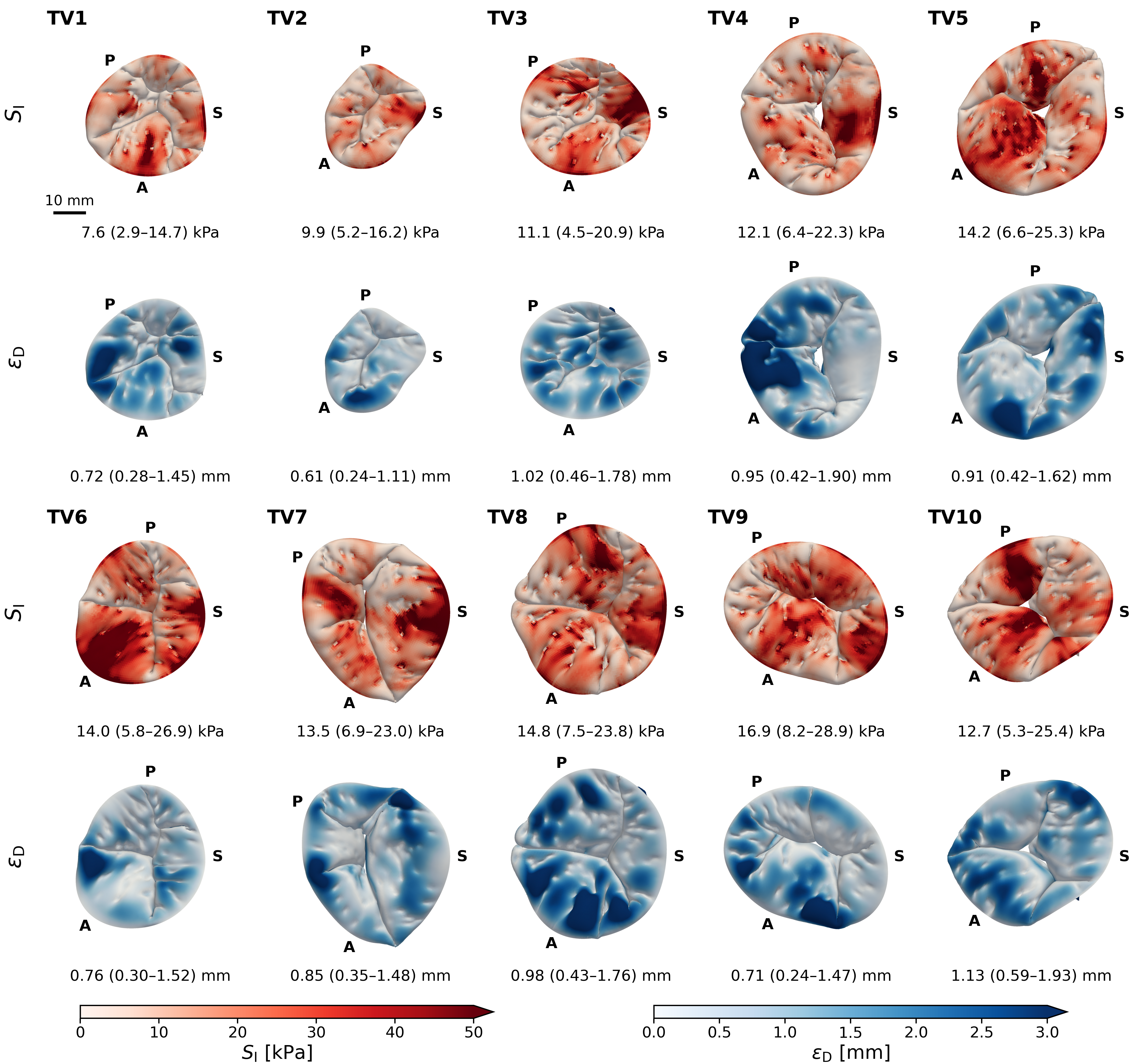}
  \caption{Simulated closure of the ten TVs at ES, atrial view, septal leaflet on the right, all valves at the same scale. For each valve, top: maximum principal stress $S_\mathrm{I}$ at the leaflet mid-surface; bottom: geometric error $\varepsilon_\mathrm{D}$. Values below each map are median (IQR); $S_\mathrm{I}$ is weighted by element area, $\varepsilon_\mathrm{D}$ is computed over the nodes. A: anterior, P: posterior, S: septal leaflet.}
  \label{fig_cohort_maps}
\end{figure*}

\begin{figure*}
  \centering
  \includegraphics[width=.9\textwidth]{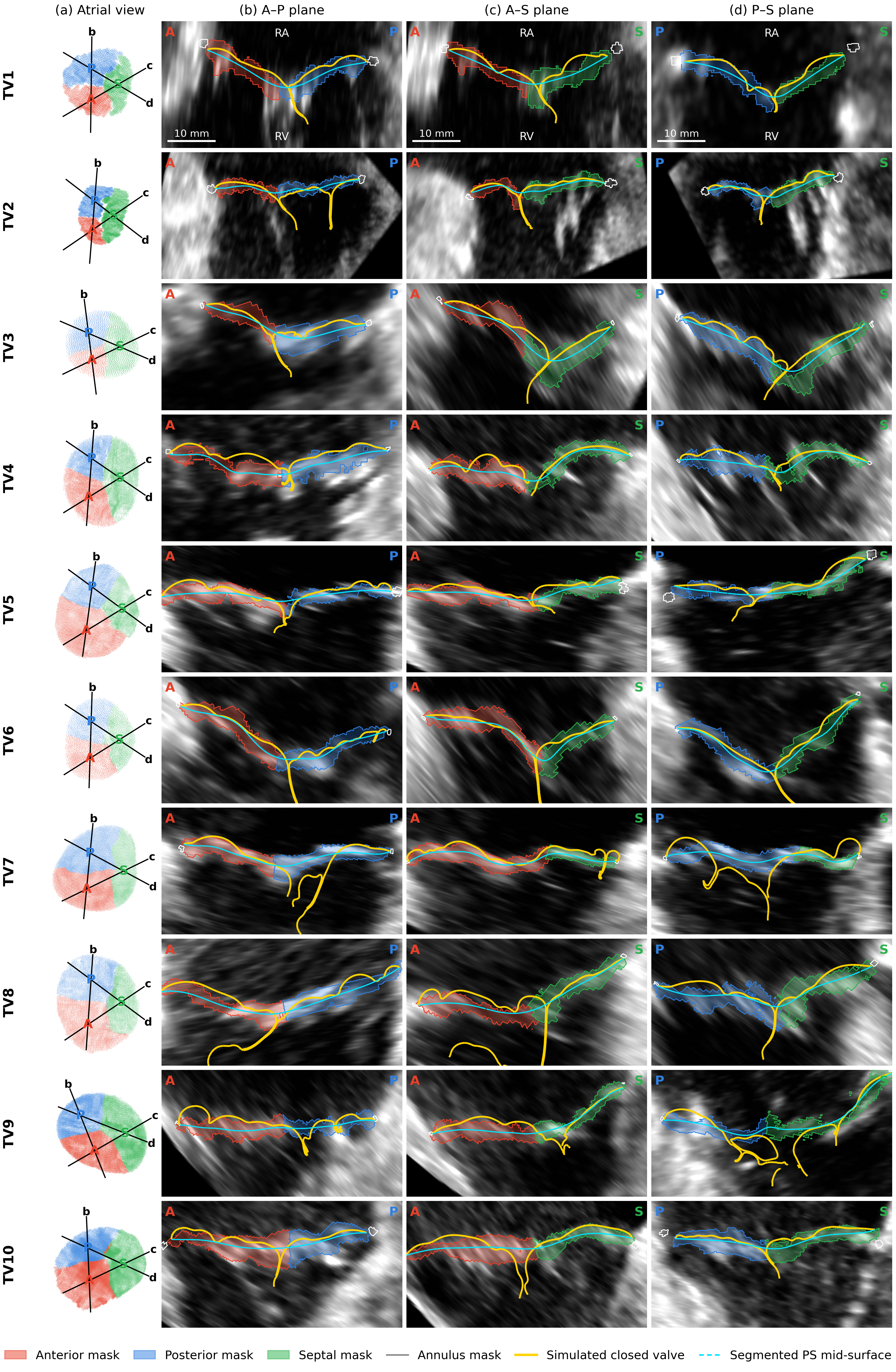}
  \caption{Simulated ES configuration against the 3D-TEE data of the ten TVs. \textbf{(a)} ES segmentation in atrial view, septal leaflet on the right, with the traces of the three cut planes. \textbf{(b)}-\textbf{(d)} Echo cut planes containing the valve axis and the centroids of two leaflets: A-P \textbf{(b)}, A-S \textbf{(c)}, and P-S \textbf{(d)}, atrium at the top. Overlaid: leaflet labels (fill and outline), annulus label (white), simulated ES configuration (yellow), and $\Omega^\mathrm{ES}$ (cyan, dashed). All planes share the same scale and vertical window.}
  \label{fig_cohort_sections}
\end{figure*}

In the ten baseline simulations (Fig. \ref{fig_cohort_maps}, Table \ref{tab:cohort}), the per-valve median of $\varepsilon_\mathrm{D}$ ranged from 0.61 to 1.13 mm (cohort median 0.88 mm), i.e. 0.7--1.6 voxel diagonals of the 3D-TEE volumes, and was highest in the band 3 to 6 mm from the annulus in all ten valves (Section \ref{supp:eps}). Agreement with the image-derived ES surface was similar: the per-valve medians of $\varepsilon_\mathrm{ES}$ were 0.55--0.95 mm (Fig. \ref{figS_epsES}). The simulated leaflets followed the segmented geometry from the annulus through the leaflet bellies, while their coapting portions extended beyond the region resolved by 3D-TEE (Fig. \ref{fig_cohort_sections}). Outside the coaptation region, they lay on the atrial side of the segmented surface, with median signed distances of $+0.33$ to $+0.90$ mm (Table \ref{tab:S_position_baseline}).

Six valves closed with residual gaps below 0.1 mm$^2$ (TV1, TV2, TV3, TV6, TV8, TV9); the other four had a small central orifice of 1.7--11.8 mm$^2$, at most 0.7\% of the projected annular area (TV4). Coaptation covered 20--42\% of the leaflet area at ES (cohort median 32\%). The per-valve median of $S_\mathrm{I}$ ranged from 7.6 to 16.9 kPa (cohort median 13.1 kPa), with 95\textsuperscript{th} percentiles of 30--57 kPa and local peaks at the CT insertions; no leaflet was consistently the most stressed. Larger valves were more stressed: the median of $S_\mathrm{I}$ correlated with the projected annular area at ES (Spearman $\rho = 0.79$, $p = 0.006$). The chordal apparatus carried 74--92\% of the transvalvular load, with a total chordal force of 1.8--5.9 N per valve (median 4.6 N). Leaflet stretch, regional stress, and chordal loading are reported in Section \ref{supp:bio}.

\begin{table}[t]
\centering
\resizebox{\textwidth}{!}{%
\begin{tabular}{lcccccc}
\toprule
Valve & $S_\mathrm{I}$ [kPa] & $\varepsilon_\mathrm{D}$ [mm] & $A_\mathrm{C}$ [mm$^2$ (\%)] & $N_\mathrm{CT}$ & $F_\mathrm{CT}$ [mN] & CT load [\%] \\
\midrule
TV1 & 7.6 (2.9--14.7) & 0.72 (0.28--1.45) & 858 (35) & 232 & 7.7 (1.1--19.3) & 83 \\
TV2 & 9.9 (5.2--16.2) & 0.61 (0.24--1.11) & 461 (31) & 127 & 10.9 (3.4--20.5) & 74 \\
TV3 & 11.1 (4.5--20.9) & 1.02 (0.46--1.78) & 981 (35) & 270 & 7.7 (0.3--21.9) & 92 \\
TV4 & 12.1 (6.4--22.3) & 0.95 (0.42--1.90) & 604 (20) & 250 & 10.2 (0.3--24.8) & 79 \\
TV5 & 14.2 (6.6--25.3) & 0.91 (0.42--1.62) & 761 (24) & 257 & 12.3 (4.3--28.7) & 82 \\
TV6 & 14.0 (5.8--26.9) & 0.76 (0.30--1.52) & 1674 (42) & 370 & 5.2 (0.0--17.1) & 89 \\
TV7 & 13.5 (6.9--23.0) & 0.85 (0.35--1.48) & 649 (20) & 295 & 10.9 (1.0--24.4) & 85 \\
TV8 & 14.8 (7.5--23.8) & 0.98 (0.43--1.76) & 1451 (33) & 344 & 7.7 (0.6--22.1) & 87 \\
TV9 & 16.9 (8.2--28.9) & 0.71 (0.24--1.47) & 899 (26) & 247 & 13.8 (2.2--26.4) & 79 \\
TV10 & 12.7 (5.3--25.4) & 1.13 (0.59--1.93) & 1367 (36) & 320 & 6.6 (0.4--22.0) & 85 \\
\midrule
Cohort median & 13.1 & 0.88 & 879 (32) & 264 & 9.0 & 84 \\
\bottomrule
\end{tabular}
}
\caption{Baseline simulations. $S_\mathrm{I}$, $\varepsilon_\mathrm{D}$, $F_\mathrm{CT}$: median (IQR) ($S_\mathrm{I}$ weighted by element area, $\varepsilon_\mathrm{D}$ over the nodes, $F_\mathrm{CT}$ over the CTs); $A_\mathrm{C}$: coaptation area, in mm$^2$ and as a percentage of the leaflet area at ES; $N_\mathrm{CT}$: number of CTs; $F_\mathrm{CT}$: force per CT; CT load: share of the transvalvular load on the leaflets carried by the chordal apparatus. Cohort median: median across valves of the per-valve values.}
\label{tab:cohort}
\end{table}

\subsection{Segmentation affected coaptation more than the other outputs}\label{res:operators}
\begin{figure*}
  \centering
  \includegraphics[width=\textwidth]{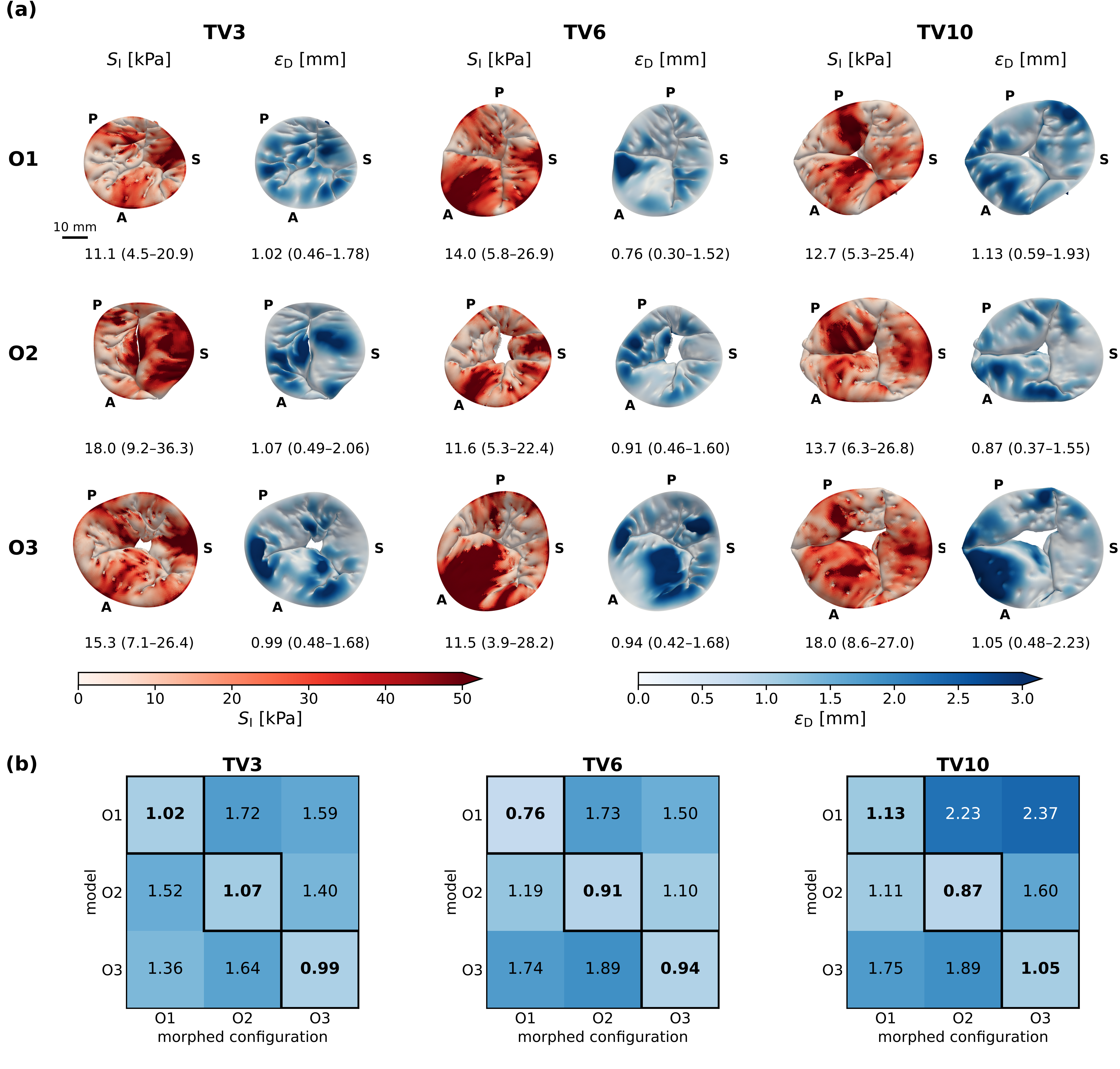}
  \caption{Inter-observer variability of the simulated closure of TV3, TV6, and TV10. \textbf{(a)} Maximum principal stress $S_\mathrm{I}$ at the leaflet mid-surface and geometric error $\varepsilon_\mathrm{D}$ at ES, for the models built from the segmentations of the reference observer (O1) and of the two additional observers (O2, O3); atrial view, septal leaflet on the right, all models at the same scale. A: anterior, P: posterior, S: septal leaflet. \textbf{(b)} For each valve, median distance of the nodes of the simulated ES configuration of each model (rows) from the morphed configuration $\Omega_\mathrm{M}^\mathrm{ES}$ obtained from the segmentation of each observer (columns), with all configurations expressed in the common 3D-TEE image space. Diagonal entries (bold) compare each model with its own morphed configuration and coincide with the median $\varepsilon_\mathrm{D}$ of panel (a); off-diagonal entries compare it with the morphed configurations of the other two observers.}
  \label{fig_operators}
\end{figure*}

All nine models of the sensitivity subset built from the segmentations of the three observers reached the ES configuration with the leaflets in contact (Fig. \ref{fig_operators}, Table \ref{tab:operators}). The observers segmented similar leaflet surfaces, 0.54--1.11 mm apart at ES (Dice similarity coefficient of the merged leaflet labels 0.64--0.73; Table \ref{tab:S_seg_pairs}), but differed in the extent of the leaflets and in the borders between them. At ED, the leaflet area varied more across observers (CV 10--19\%) than the annular diameters, perimeter, and projected area (CV 0--5\%); at ES, the CV of the annular dimensions reached 7\% for the diameters and perimeter and 13\% for the projected area (Tables \ref{tab:operators} and \ref{tab:S_seg_operators}).

The coaptation area was the output most affected by segmentation (CV 20--37\%, against 11--23\% for the median of $S_\mathrm{I}$ and 4--13\% for the median of $\varepsilon_\mathrm{D}$). In TV6, the range of $A_\mathrm{C}$ across observers covered 72\% of the range across the ten baseline valves. Within valves, $A_\mathrm{C}$ correlated with the ratio of the leaflet area at ED to the projected annular area at ES (repeated-measures $r = 0.89$, $p = 0.007$), less strongly with the leaflet area alone ($r = 0.79$, $p = 0.03$), and not with the projected annular area ($r = -0.12$, $p = 0.79$). Three models closed with a central orifice larger than any baseline model, up to 62 mm$^2$ (3.2\% of the projected annular area). The geometric error varied little across observers (median $\varepsilon_\mathrm{D}$ 0.76--1.13 mm) because each model is compared with the morphed configuration of its own observer; measured against the morphed configurations of the other observers, the median error was 1.10--2.37 mm (Fig. \ref{fig_operators}b). Model-level details and the segmentation-level comparison are reported in Sections \ref{supp:res_seg} and \ref{supp:seg}.

\begin{table}[t]
\centering
\resizebox{\textwidth}{!}{%
\begin{tabular}{llcccccccc}
\toprule
 & & & & & & \multicolumn{2}{c}{Annulus, ED [mm]} & \multicolumn{2}{c}{Annulus, ES [mm]} \\
\cmidrule(lr){7-8}\cmidrule(lr){9-10}
Valve & Observer & $S_\mathrm{I}$ [kPa] & $\varepsilon_\mathrm{D}$ [mm] & $A_\mathrm{C}$ [mm$^2$ (\%)] & $A^\mathrm{ED}$ [mm$^2$] & $D_\mathrm{SL}$ & $D_\mathrm{AP}$ & $D_\mathrm{SL}$ & $D_\mathrm{AP}$ \\
\midrule
TV3 & O1 & 11.1 (4.5--20.9) & 1.02 (0.46--1.78) & 981 (35) & 2260 & 44.7 & 43.6 & 41.8 & 37.9 \\
 & O2 & 18.0 (9.2--36.3) & 1.07 (0.49--2.06) & 683 (28) & 1923 & 45.0 & 44.2 & 42.6 & 42.0 \\
 & O3 & 15.3 (7.1--26.4) & 0.99 (0.48--1.68) & 742 (23) & 2515 & 47.0 & 47.2 & 47.6 & 43.3 \\
 & CV [\%] & 23 & 4 & 20 & 13 & 3 & 4 & 7 & 7 \\
\midrule
TV6 & O1 & 14.0 (5.8--26.9) & 0.76 (0.30--1.52) & 1674 (42) & 3137 & 48.8 & 43.8 & 38.9 & 46.0 \\
 & O2 & 11.6 (5.3--22.4) & 0.91 (0.46--1.60) & 804 (31) & 2133 & 46.6 & 44.4 & 41.9 & 42.0 \\
 & O3 & 11.5 (3.9--28.2) & 0.94 (0.42--1.68) & 1106 (31) & 2846 & 45.1 & 43.6 & 42.3 & 47.2 \\
 & CV [\%] & 11 & 11 & 37 & 19 & 4 & 1 & 5 & 6 \\
\midrule
TV10 & O1 & 12.7 (5.3--25.4) & 1.13 (0.59--1.93) & 1367 (36) & 2937 & 53.6 & 49.0 & 51.0 & 45.2 \\
 & O2 & 13.7 (6.3--26.8) & 0.87 (0.37--1.55) & 874 (28) & 2408 & 54.8 & 47.6 & 51.7 & 46.4 \\
 & O3 & 18.0 (8.6--27.0) & 1.05 (0.48--2.23) & 916 (25) & 2754 & 52.3 & 48.1 & 51.9 & 47.5 \\
 & CV [\%] & 19 & 13 & 26 & 10 & 2 & 1 & 1 & 2 \\
\bottomrule
\end{tabular}
}
\caption{Simulations of TV3, TV6, and TV10 built from the segmentations of the reference observer (O1) and of the two additional observers (O2, O3). $S_\mathrm{I}$, $\varepsilon_\mathrm{D}$: median (IQR) ($S_\mathrm{I}$ weighted by element area, $\varepsilon_\mathrm{D}$ over the nodes); $A_\mathrm{C}$: coaptation area, in mm$^2$ and as a percentage of the leaflet area at ES; $A^\mathrm{ED}$: leaflet area of the model at ED; $D_\mathrm{SL}$, $D_\mathrm{AP}$: septal-lateral and anteroposterior diameters of the segmented annulus at ED and ES, i.e. the chords of the annular curve, projected on the common annular plane of the valve, through its centroid, along the direction of the middle of the septal annular arc of O1 and perpendicular to it. CV: coefficient of variation across the three observers (for $S_\mathrm{I}$ and $\varepsilon_\mathrm{D}$, of the medians). O1 rows of $S_\mathrm{I}$, $\varepsilon_\mathrm{D}$, and $A_\mathrm{C}$ coincide with Table \ref{tab:cohort}.}
\label{tab:operators}
\end{table}

\subsection{Insertion density and leaflet stiffness had the largest effects on geometric error}\label{res:density}
Measured as the difference between the extreme models of each valve, and compared with the spread across observers (Table \ref{tab:operators}), insertion density and leaflet stiffness had the largest effects on the geometric error: they changed the median of $\varepsilon_\mathrm{D}$ by 0.45--0.58 and 0.36--0.56 mm, respectively, against 0.08--0.26 mm across observers. Their effects on the other outputs were smaller or comparable: insertion density changed the median of $S_\mathrm{I}$ more than segmentation only in TV6 and TV10 (5.9--7.5 vs. 2.5--5.3 kPa) and $A_\mathrm{C}$ much less (68--149 vs. 298--870 mm$^2$), whereas leaflet stiffness changed $A_\mathrm{C}$ by an amount of the same order (391--612 mm$^2$) and the median of $S_\mathrm{I}$ more than segmentation only in TV6. Leaflet thickness changed the median of $S_\mathrm{I}$ as much as segmentation did (4.6--6.5 vs. 2.5--6.9 kPa) but $\varepsilon_\mathrm{D}$ and $A_\mathrm{C}$ far less (at most 0.08 mm and 66 mm$^2$), and chordal size and material changed all three outputs far less (at most 0.03 mm, 1.6 kPa, and 17 mm$^2$).

\begin{figure*}
  \centering
  \includegraphics[width=.7\textwidth]{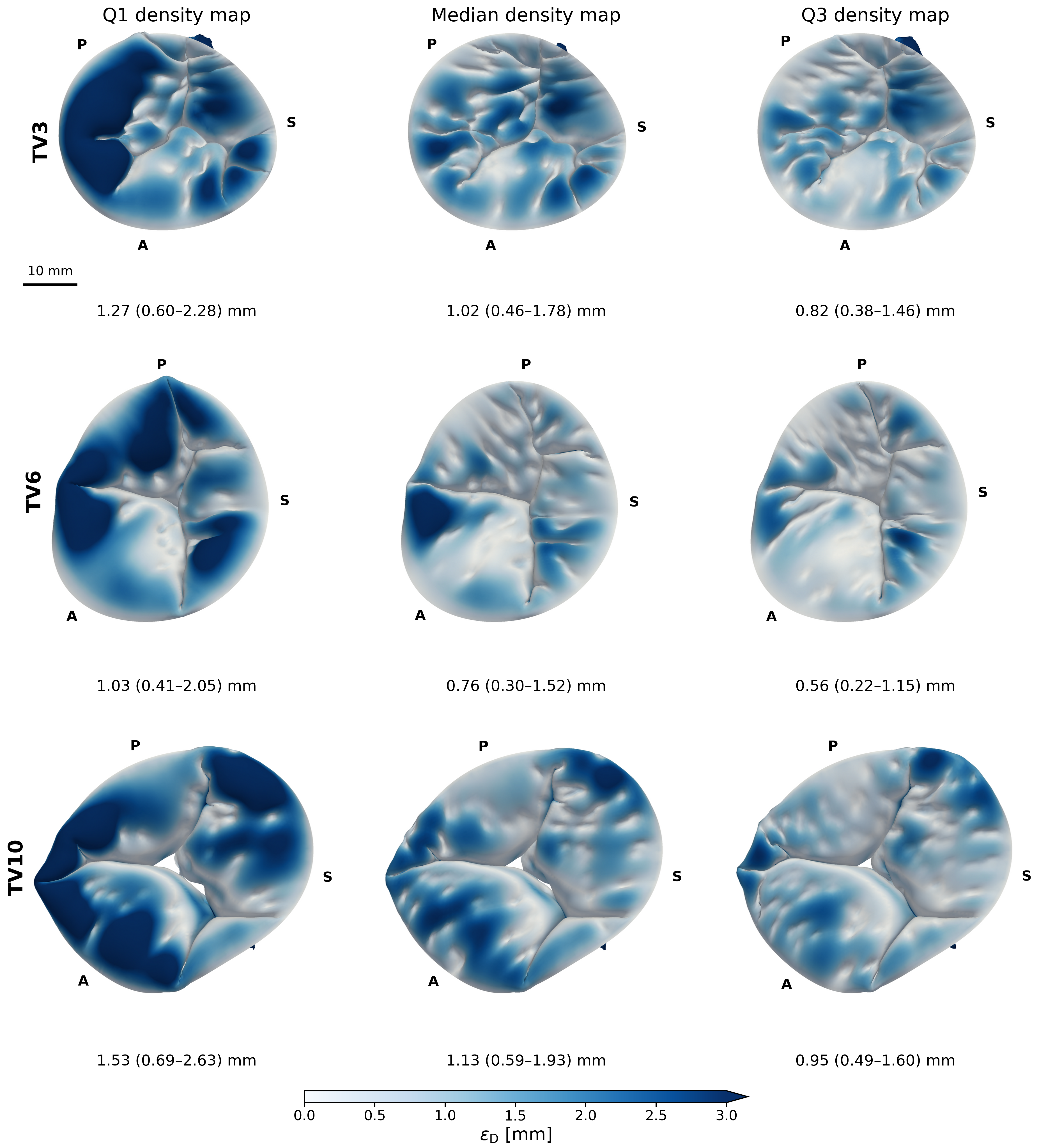}
  \caption{Geometric error $\varepsilon_\mathrm{D}$ at ES of TV3, TV6, and TV10 (rows) simulated with the first-quartile ($Q_1$), median, and third-quartile ($Q_3$) chordal insertion density maps (columns); atrial view, septal leaflet on the right, all models at the same scale. Values below each map are median (IQR) over the nodes. A: anterior, P: posterior, S: septal leaflet.}
  \label{fig_density}
\end{figure*}

\paragraph{Insertion density.}
The $Q_1$ maps yielded 162--224 CTs per model and the $Q_3$ maps 379--508 (Table \ref{tab:density}). Increasing insertion density reduced the geometric error, the leaflet stress, and the force per CT in all three valves (Fig. \ref{fig_density}). Relative to baseline, the $Q_1$ maps increased the median of $\varepsilon_\mathrm{D}$ by 24--36\% and that of $S_\mathrm{I}$ by 18--40\%, with the error concentrated in the leaflet bellies, whereas the $Q_3$ maps reduced them by 16--26\% and 13--24\%; the $Q_3$ maps also reduced $\varepsilon_\mathrm{ES}$ by 19--23\% (Table \ref{tab:S_position_sens}). The share of the transvalvular load carried by the chordal apparatus increased from 73--80\% with the $Q_1$ maps to 87--100\% with the $Q_3$ maps, whereas $A_\mathrm{C}$ changed by less than 8\%. Since each insertion carried its own CT, the total chordal cross-section also increased with density, by 2.4--2.6 times from the $Q_1$ to the $Q_3$ maps (Section \ref{supp:res_density}).

\begin{table}[t]
\centering
\resizebox{\textwidth}{!}{%
\begin{tabular}{llcccccc}
\toprule
Valve & Density & $N_\mathrm{CT}$ & $S_\mathrm{I}$ [kPa] & $\varepsilon_\mathrm{D}$ [mm] & $A_\mathrm{C}$ [mm$^2$ (\%)] & $F_\mathrm{CT}$ [mN] & CT load [\%] \\
\midrule
TV3  & $Q_1$     & 162 & 14.6 (6.0--25.1) & 1.27 (0.60--2.28) & 964 (34)  & 14.3 (2.5--28.1) & 80 \\
     & Median & 270 & 11.1 (4.5--20.9) & 1.02 (0.46--1.78) & 981 (35)  & 7.7 (0.3--21.9)  & 92 \\
     & $Q_3$     & 379 & 9.7 (3.5--19.5)  & 0.82 (0.38--1.46) & 1054 (38) & 4.5 (0.1--16.2)  & 100 \\
\midrule
TV6  & $Q_1$     & 224 & 16.5 (6.6--33.7) & 1.03 (0.41--2.05) & 1650 (42) & 8.1 (0.6--17.8)  & 75 \\
     & Median & 370 & 14.0 (5.8--26.9) & 0.76 (0.30--1.52) & 1674 (42) & 5.2 (0.0--17.1)  & 89 \\
     & $Q_3$     & 508 & 10.6 (4.7--21.8) & 0.56 (0.22--1.15) & 1718 (44) & 3.8 (0.0--11.9)  & 90 \\
\midrule
TV10 & $Q_1$     & 197 & 17.8 (7.2--30.7) & 1.53 (0.69--2.63) & 1282 (34) & 10.6 (2.3--28.7) & 73 \\
     & Median & 320 & 12.7 (5.3--25.4) & 1.13 (0.59--1.93) & 1367 (36) & 6.6 (0.4--22.0)  & 85 \\
     & $Q_3$     & 467 & 10.3 (4.3--21.0) & 0.95 (0.49--1.60) & 1431 (39) & 3.1 (0.0--15.3)  & 87 \\
\bottomrule
\end{tabular}
}
\caption{Simulations of TV3, TV6, and TV10 with the first-quartile ($Q_1$), median, and third-quartile ($Q_3$) chordal insertion density maps. $N_\mathrm{CT}$: number of CTs; $S_\mathrm{I}$, $\varepsilon_\mathrm{D}$, $F_\mathrm{CT}$: median (IQR) ($S_\mathrm{I}$ weighted by element area, $\varepsilon_\mathrm{D}$ over the nodes, $F_\mathrm{CT}$ over the CTs); $A_\mathrm{C}$: coaptation area, in mm$^2$ and as a percentage of the leaflet area at ES; CT load: share of the transvalvular load on the leaflets carried by the chordal apparatus. Median rows coincide with Table \ref{tab:cohort}.}
\label{tab:density}
\end{table}

\begin{figure*}
  \centering
  \includegraphics[width=.8\textwidth]{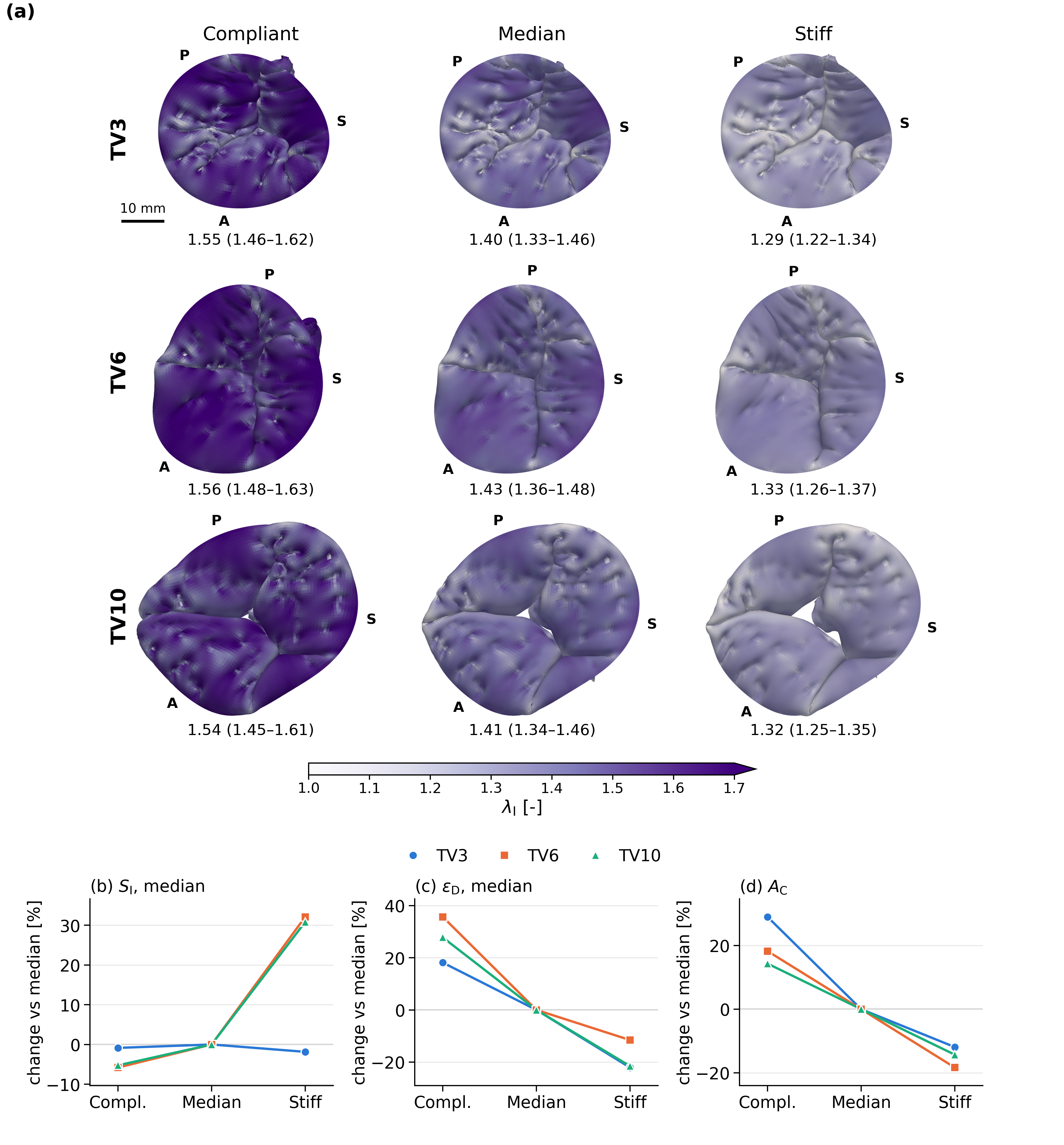}
  \caption{Leaflet stiffness sensitivity of TV3, TV6, and TV10. \textbf{(a)} Major in-plane principal stretch $\lambda_\mathrm{I}$ from ED to ES of each valve (rows) with the compliant, median, and stiff leaflet sets (columns); atrial view, septal leaflet on the right, all models at the same scale. Values below each map are median (IQR), weighted by element area. A: anterior, P: posterior, S: septal leaflet. \textbf{(b)}-\textbf{(d)} Percentage change, relative to the median set of the same valve, of \textbf{(b)} the median of the maximum principal stress $S_\mathrm{I}$, \textbf{(c)} the median of the geometric error $\varepsilon_\mathrm{D}$, and \textbf{(d)} the coaptation area $A_\mathrm{C}$.}
  \label{fig_leaflet_stiffness}
\end{figure*}

\paragraph{Leaflet stiffness.}\label{res:leafletstiff}
Increasing leaflet stiffness reduced leaflet stretch, geometric error, and coaptation area (Fig. \ref{fig_leaflet_stiffness}, Table \ref{tab:leaflet_stiffness}). The median of $\lambda_\mathrm{I}$ was 1.54--1.56 with the compliant set and 1.29--1.33 with the stiff set, against 1.40--1.43 at baseline. Relative to baseline, the compliant leaflets billowed towards the atrium and increased the median of $\varepsilon_\mathrm{D}$ by 18--36\%, whereas the stiff leaflets moved towards the ventricle and reduced it by 12--23\%; the stiff set also reduced $\varepsilon_\mathrm{ES}$ by 26--29\% (Table \ref{tab:S_position_sens}). $A_\mathrm{C}$ increased by 14--29\% with the compliant set and decreased by 12--18\% with the stiff set, a change of the same order as that across observers. The median of $S_\mathrm{I}$ changed by $-6$ to $-1$\% with the compliant set, whereas the stiff set raised it by 31--32\% in TV6 and TV10, increasing it on the anterior and septal leaflets and decreasing it on the posterior leaflet, and left it unchanged in TV3. The total chordal force changed by $-8$ to $+5$\% (Section \ref{supp:res_stiff}).

\begin{table}[t]
\centering
\resizebox{\textwidth}{!}{%
\begin{tabular}{llccccc}
\toprule
Valve & Leaflet set & $\lambda_\mathrm{I}$ [-] & $S_\mathrm{I}$ [kPa] & $\varepsilon_\mathrm{D}$ [mm] & $A_\mathrm{C}$ [mm$^2$ (\%)] & CT load [\%] \\
\midrule
TV3  & Compliant  & 1.55 (1.46--1.62) & 11.0 (4.1--22.7) & 1.20 (0.54--2.17) & 1265 (41) & 86 \\
     & Median         & 1.40 (1.33--1.46) & 11.1 (4.5--20.9) & 1.02 (0.46--1.78) & 981 (35)  & 92 \\
     & Stiff     & 1.29 (1.22--1.34) & 10.9 (4.8--20.3) & 0.79 (0.35--1.38) & 864 (33)  & 99 \\
\midrule
TV6  & Compliant  & 1.56 (1.48--1.63) & 13.2 (4.6--28.1) & 1.03 (0.41--1.93) & 1980 (46) & 84 \\
     & Median         & 1.43 (1.36--1.48) & 14.0 (5.8--26.9) & 0.76 (0.30--1.52) & 1674 (42) & 89 \\
     & Stiff     & 1.33 (1.26--1.37) & 18.5 (6.9--31.9) & 0.67 (0.30--1.21) & 1368 (37) & 96 \\
\midrule
TV10 & Compliant  & 1.54 (1.45--1.61) & 12.0 (4.6--24.8) & 1.44 (0.73--2.42) & 1563 (38) & 86 \\
     & Median         & 1.41 (1.34--1.46) & 12.7 (5.3--25.4) & 1.13 (0.59--1.93) & 1367 (36) & 85 \\
     & Stiff     & 1.32 (1.25--1.35) & 16.6 (7.2--29.8) & 0.88 (0.44--1.57) & 1172 (34) & 85 \\
\bottomrule
\end{tabular}
}
\caption{Simulations of TV3, TV6, and TV10 with the compliant, median, and stiff leaflet sets of Table \ref{tab:leaflet_params}. $\lambda_\mathrm{I}$: major in-plane principal stretch from ED to ES; $\lambda_\mathrm{I}$, $S_\mathrm{I}$, $\varepsilon_\mathrm{D}$: median (IQR) ($\lambda_\mathrm{I}$ and $S_\mathrm{I}$ weighted by element area, $\varepsilon_\mathrm{D}$ over the nodes); $A_\mathrm{C}$: coaptation area, in mm$^2$ and as a percentage of the leaflet area at ES; CT load: share of the transvalvular load on the leaflets carried by the chordal apparatus. Median rows coincide with Table \ref{tab:cohort}.}
\label{tab:leaflet_stiffness}
\end{table}

\subsection{Leaflet thickness mainly affected stress; chordal size and material had small effects after length calibration}\label{res:thickness}

\paragraph{Leaflet thickness.}
Relative to the median field, the area-weighted mean thickness at ED decreased by 17--19\% with the $Q_1$ field and increased by 22--26\% with the $Q_3$ field (Table \ref{tab:thickness}). The median of $S_\mathrm{I}$ increased by 7--25\% with the $Q_1$ field and decreased by 23--33\% with the $Q_3$ field, approximately in inverse proportion to thickness, so that thickness changed the median stress as much as segmentation did. The membrane tension, i.e. $S_\mathrm{I}$ multiplied by the current thickness, changed little (median element-wise ratio to baseline 0.95--1.01), and so did the kinematics: the median of $\lambda_\mathrm{I}$ changed by at most 0.03, that of $\varepsilon_\mathrm{D}$ by at most 0.08 mm, and $A_\mathrm{C}$ by 2--3\%. The total chordal force changed by less than 3\% (Section \ref{supp:res_thick}).

\begin{table}[t]
\centering
\resizebox{\textwidth}{!}{%
\begin{tabular}{llccccccc}
\toprule
Valve & Thickness & $\bar{t}$ [mm] & $S_\mathrm{I}$ [kPa] & $T$ [N/m] & $\lambda_\mathrm{I}$ [-] & $\varepsilon_\mathrm{D}$ [mm] & $A_\mathrm{C}$ [mm$^2$ (\%)] & CT load [\%] \\
\midrule
TV3  & $Q_1$     & 0.57 & 13.9 (5.8--26.4) & 5.4 (2.5--10.6) & 1.42 (1.35--1.47) & 1.08 (0.49--1.85) & 1012 (36) & 92 \\
     & Median & 0.70 & 11.1 (4.5--20.9) & 5.3 (2.4--10.4) & 1.40 (1.33--1.46) & 1.02 (0.46--1.78) & 981 (35)  & 92 \\
     & $Q_3$     & 0.88 & 7.4 (3.2--15.8)  & 4.7 (2.1--9.8)  & 1.38 (1.30--1.44) & 1.00 (0.46--1.72) & 960 (35)  & 93 \\
\midrule
TV6  & $Q_1$     & 0.57 & 14.9 (6.1--29.5) & 5.8 (2.5--13.0) & 1.44 (1.36--1.48) & 0.73 (0.30--1.31) & 1709 (43) & 88 \\
     & Median & 0.69 & 14.0 (5.8--26.9) & 6.3 (3.0--13.3) & 1.43 (1.36--1.48) & 0.76 (0.30--1.52) & 1674 (42) & 89 \\
     & $Q_3$     & 0.85 & 10.3 (4.1--20.6) & 6.0 (2.6--13.5) & 1.41 (1.33--1.47) & 0.68 (0.28--1.29) & 1643 (43) & 89 \\
\midrule
TV10 & $Q_1$     & 0.57 & 15.7 (6.1--31.1) & 6.7 (2.8--12.7) & 1.42 (1.36--1.47) & 1.10 (0.59--1.94) & 1392 (37) & 85 \\
     & Median & 0.69 & 12.7 (5.3--25.4) & 6.8 (3.0--12.6) & 1.41 (1.34--1.46) & 1.13 (0.59--1.93) & 1367 (36) & 85 \\
     & $Q_3$     & 0.84 & 9.7 (4.3--19.4)  & 6.5 (3.0--12.2) & 1.40 (1.33--1.45) & 1.09 (0.56--1.83) & 1337 (36) & 85 \\
\bottomrule
\end{tabular}
}
\caption{Simulations of TV3, TV6, and TV10 with the first-quartile ($Q_1$), median, and third-quartile ($Q_3$) leaflet thickness fields. $\bar{t}$: area-weighted mean thickness at ED; $T$: membrane tension, i.e. $S_\mathrm{I}$ multiplied by the current thickness; $\lambda_\mathrm{I}$: major in-plane principal stretch from ED to ES; $S_\mathrm{I}$, $T$, $\lambda_\mathrm{I}$, $\varepsilon_\mathrm{D}$: median (IQR) ($S_\mathrm{I}$, $T$, and $\lambda_\mathrm{I}$ weighted by element area, $\varepsilon_\mathrm{D}$ over the nodes); $A_\mathrm{C}$: coaptation area, in mm$^2$ and as a percentage of the leaflet area at ES; CT load: share of the transvalvular load on the leaflets carried by the chordal apparatus. Median rows coincide with Table \ref{tab:cohort}.}
\label{tab:thickness}
\end{table}

\paragraph{Chordal size and material properties.}\label{res:ctstiff}
With the $Q_1$ and $Q_3$ CT sets, the initial axial stiffness of a CT was 0.67--0.88 and 1.19--1.47 times the baseline value, and the recalibrated stress-free lengths differed from the baseline ones by at most 0.3\% (Table \ref{tab:ctstiff}). After the length calibration, the CTs worked at median axial strains below 0.2\% (at most 1.8\%). The median of $\varepsilon_\mathrm{D}$ changed by at most 0.03 mm, $A_\mathrm{C}$ by less than 2\%, and the total chordal force by less than 1\%, except in TV3 with the $Q_1$ set ($+5$\%); the median of $S_\mathrm{I}$ changed by $-2$ to $+5$\%, except for an 11\% decrease in TV6 with the $Q_3$ set, associated with a locally different fold of the posterior leaflet. The position of the leaflets relative to $\Omega^\mathrm{ES}$ was unchanged (Table \ref{tab:S_position_sens}), although the total chordal cross-section changed by a factor comparable to that of the density maps (Section \ref{supp:res_ct}).

\begin{table}[t]
\centering
\resizebox{\textwidth}{!}{%
\begin{tabular}{llcccccc}
\toprule
Valve & CT set & $S_\mathrm{I}$ [kPa] & $\varepsilon_\mathrm{D}$ [mm] & $A_\mathrm{C}$ [mm$^2$ (\%)] & $F_\mathrm{CT}$ [mN] & CT strain [\%] & CT load [\%] \\
\midrule
TV3  & $Q_1$     & 11.7 (4.7--22.2) & 1.02 (0.47--1.81) & 997 (35)  & 9.6 (1.7--21.4) & 0.17 (0.04--0.37) & 93 \\
     & Median & 11.1 (4.5--20.9) & 1.02 (0.46--1.78) & 981 (35)  & 7.7 (0.3--21.9) & 0.12 (0.00--0.31) & 92 \\
     & $Q_3$     & 11.1 (4.5--20.8) & 1.04 (0.48--1.80) & 984 (35)  & 8.2 (0.1--22.1) & 0.09 (0.00--0.23) & 92 \\
\midrule
TV6  & $Q_1$     & 13.8 (5.6--26.7) & 0.74 (0.29--1.48) & 1673 (42) & 5.0 (0.0--17.2) & 0.09 (0.00--0.30) & 89 \\
     & Median & 14.0 (5.8--26.9) & 0.76 (0.30--1.52) & 1674 (42) & 5.2 (0.0--17.1) & 0.07 (0.00--0.25) & 89 \\
     & $Q_3$     & 12.4 (5.2--26.0) & 0.75 (0.31--1.39) & 1689 (43) & 4.9 (0.0--17.2) & 0.05 (0.00--0.19) & 89 \\
\midrule
TV10 & $Q_1$     & 12.7 (5.3--25.0) & 1.13 (0.60--1.93) & 1374 (37) & 7.2 (0.7--21.6) & 0.13 (0.01--0.37) & 85 \\
     & Median & 12.7 (5.3--25.4) & 1.13 (0.59--1.93) & 1367 (36) & 6.6 (0.4--22.0) & 0.09 (0.01--0.31) & 85 \\
     & $Q_3$     & 12.5 (5.2--24.9) & 1.11 (0.58--1.90) & 1380 (37) & 6.5 (0.6--21.8) & 0.07 (0.01--0.23) & 85 \\
\bottomrule
\end{tabular}
}
\caption{Simulations of TV3, TV6, and TV10 with the CT sets at the first-quartile ($Q_1$), median, and third-quartile ($Q_3$) cross-sectional area of Table \ref{tab:material_params}. $S_\mathrm{I}$, $\varepsilon_\mathrm{D}$, $F_\mathrm{CT}$, CT strain: median (IQR) ($S_\mathrm{I}$ weighted by element area, $\varepsilon_\mathrm{D}$ over the nodes, $F_\mathrm{CT}$ and CT strain over the CTs); CT strain: axial strain at ES relative to the calibrated stress-free length; $A_\mathrm{C}$: coaptation area, in mm$^2$ and as a percentage of the leaflet area at ES; CT load: share of the transvalvular load on the leaflets carried by the chordal apparatus. Median rows coincide with Table \ref{tab:cohort}.}
\label{tab:ctstiff}
\end{table}

\section{Discussion}\label{discussion}

In this study, we adapted our previous modeling framework to construct and simulate patient-specific TVs using clinical 3D-TEE as the only source of patient-specific anatomical data \cite{Mathur2026,tondi2025magnetic}. A central challenge in image-based TV modeling is that clinical images do not fully resolve the chordal apparatus or provide tissue mechanical properties. By combining image-derived leaflet and annular geometry with synthetic chordae and population-based human tissue measurements, we reproduced the closed configurations of ten functionally competent valves. This extends our previous experimental applications to a cohort of clinically imaged human valves and supports the use of routinely acquired images to build models for investigating valve mechanics. The sensitivity analyses establish how segmentation choices and assumed properties influence the resulting simulations and point to where further patient-specific information would be most valuable.

The simulated closed configurations agreed with the morphed references to within per-valve median distances of 0.61–1.13 mm, comparable to the voxel diagonals of the clinical 3D-TEE volumes. Median distances to the image-derived ES surfaces were similar, at 0.55–0.95 mm, supporting agreement over the visible leaflet geometry. Since the CT lengths were calibrated on $\Omega_\mathrm{M}^\mathrm{ES}$, $\varepsilon_\mathrm{D}$ is partly a calibration residual; $\varepsilon_\mathrm{ES}$, which does not depend on the morphing, was of the same size. Each model was also closer to the morphed configuration of its own observer than to those of the other two (0.76--1.13 vs. 1.10--2.37 mm), so that an error of about 1 mm is close to the limit set by segmentation rather than a model deficiency to be reduced further. Kong et al. reported smaller average errors of 0.55–0.60 mm in three CT-based human TV models, but evaluated selected leaflet-belly points rather than the full leaflet surface \cite{kong2018tv}. The present findings demonstrate that close geometric agreement can also be obtained across a cohort built from 3D-TEEs, despite their incomplete visualization of the structures required to build the FE model.

Beyond just reproducing the closed geometry, the models also provided estimates of leaflet deformation, stress, coaptation, and chordal loading that are not measurable from routine clinical images. Our models broadly agreed with previous experimental and computational observations. Per-valve median leaflet stresses of 7.6–16.9 kPa were comparable in magnitude to leaflet belly stresses in the Texas TriValve and in the porcine model of Singh-Gryzbon et al. \cite{mathur2022trivalve,singhgryzbon2019abme}, and larger valves were more stressed, consistent with Laplace's law: at equal pressure and similar thickness, leaflet tension scales with the radius of curvature. Higher values reported in other studies included local peaks or measurements from selected regions which may also reflect differences in loading, thickness, and material assumptions \cite{stevanella2010tv,kong2018tv}. Deformation was predominantly radial (Section \ref{supp:bio}), and coaptation covered 20–42\% of the ES leaflet area, compared with 43\% in the Texas TriValve \cite{mathur2022trivalve}. The chordae carried 74–92\% of the pressure load, with greater forces in chordae inserting on the leaflet body than near the free edge, consistent with previous computational and experimental observations \cite{kong2018tv,troxler2012strut}.

We found that manual segmentation introduced substantial variability into the predicted valve mechanics. The observers reconstructed similar leaflet surfaces, 0.54--1.11 mm apart, close to the distance reported for manual 3D-TEE segmentation of normal mitral valves (0.6 $\pm$ 0.2 mm \cite{jassar2011mv}); differences in segmented leaflet extent, however, substantially affected coaptation area, which varied across observers with CVs of 20--37\%, and in TV6 over 72\% of the range across the ten valves. The strong association of coaptation area with the ratio of ED leaflet area to ES annular area indicates that the amount of available leaflet tissue relative to the orifice was a major determinant of predicted contact. Geometric error relative to each observer’s own reference varied comparatively little, showing that similar agreement with the calibration target can coexist with different functional predictions. For applications concerned with valve competence, segmentation assessment should therefore include the marked leaflet area and its relationship to annular size. Standardized or automated segmentation \cite{pouch2017tv} could improve consistency. More generally, uncertainty analyses of biomedical FE models have mostly considered material properties and loading rather than geometry \cite{mangado2016uncertainty}; here, segmentation was the dominant source of uncertainty for coaptation.

The sensitivity analyses also distinguished the importance of our modeling assumptions. Across the tested sensitivities, leaflet stiffness and chordal insertion density produced larger changes in median geometric error than differences in segmentation between observers. This suggests that knowledge of the closed geometry alone may not uniquely constrain these inputs. Neither factor, however, removed the atrial offset near the annulus (78--98\% of the non-contact nodes on the atrial side in every run; Table \ref{tab:S_position_sens}), which suggests a geometric origin, e.g. in the reconstruction of the hinge region. Further, we observed that increasing insertion density improved agreement with the closed geometry, consistent with our previous study \cite{Mathur2026}. We also found that more compliant leaflets exhibited greater atrial billowing and larger errors, consistent with Kong et al. \cite{kong2018tv}. Both denser insertions and stiffer leaflets improved geometric agreement, but their effects on stress differed. Leaflet thickness primarily affected stress, while membrane tension and closure geometry changed comparatively little. Chordal size and material properties had limited effects after recalibration of stress-free lengths, consistent with the small  strains observed in the chordae and with previous models that also calibrated chordal lengths \cite{kong2018tv}. These findings support concentrating further characterization on native leaflet size, insertion density, and leaflet stiffness. These analyses also identify which inputs warrant additional simulations to quantify uncertainty and which can reasonably remain at representative population values. 

The workflow also addresses practical barriers to building FE models across different patients. Following manual segmentation of the TEE images, our model construction, chordal calibration, and closure simulations were fully automated. Thus, we have a repeatable process for converting clinical images into biomechanical predictions. At the selected 0.5 mm mesh resolution, which met the prescribed whole-valve tolerances, closure simulations required approximately 31–37 min on 20 CPU cores. This runtime covers closure simulation alone and excludes segmentation, model preparation, and chordal calibration. Automation allows the same modeling procedure to be applied consistently across patients. 

Our methods and models developed in this study are particularly relevant to patient-specific repair planning. They can establish a digital representation of the individual valve, which can subsequently be modified to investigate candidate interventions. Such models could compare repair procedures or investigate repair versus replacement strategies. The present sensitivity findings provide guidance for such applications: coaptation comparisons should account for segmentation-dependent leaflet extent, geometric predictions should include plausible variation in insertion density and leaflet stiffness, and stress predictions should retain their dependence on assumed thickness. Further verification and validation of our methods on regurgitant valves and against intervention outcomes would then determine how reliably these simulations can distinguish treatment effects and inform repair selection.

\subsection{Limitations and future developments}\label{limitations}
This study has several limitations arising from the assumptions made and methodology used. 

The models rely on manual ED and ES segmentations, which differed between observers mainly in leaflet extent and in the borders between leaflets; automated segmentation \cite{pouch2017tv} could reduce this variability. 

We reduced the uncertainty in the subvalvular apparatus as far as possible by informing chordal insertion sites, cross-sectional areas, and material properties with \textit{ex vivo} human data. Some assumptions nonetheless remain. The PM origins were idealized as arcs beneath the commissures, whereas the papillary anatomy of the human TV is more variable \cite{silver1971tv}. Their motion from ED to ES was derived from ovine sonomicrometry \cite{gaweda2023papillary,jazwiec2018tricuspid}. The assignment of each CT to a PM was also assumed, because the PM origins of the chordae were not recorded \textit{ex vivo} \cite{haese2027chordae}. Each terminal insertion was connected to its own unbranched CT, so that the total chordal cross-section is probably overestimated; after the length calibration, however, chordal cross-section and material had a negligible effect on closure. The median insertion-density maps, pooled across control and heart-failure valves, corresponded to 9.3--12.2 insertions/cm$^2$, below the pooled \textit{ex vivo} density of 12.57 insertions/cm$^2$ \cite{haese2027chordae}, mainly because of the anterior leaflet (Table \ref{tab:S_position_baseline}). Since insertion density was the most influential of the assumed inputs, \textit{ex vivo} data that also record PM origins and chordal branching, together with a calibration of the insertion density on the imaged ES configuration, are the most promising ways to reduce the uncertainty of the model.

The leaflet material was isotropic and pooled across leaflets, which precludes differences among leaflets, such as the larger strain of the anterior leaflet measured \textit{in vivo} \cite{mathur2019strains}. Its parameters were identified on 28 specimens from only 5 subjects, and $C_2$ is reported conditionally on the bound of its hyperprior. Leaflet-specific and anisotropic descriptions, identified on a larger set of human donors, would allow the stiffness band used here to be narrowed and differences among leaflets to be represented.

Our sensitivity analysis was performed on three valves, one per image-quality pool, varying one factor at a time between inter-quartile levels, instead of a full-factorial design \cite{mangado2016uncertainty} to limit the number of simulations. The levels reflect the spread of the available \textit{ex vivo} data rather than the full uncertainty of each input, and three observers sample, but do not bound, the uncertainty of segmentation \cite{krygier2021segmentation}. By identifying chordal insertion density and leaflet stiffness as the dominant assumptions, the present results narrow the space that a global uncertainty analysis would need to explore, and make such an analysis possible.

In the absence of synchronized electrocardiogram data, we selected the ED and ES frames by visual inspection and, when the leaflets could not be distinguished from the endocardium, shifted the ED frame to a later time point (Section~\ref{Imseg}). ECG-gated acquisitions and automated frame selection could remove this source of variability in future studies.

Finally, the cohort of ten valves was functionally competent, so the models' accuracy and the ranking of the modeling assumptions apply to non-regurgitant anatomies. Repeating the present analysis on valves with TR is the natural next step.

Taken together, some of these limitations concern inputs that clinical images do not provide. The approach introduced here, in which each assumption is judged against the variability of the image-derived geometry, offers a criterion for deciding which of them is worth refining: an assumption deserves better data only if its effect exceeds that of segmentation.



\clearpage


\backmatter

\bibliography{bibliography}

\clearpage
\setcounter{section}{0}\setcounter{figure}{0}\setcounter{table}{0}\setcounter{equation}{0}
\renewcommand{\thesection}{S\arabic{section}}
\renewcommand{\thefigure}{S\arabic{figure}}
\renewcommand{\thetable}{S\arabic{table}}
\renewcommand{\theequation}{S\arabic{equation}}
\renewcommand{\theHsection}{S\arabic{section}}
\renewcommand{\theHfigure}{S\arabic{figure}}
\renewcommand{\theHtable}{S\arabic{table}}
\renewcommand{\theHequation}{S\arabic{equation}}
\section*{Supplementary Material}

\noindent Contents of the Supplementary Material. Model construction and simulation: \ref{supp:geom} Surface reconstruction, partitioning, and meshing; \ref{supp:thick} Leaflet thickness maps; \ref{supp:chordae} Chordal architecture and material properties; \ref{supp:mlm} Leaflet tissue: tests and Bayesian multilevel model; \ref{supp:numerics} Chordal length calibration, closure simulation, and numerical verification; \ref{supp:metrics} Output metrics; \ref{supp:sens} Sensitivity analysis; \ref{supp:sensres} Sensitivity analysis: detailed results. Additional results: \ref{supp:eps} Data-only geometric error $\varepsilon_\mathrm{ES}$; \ref{supp:pos} Position of the leaflets relative to $\Omega^\mathrm{ES}$ and chordal insertion density; \ref{supp:seg} Segmentation-level inter-observer agreement; \ref{supp:bio} Baseline biomechanics of the ten valves. Section, table, figure, and equation numbers without the prefix S refer to the main text.

\section{Surface reconstruction, partitioning, and meshing}\label{supp:geom}

We extracted each leaflet label boundary with the marching-cubes algorithm to obtain the isosurface representing the label boundary. We retained only the largest simply connected component of each leaflet and discarded fragments arising from segmentation noise. Then, for anatomical consistency across patients, we used a least-squares plane fit and principal component analysis of the annular point cloud to define a local cylindrical coordinate system. The axial direction ($z$) pointed toward the right atrium, and the circumferential datum aligned with the centroid of the septal leaflet. All surfaces were expressed in the associated cylindrical coordinate system $(\rho, \theta, z)$. Further, we regularized the annulus with fourth-order Fourier series.

At ED, we sliced the surface of each leaflet label with 40 radial planes, equally spaced in angle about the valve axis between the boundaries of the leaflet. In each slicing plane, we extracted the leaflet cross-section and generated a medial profile from the annulus to the free edge. The medial profile was regularized using a cubic B-spline. The profiles were then lofted into a single continuous leaflet midsurface, where leaflet boundaries were taken directly from the segmentation labels. The three main commissures, anteroseptal (A--S), anteroposterior (A--P), and posteroseptal (P--S), were identified where adjacent leaflet surfaces came closest. Accessory commissures, if present, were identified as local minima of the distance between the free edge and the annulus, measured along the axial direction, i.e. as the points where the free edge indents towards the annulus.
A leaflet bounded by two main commissures therefore consists of a single scallop, whereas a leaflet containing an accessory commissure consists of two scallops. Two valves had an accessory posterior commissure and one had an accessory septal commissure. The other seven valves, including the three of the sensitivity subset, only had three commissures.

We divided each scallop into three radial bands using normalized geodesic distances of 1/3 and 2/3 from the annulus. To define the circumferential bands, we split the scallop free edge into three equal arc lengths and extended the corresponding angular divisions radially inward toward the annulus so that each sector is bounded by two angular wedges. This produced a 3 $\times$ 3 grid on each scallop. We meshed the ED mid-surface ($\Omega^\mathrm{ED}$) using a custom tool based on HelloTriangle\footnote{\url{https://www.hellotriangle.io/}}, with a reference element size of 0.5 mm, into four-node shell elements with reduced integration (S4R, Abaqus/Explicit 2023). The ES reconstruction used the ED coordinate frame. Since the leaflets are in contact at ES, the profile-based mid-surface reconstruction adopted at ED cannot be applied: we fitted the ES profiles with periodic NURBS, lofted them into one continuous surface ($\Omega^\mathrm{ES}$), and triangulated it at a target edge length of 1 mm. 
This surface served as the rigid S3R target of the morphing simulation (Section \ref{supp:numerics}).

\section{Leaflet thickness maps}\label{supp:thick}
Leaflet thickness maps were measured on 28 explanted human TV leaflets \cite{kostelnik_leaflet_data_and_code_2026}, following a technique used previously to measure TV thicknesses \cite{Kostelnik2026sexspecific}. Briefly, leaflets were coated with talcum powder for optical contrast and imaged at 50× using an optical profilometer (VHX-5000, Keyence, Osaka, Japan) \cite{Kostelnik2026sexspecific,lin2023thickness}. We grouped maps from the healthy valves by leaflet and scallop configuration and computed the median, first, and third quartiles across specimens in each cell of a 100 × 100 grid. We only used healthy valves to reflect the functional nature of the valves used in this study (Section \secDatacollection{} of the main text), and to stay consistent with the healthy leaflet material properties used. Across the five leaflet-scallop combinations used in the models, cell-area-weighted mean thickness ranged from 0.59 to 0.84 mm for median maps, 0.50 to 0.75 mm for first-quartile maps and 0.74 to 1.06 mm for third-quartile maps.

We floored cell values at 0.25 mm. The floor affected at most 7.5\% of cells in any median map and 15.3\% in any first-quartile map, both in the single-scallop posterior group. It changed the area-weighted mean of any map by at most 2.1\%. We used the single-scallop septal map across the full septal leaflet of the one model with two septal scallops because a matching \textit{ex vivo} map was unavailable. 
For each leaflet node, we assigned a radial and a circumferential coordinate equal to the normalized arc length along the free edge of the scallop (of the whole leaflet, for the two-scallop septal leaflet) at the circumferential angle of the node. We bilinearly interpolated the maps between cell centers at the nodal coordinates. We held outer cell values constant to the map boundary and averaged the interpolated values at nodes shared by adjacent scallops. The shell elements were assigned a thickness field interpolated from these nodal values. The median maps define the reference thickness field used in the baseline simulations, and the first- and third-quartile maps define the perturbed levels of the sensitivity analysis (Section \ref{supp:sens}; Fig. \figInputs{}c of the main text). Being computed cell by cell, the quartile maps do not correspond to any individual specimen but to a leaflet thinner or thicker than the median at every location.

\section{Chordal architecture and material properties}\label{supp:chordae}

\subsection{Papillary muscle origins}\label{supp:pm}
Since the PMs could not be reliably identified in the 3D-TEE images, the origins of the CTs were not reconstructed from the images but were prescribed geometrically. We assumed three papillary bodies, consistent with the anatomy of the right ventricle (RV) \cite{nigri2001papillary}. Each was modeled as a circular arc lying on a plane ($\mathcal{P}_{PM}$) parallel to the annular best-fit plane ($\mathcal{P}_{ann}$). The best-fit plane itself was obtained by singular value decomposition of the annular point cloud, and each arc was centered at the projection of a main commissural annular point onto that plane. When an accessory commissure was present, an additional arc was placed beneath the annular point at the same circumferential angle and appended to the closest of the three papillary bodies. The angular extent of each arc was derived from a 6 mm chordal-origin footprint, consistent with reported right PM breadth \cite{hosapatna2021morphologyPMs}, giving half-widths of 8--10$^{\circ}$.
The apical offset of the papillary plane was set so that the mean commissural chordal length equaled 16 mm and was constrained to remain at least 5 mm apical to the deepest point of the free margin.

The same construction was used in both ED and ES frames with their respective annular point clouds and the same fixed distance between $\mathcal{P}_{PM}$ and $\mathcal{P}_{ann}$ planes. This assumption is based on sonomicrometry measurements in healthy sheep, where the distance of all papillary tips from the least-squares tricuspid annular plane changed by at most 1 mm between end-diastole and end-systole (median values of 20--21 mm) \cite{gaweda2023papillary}. 
Accordingly, PM motion from ED to ES follows the systolic excursion of the annulus toward the apex and the in-plane displacement of the commissures, without additional shortening of the annulus-to-PM distance. 

\subsection{Insertion sampling and attachment}\label{supp:insertion}

The distribution of CT insertions on the leaflets was derived from the chordal insertion density maps measured on 25 explanted human TVs \cite{haese2027chordae}. Therein, all chordal insertion sites were segmented using calibrated images of the valve ventricular surface. After segmentation, each scallop was identified and partitioned into a $3 \times 3$ radial-by-circumferential grid. Then chordal insertion density was computed as the number of insertions per unit area of each cell in the grid, i.e., as a matrix $\boldsymbol{\rho}$ with entries $\rho_{mn}$ [insertions/cm\textsuperscript{2}], $m,n = 1,2,3$. Data for each cell of the matrix were pooled across specimens with the same leaflet and number of scallops in that leaflet (one or two), as in Haese et al. \cite{haese2027chordae}. We then calculated the cell-wise median, first quartile, and third quartile of $\rho_{mn}$ for use in the current work. We used maps based on data pooled across control and heart-failure valves. The median maps computed on the control valves alone, with the same procedure, lay within the first-to-third quartile range of the pooled maps in every cell, with a whole-scallop density between $-14\%$ and $+21\%$ of the pooled median, i.e. within the range explored in the sensitivity analysis (Section \ref{supp:sens}).

Each scallop identified in the present models (Section \ref{supp:geom}) received the median map of the corresponding leaflet-and-scallop group in the baseline simulations. The quartile maps define the perturbed levels of the sensitivity analysis (Section \ref{supp:sens}; Fig. \figInputs{}d of the main text). For each cell, the target number of insertions was computed as $n_{mn} = \rho_{mn} A_{mn}$, where $A_{mn}$ is the area of the corresponding cell on the discretized surface, expressed in cm\textsuperscript{2}. Non-integer targets were rounded using the largest-remainder method, and insertions were sampled with a fixed random seed so that the chordal apparatus of each model is reproducible. In the cells adjacent to the free margin, a fraction $f_{\mathrm{frm}} = 0.4$ of the insertions was placed equidistantly along the free edge, while the remaining insertions were sampled immediately below it. This procedure yielded 264 [248--314] CTs per model (median [IQR] across the 10 models; range 127--370), of which 52 [48--59] were marginal and 212 [200--255] were non-marginal, with marginal chordae thus accounting for 18--21\% of the CTs of each model.

For each CT, the origin on the corresponding PM arc was defined as the arc point with the minimum Euclidean distance to the leaflet insertion point of that chorda. Each CT was modeled as non-branched strings such that each terminal insertion corresponded to one CT running from its PM arc to the leaflet. Each CT was discretized into linear truss elements (T3D2, ABAQUS) with a characteristic length of 2 mm.

To avoid applying each chordal force at a single leaflet node, every CT was connected to the leaflet through a distributing coupling constraint (Abaqus/Explicit). The reference node of the constraint was the end node of the CT, placed at the insertion point but not merged with the leaflet mesh, and its translational degrees of freedom were coupled to the leaflet nodes surrounding the insertion. The chordal force was distributed over these nodes with uniform weights, so that the nodal forces were statically equivalent to the chordal force. Because the coupling transmits forces without adding stiffness or mass, the leaflet tissue around each insertion deforms freely and carries the full chordal load. The coupling region included the nodes of the elements within 0.4 mm of the insertion, was truncated at the leaflet edge, and excluded the annular nodes. This radius approximated the largest median chordal radius, 0.38 mm, calculated from the \textit{ex vivo} cross-sectional areas (Table \ref{tab:material_params}), so that each coupling region approximates the footprint of the chordal cross-section on the leaflet.

\subsection{Chordal constitutive model}\label{supp:ctmat}
The mechanical behavior of CTs was described using a one-term Ogden model \cite{Ogden1972, haese2027chordae}, in the form implemented in ABAQUS:
\begin{equation}
U = \frac{2\mu_1}{\alpha_1^2}\left(\bar{\lambda}_1^{\alpha_1} + \bar{\lambda}_2^{\alpha_1} + \bar{\lambda}_3^{\alpha_1} - 3\right) + \frac{1}{D_1}\left(J - 1\right)^2,
\label{eq:W_ogden}
\end{equation}
where $\bar{\lambda}_i = J^{-1/3}\lambda_i$ are the deviatoric principal stretches, $J$ is the volume ratio, $\mu_1$ [MPa] is the initial shear modulus, $\alpha_1$ [-] controls the nonlinearity of the response, and $D_1$ [MPa$^{-1}$] governs compressibility, with initial bulk modulus $K_0 = 2/D_1$. We set $D_1 = 2/(1000\,\mu_1)$, giving $K_0/\mu_0 = 1000$ (where $\mu_0$ is the shear modulus) which corresponds to an initial Poisson ratio $\nu_0 = (3K_0/\mu_0 - 2)/(6K_0/\mu_0 + 2) \approx 0.4995$. In the incompressible limit ($J = 1$) and under uniaxial extension, with axial stretch $\lambda$ and transverse stretches $\lambda^{-1/2}$, the nominal (first Piola-Kirchhoff) axial stress is
\begin{equation}
P(\lambda) = \frac{2\mu_1}{\alpha_1}\left(\lambda^{\alpha_1 - 1} - \lambda^{-\alpha_1/2 - 1}\right),
\label{eq:P_ogden}
\end{equation}
and the corresponding axial Cauchy stress is $\sigma(\lambda) = \lambda P(\lambda)$.\\

Constitutive parameters were taken from Haese et al. \cite{haese2027chordae}, who tested 265 CT specimens from 12 human valves. They identified the area-dependent Ogden parameter pairs $(\mu_1, \alpha_1)$ through a hierarchical Bayesian model. They found that mechanical differences between anatomical positions arose mainly from differences in cross-sectional areas, with smaller chordae being stiffer. For each leaflet-by-radial-band group we evaluated the posterior at the relevant area quantile with valve and specimen effects set to zero. The closest representative joint draw was selected nearest to the posterior median stress–stretch response, resulting in the parameters given in Table \ref{tab:material_params}. Since the circumferential position of the chordae was not recorded in that study, circumferential sectors and multiple scallops within a leaflet shared the corresponding radial-band properties.

\begin{table}[t]
\centering
\resizebox{\textwidth}{!}{%
\begin{tabular}{llccccccccc}
\toprule
 & & \multicolumn{3}{c}{Median area} & \multicolumn{3}{c}{First quartile of area} & \multicolumn{3}{c}{Third quartile of area} \\
\cmidrule(lr){3-5}\cmidrule(lr){6-8}\cmidrule(lr){9-11}
Leaflet & Band & $A$ [mm\textsuperscript{2}] & $\mu_1$ [MPa] & $\alpha_1$ [-] & $A$ [mm\textsuperscript{2}] & $\mu_1$ [MPa] & $\alpha_1$ [-] & $A$ [mm\textsuperscript{2}] & $\mu_1$ [MPa] & $\alpha_1$ [-] \\
\midrule
\multirow{3}{*}{Anterior}  & NA  & 0.372 & 8.484  & 45.148 & 0.276 & 9.560  & 48.650 & 0.727 & 6.371  & 38.023 \\
                           & Bel & 0.454 & 7.716  & 42.940 & 0.249 & 9.971  & 49.973 & 0.644 & 6.728  & 39.109 \\
                           & FRM & 0.213 & 10.653 & 52.261 & 0.134 & 13.068 & 58.382 & 0.340 & 8.785  & 46.103 \\
\midrule
\multirow{3}{*}{Posterior} & NA  & 0.176 & 11.598 & 54.627 & 0.120 & 13.473 & 60.348 & 0.300 & 9.261  & 47.708 \\
                           & Bel & 0.234 & 10.252 & 50.822 & 0.119 & 13.556 & 60.510 & 0.365 & 8.552  & 45.284 \\
                           & FRM & 0.255 & 9.951  & 49.704 & 0.153 & 12.210 & 56.742 & 0.353 & 8.549  & 45.964 \\
\midrule
\multirow{3}{*}{Septal}    & NA  & 0.135 & 13.056 & 58.357 & 0.087 & 15.448 & 65.719 & 0.201 & 10.787 & 53.052 \\
                           & Bel & 0.198 & 10.856 & 53.215 & 0.155 & 12.143 & 56.479 & 0.304 & 9.212  & 47.543 \\
                           & FRM & 0.132 & 13.226 & 58.706 & 0.101 & 14.573 & 63.136 & 0.222 & 10.407 & 51.688 \\
\bottomrule
\end{tabular}
}
\caption{Chordal cross-sectional area $A$ and Ogden-1 parameters for each leaflet and radial band (NA: near-annulus; Bel: belly; FRM: free margin), from Haese et al. \cite{haese2027chordae}. Parameters are population-level predictions of the area-conditioned hierarchical Bayesian model, evaluated at the median, first, and third quartile of the cross-sectional area measured \textit{ex vivo} in each cell. Parameters follow the ABAQUS convention of Eq. (\ref{eq:W_ogden}), so $\mu_1$ and $\alpha_1$ correspond directly to the ABAQUS Ogden coefficients. Median values are used in the baseline simulations; quartile values define the perturbed levels of the chordal stiffness sensitivity (Section \ref{supp:sens}).}
\label{tab:material_params}
\end{table}

\section{Leaflet material properties}\label{supp:mlm}

\subsection{Equibiaxial tests}\label{supp:mlm_tests}
Leaflet mechanical properties were determined from equibiaxial tensile tests of the leaflet belly region of 28 specimens from 5 healthy subjects (11 anterior, 9 posterior, and 8 septal specimens; 5--6 specimens per subject) \cite{kostelnik_leaflet_data_and_code_2026}. We excised $7 \times 7$ mm specimens from the central belly of the septal, anterior, and posterior leaflets. The thickness of each specimen was manually measured with a digital thickness gauge (547-500S, Mitutoyo Corp.), and four fiducial markers were applied to the atrial surface to enable discrete strain tracking throughout the mechanical test. The specimens were then individually mounted on a commercial planar biaxial tester (BioTester, Cellscale, Waterloo, ON, Canada). Each specimen was subjected to ten equibiaxial preconditioning cycles to 400 mN \cite{Kostelnik2026sexspecific}. After preconditioning, a 10 mN preload was applied and data were continuously recorded for two force-controlled equibiaxial cycles to 400 mN at 23 mN/s. Data were recorded at a rate of 5 Hz and strain tracking was performed with Labjoy (CellScale Biomaterials Testing, Waterloo, ON, Canada). The data from the final equibiaxial unloading cycle of each specimen was analyzed relative to its respective stress-free state. For each specimen, the data consist of the two in-plane stretches $(\lambda_1, \lambda_2)$ and the corresponding Cauchy stresses $(\sigma_{11}, \sigma_{22})$, for a total of 1819 loading states and 3638 stress observations.

\subsection{Constitutive model}\label{supp:mlm_model}
TV leaflet tissue was described as a nearly incompressible, isotropic hyperelastic material through the strain-energy function of Eq.~(\eqWleaflet) of the main text, which combines a neo-Hookean term and an exponential stiffening term. Therein, $I_1 = \mathrm{tr}\,\mathbf{C}$ is the first invariant of the right Cauchy-Green deformation tensor, $C_0$ [MPa] is the initial shear modulus of the ground matrix, $C_1$ [MPa] scales the collagen-driven stiffening, and $C_2$ [-] controls how sharply that stiffening engages. The model was implemented through the ABAQUS/Explicit user subroutine for invariant-based hyperelasticity, with initial bulk modulus $K_0 = 2C_0(1+\nu_0)/[3(1-2\nu_0)]$, $\nu_0 = 0.495$, and compressibility parameter $D = 2/K_0$.

For given parameters, the model predicts the measured Cauchy stresses from the measured stretches; these predictions are compared with the measured stresses in the likelihood below. Under the assumptions of incompressibility and planar biaxial loading, $\lambda_3 = (\lambda_1 \lambda_2)^{-1}$ and $I_1 = \lambda_1^2 + \lambda_2^2 + \lambda_3^2$, and the strain-energy function of Eq.~(\eqWleaflet) of the main text gives the in-plane Cauchy stresses
\begin{equation}
\sigma_{\alpha\alpha} = 2\left(\lambda_\alpha^2 - \lambda_3^2\right) \left[ \tfrac{1}{2} C_0 + C_1 C_2 (I_1 - 3)\, \exp\!\left( C_2 (I_1 - 3)^2 \right) \right], \qquad \alpha = 1, 2.
\label{eq:S_sigma}
\end{equation}
During sampling, the exponent $C_2 (I_1-3)^2$ was clipped at 30 to prevent numerical overflow for extreme parameter values proposed by the sampler; this does not affect the representative sets of Table \ref{tab:leaflet_params}, whose exponent is 3.1--4.0 at the largest tested stress (192 kPa).

\subsection{Bayesian multilevel model}\label{supp:mlm_bayes}
For each parameter $\theta \in \{C_0, C_1, C_2\}$, the value of specimen $j$ was modeled as log-normally distributed about a population median $m_\theta$, with log-scale standard deviation $s_\theta$, using a non-centered parameterization \cite{betancourt2015hmc} as given in Eq. \ref{eq:S_noncentered}. 
\begin{equation}
\theta_j = m_\theta \exp\!\left( s_\theta\, z_{\theta,j} \right), \qquad z_{\theta,j} \sim \mathcal{N}(0, 1), \qquad j = 1, \dots, 28.
\label{eq:S_noncentered}
\end{equation}
This partial-pooling structure \cite{bol2019bayesian, pensalfini2023} shrinks the estimates of poorly constrained specimens towards the population median while retaining specimen-to-specimen variability. Specimens were treated as exchangeable within the population, without a subject level, and the specimens of the three leaflets were pooled so that the same properties were assigned to all leaflets. The observed stresses of both components were modeled jointly with a single homoscedastic Gaussian error,
\begin{equation}
\sigma^\mathrm{obs}_{\alpha\alpha,k} \sim \mathcal{N}\!\left( \sigma_{\alpha\alpha}\!\left(\lambda_{1,k}, \lambda_{2,k}; C_{0,j(k)}, C_{1,j(k)}, C_{2,j(k)}\right),\; \sigma_m^2 \right),
\label{eq:S_lik}
\end{equation}
where $j(k)$ is the specimen of observation $k$ and $\sigma_{\alpha\alpha}$ is given by Eq.~(\ref{eq:S_sigma}). The hyperpriors are listed in Table~\ref{tab:S_mlm_priors}. The uniform priors on the population medians bound each median within physically plausible limits. The upper bound of $m_{C_2}$ acts as an identifiability constraint. Relaxing the upper bound shifted $m_{C_2}$ towards the new upper bound, whereas $m_{C_0}$ and $m_{C_1}$ changed by about 1\% or less with a change in bounds. Therefore, $C_2$ is reported conditionally on the bound.

\begin{table}[htbp]
\centering
\begin{tabular}{lll}
\toprule
Parameter & Prior & Meaning \\
\midrule
$m_{C_0}$ [MPa]    & $\mathrm{U}(10^{-4},\, 5\times10^{-2})$ & population median of $C_0$ \\
$m_{C_1}$ [MPa]    & $\mathrm{U}(10^{-6},\, 10^{-2})$        & population median of $C_1$ \\
$m_{C_2}$ [-]      & $\mathrm{U}(0.1,\, 8)$                  & population median of $C_2$ \\
$s_{C_0}$ [-]      & $\mathrm{Exp}(1)$                       & log-scale spread of $C_0$ \\
$s_{C_1}$ [-]      & $\mathrm{Exp}(1)$                       & log-scale spread of $C_1$ \\
$s_{C_2}$ [-]      & $\mathrm{Exp}(2)$                       & log-scale spread of $C_2$ \\
$\sigma_m$ [MPa]   & $\mathrm{Exp}(300)$                     & measurement error (prior mean 3.3 kPa) \\
$z_{\theta,j}$ [-] & $\mathcal{N}(0,1)$                      & standardized specimen effects \\
\bottomrule
\end{tabular}
\caption{Hyperpriors of the multilevel model. $\mathrm{U}(a,b)$: uniform on $[a,b]$; $\mathrm{Exp}(\lambda)$: exponential with rate $\lambda$ (mean $1/\lambda$). Stresses are in MPa.}
\label{tab:S_mlm_priors}
\end{table}

The posterior was sampled with the No-U-Turn sampler \cite{hoffman2014nuts} as implemented in nutpie (v0.16.8) \cite{nutpie}, through PyMC (v5.28.5) \cite{abrilpla2023pymc}. Four chains were run, each with 8,000 tuning iterations followed by 30,000 retained draws ($1.2 \times 10^5$ draws in total), with a target acceptance rate of 0.99, a maximum tree depth of 10, and random seed 42. No divergent transitions occurred. The rank-normalized split-$\hat{R}$ \cite{vehtari2021rank} was at most 1.013 over the hyperparameters (Table~\ref{tab:S_mlm_post}) and at most 1.0003 over the 84 specimen-level parameters. The smallest bulk and tail effective sample sizes (ESS) were 356 and 491 ($s_{C_2}$) for the hyperparameters, and above 50,000 for the specimen-level parameters. The estimated Bayesian fraction of missing information (E-BFMI) was 0.71--0.73 in every chain, and the mean acceptance rate was 0.98-0.99. The maximum tree depth was reached in 75\% of the post-warm-up iterations (60-98\% depending on the chain), which does not bias the draws but reduces sampling efficiency. This is consistent with a strongly curved posterior along the $m_{C_2}$--$s_{C_2}$ direction near the upper bound of $m_{C_2}$. Correspondingly, these two hyperparameters have the lowest effective sample sizes. The posterior of $m_{C_2}$ is concentrated against its upper bound, with 57\% of the draws exceeding $0.9 \times 8$.

\begin{table}[htbp]
\centering
\begin{tabular}{lccccc}
\toprule
Parameter & Median & 95\% HDI & ESS$_\mathrm{bulk}$ & ESS$_\mathrm{tail}$ & $\hat{R}$ \\
\midrule
$m_{C_0}$ [$10^{-3}$ MPa] & 0.949 & 0.360--1.674 & 18\,420  & 29\,030 & 1.000 \\
$m_{C_1}$ [$10^{-4}$ MPa] & 0.666 & 0.211--1.444 & 791      & 1\,246  & 1.003 \\
$m_{C_2}$ [-]             & 7.34  & 5.77--8.00   & 481      & 563     & 1.007 \\
$s_{C_0}$ [-]             & 1.52  & 0.95--2.28   & 20\,177  & 32\,631 & 1.000 \\
$s_{C_1}$ [-]             & 2.30  & 1.76--2.99   & 1\,767   & 2\,856  & 1.006 \\
$s_{C_2}$ [-]             & 1.33  & 1.04--1.69   & 356      & 491     & 1.013 \\
$\sigma_m$ [kPa]          & 3.07  & 2.99--3.14   & 114\,105 & 75\,011 & 1.000 \\
\bottomrule
\end{tabular}
\caption{Posterior of the hyperparameters: median, 95\% highest-density interval (HDI), bulk and tail effective sample size (ESS), and rank-normalized split-$\hat{R}$.}
\label{tab:S_mlm_post}
\end{table}

\subsection{Representative parameter sets}\label{supp:mlm_sets}
For each posterior draw $i$, the parameters of a new, unobserved specimen were obtained independently for the three parameters as
\begin{equation}
\theta^{\mathrm{pop}}_i = m_{\theta,i} \exp\!\left( s_{\theta,i}\, \tilde{z}_{\theta,i} \right), \qquad \tilde{z}_{\theta,i} \sim \mathcal{N}(0,1).
\label{eq:S_pop}
\end{equation}
Since the response depends jointly and nonlinearly on the three parameters, combining the marginal quantiles of each parameter does not yield a quantile of the response. Representative parameter sets were therefore selected based on the predicted mechanical response, following the approach of Haese et al. \cite{haese2027chordae}. For each joint draw, the equibiaxial stretch was computed at ten Cauchy stress levels between 1 and 200 kPa (1, 2, 5, 10, 25, 50, 75, 100, 150, 200 kPa). The pointwise median and the first and third quartiles of the stretch were computed across draws, and for each of them, the joint draw with the smallest root-mean-square stretch difference from that curve was retained (draws 35445, 12519, and 44150, in chain-major order). The draw on the median curve defines the median set, adopted in the baseline simulations; the draws on the first and third quartiles of the stretch define the stiff and compliant sets, used as the perturbed levels of the leaflet stiffness sensitivity analysis (Table \ref{tab:leaflet_params}; Section \ref{supp:sens}; Fig. \figInputs{}e of the main text). At an equibiaxial Cauchy stress of 10 kPa, the three sets reach a stretch of 1.165 (stiff), 1.238 (median), and 1.333 (compliant). The healthy population was adopted since all patients considered in this study presented with a functionally competent TV (Section \secDatacollection{} of the main text). Because $\tilde{z}_{\theta,i}$ is drawn independently for each parameter, a predicted specimen can have $C_2$ above the bound on $m_{C_2}$, as the stiff set does: the bound applies to the population median, not to individual specimens.

\begin{table}[t]
\centering
\begin{tabular}{lccc}
\toprule
Set & $C_0$ [$10^{-3}$ MPa] & $C_1$ [$10^{-4}$ MPa] & $C_2$ [-] \\
\midrule
Stiff     & 3.149 & 6.077 & 12.74 \\
Median    & 1.205 & 4.125 & 5.310 \\
Compliant & 0.575 & 3.081 & 2.209 \\
\bottomrule
\end{tabular}
\caption{Leaflet constitutive parameters of the healthy population, pooled across the three leaflets. Each set is a single joint draw from the population-predictive distribution of the multilevel fit (Eq.~\ref{eq:S_pop}), whose equibiaxial response lies closest to the pointwise first quartile (stiff), median, or third quartile (compliant) of the predicted stretch between 1 and 200 kPa. The median set is used in the baseline simulations; the stiff and compliant sets define the perturbed levels of the leaflet stiffness sensitivity analysis.}
\label{tab:leaflet_params}
\end{table}

\section{Chordal length calibration, closure simulation, and numerical verification}\label{supp:numerics}

\paragraph{Chordal length calibration.}
Since CTs are not directly visualized by standard clinical imaging modalities \cite{Hahn2016revimagingTV}, a dedicated strategy was implemented to model CT lengths while reproducing ground-truth ultrasound-derived segmentations. As in \cite{tondi2025magnetic}, prior to simulating valve closure, CT stress-free length at ED was calibrated through a three-step process (Fig. \figWorkflow{} of the main text):
\begin{enumerate}
\item \textit{Morphing simulation} - $\Omega^\mathrm{ED}$ was driven to approximate $\Omega^\mathrm{ES}$, yielding an estimated closed-valve configuration ($\Omega_{\mathrm{M}}^\mathrm{ES}$). In a first step, lasting 1 s, $\Omega^\mathrm{ES}$ was modeled as a rigid surface, initially offset toward the atrium and progressively translated back to its reconstructed position, while a pressure of up to 30 mmHg was applied on the ventricular side of the leaflets, pressing them against it. Concurrently, a total force of 1N was distributed uniformly over the nodes of the free margin and directed along the axial direction of the reference frame and towards the RV apex, following the same time history as the pressure \cite{Mathur2026}. Contextually, annular nodes and CT origins were driven from their ED to their ES positions. CTs were modeled as linear elastic and compliant relative to CT tissue ($E = 2$ MPa, $\nu = 0.3$), so as to follow leaflet motion. 
Over the portion of the valve visible in the images, $\Omega_\mathrm{M}^\mathrm{ES}$ reproduced $\Omega^\mathrm{ES}$ closely, the distance from the nodes of $\Omega^\mathrm{ES}$ to $\Omega_\mathrm{M}^\mathrm{ES}$ having per-valve medians of 0.02-0.08 mm.
Based on the position of the insertion nodes on $\Omega_{\mathrm{M}}^\mathrm{ES}$, the loaded length of each CT at ES ($l_i^\mathrm{ES}$) was estimated as the distance between its insertion node and its origin on the corresponding PM arc at ES.
\item \textit{Pressurization simulation} - CTs of length $l_i^\mathrm{ES}$ were rebuilt on $\Omega^\mathrm{ED}$, directed from their insertion toward their origin at ED, and modeled as nearly inextensible linear elastic elements ($E = 100$ GPa, $\nu = 0.475$). The simulation followed the same two steps as the closure simulation (see below): CT origins were first driven to their ED positions with the leaflet nodes fixed, and then annulus and PMs were driven from ED to ES while a transvalvular pressure up to 22 mmHg was applied on the ventricular side of the leaflets. The axial force transmitted by each CT at ES ($F_i^\mathrm{ES}$) was extracted as the reaction force at its origin at the final frame of the simulation.
\item \textit{Final CT length tuning} - the nominal axial stress of each CT was computed as $P_i = F_i^\mathrm{ES}/A_i$, with $A_i$ its cross-sectional area (Table \ref{tab:material_params}), and its axial stretch $\lambda_{\mathrm{AX},i}$ was obtained by numerically inverting Eq. (\ref{eq:P_ogden}) with the parameters assigned to that CT. Since CTs are nearly inextensible in the pressurization simulation, their length at ES coincides with $l_i^\mathrm{ES}$, and the stress-free CT length at ED was calculated as $l_i^\mathrm{ED} = l_i^\mathrm{ES}/\lambda_{\mathrm{AX},i}$. The new reference configuration of each CT was set by keeping its insertion on the TV leaflets and its direction toward its origin at ED, $\mathbf{O}_{\mathrm{CT_{ij}}}^{\mathrm{ED}}$, i.e. the point of the papillary arc of Section \ref{supp:pm} assigned to the $i$-th CT of the $j$-th papillary body, thus defining a new position of its origin ($\mathbf{O}_{\mathrm{CT_{ij}}}^{\mathrm{ED_{tuned}}}$).
\end{enumerate}

\paragraph{Closure simulation.}
Following the calibration of CT initial length, TV systolic closure from ED to ES was simulated for each patient through a two-step simulation (Fig. \figWorkflow{} of the main text), using $\Omega^\mathrm{ED}$ as the initial configuration of leaflets and annulus and $\mathbf{O}_{\mathrm{CT_{ij}}}^{\mathrm{ED_{tuned}}}$ as the initial position of CT origins.\\
In step 1, lasting 0.25 s, the nodes of the annulus and leaflets were prevented from any translation, and the CT origins were moved from $\mathbf{O}_{\mathrm{CT_{ij}}}^{\mathrm{ED_{tuned}}}$ to their ED position by imposing the nodal displacements
\begin{equation}
\mathbf{U}_{\mathrm{CT_{ij}}}=\mathbf{O}_{\mathrm{CT_{ij}}}^{\mathrm{ED}}-\mathbf{O}_{\mathrm{CT_{ij}}}^{\mathrm{ED_{tuned}}}.
\label{eq:step1}
\end{equation}
This was done only for the CTs whose stress-free length $l_i^\mathrm{ED}$ was not shorter than the distance between their insertion and $\mathbf{O}_{\mathrm{CT_{ij}}}^{\mathrm{ED}}$, so that no CT was stretched against the fixed leaflets. The origins of the remaining CTs were left at $\mathbf{O}_{\mathrm{CT_{ij}}}^{\mathrm{ED_{tuned}}}$ in step 1, and all CT origins were driven to their ES positions in step 2.
In step 2, lasting 1 s, a transvalvular pressure that increased smoothly up to 22 mmHg was applied on the ventricular side of $\Omega^\mathrm{ED}$. All contact interactions, including leaflet self-contact, were modeled through a penalty contact algorithm with a linear pressure-overclosure relationship and a friction coefficient of 0.05 \cite{stevanella2011mitral}. CTs were described by the Ogden model of Section \ref{supp:ctmat}, and connected to the leaflets through the distributing couplings of Section \ref{supp:insertion}. Contextually, nodal displacements were prescribed to the nodes on the annular profile and to the CT origins to replicate the annulus and PM motion between ED and ES as obtained from 3D-TEE. Displacements of annular nodes were prescribed by enforcing a kinematically consistent deformation of each inter-commissural segment while preserving the original nodal discretization, such that the number of annular nodes within each segment remained unchanged throughout the deformation.\\
\paragraph{Solver settings and energy balance.}
In all simulations, including the morphing and pressurization setup simulations described above, mass-proportional damping ($\alpha_R = 30$ s$^{-1}$) was applied to leaflets and CTs, and bulk viscosity was applied with linear and quadratic coefficients of 0.5 and 1.2. In the closure simulations, uniform variable mass scaling with a target stable time increment of $1 \times 10^{-5}$ s was applied in step 1, in which only the CT origins move, whereas in step 2 mass scaling was restricted to the elements whose stable time increment fell below $5 \times 10^{-7}$ s, which increased the total mass by at most 0.02\%.\\
The quasi-static character of the closure simulations was verified from the whole-model energy history. In step 2, the ratio of kinetic to internal energy peaked at 2.9--18.1\% at 0.14--0.17 s, when the leaflets moved towards closure under a transvalvular pressure below 1 mmHg. It fell below 1\% within 0.20--0.77 s, did not exceed 6\% in the second half of the step, and was 0.02--0.20\% at ES. Damping and bulk viscosity dissipated 18--52\% of the external work done in step 2; since both act on velocity, which vanishes at ES, they do not affect the computed ES configuration. The artificial strain energy associated with hourglass control was 3.2--5.4\% of the internal energy, and the total energy remained balanced within 2.2\% of the external work.

\clearpage
\section{Output metrics}\label{supp:metrics}
This section details the definitions of $\varepsilon_\mathrm{D}$ and $A_\mathrm{C}$ (Section \secMetrics{} of the main text), while $\varepsilon_\mathrm{ES}$ is defined in Section \ref{supp:eps}.

\paragraph{Geometric error.}
For each node of the simulated leaflet configuration at ES, $\varepsilon_\mathrm{D}$ is the Euclidean distance to the closest point of $\Omega_\mathrm{M}^\mathrm{ES}$, the leaflet surface obtained from the morphing simulation (Section \ref{supp:numerics}), triangulated from its shell elements. $\Omega_\mathrm{M}^\mathrm{ES}$ was preferred to $\Omega^\mathrm{ES}$ as the reference because $\Omega^\mathrm{ES}$ is a single continuous surface lofted through the leaflet profiles at ES (Section \ref{supp:geom}): where the leaflets coapt, it represents the contact region by one surface and does not contain the coapting portion of each leaflet, which is not resolved by 3D-TEE. The distance from the simulated leaflets to $\Omega^\mathrm{ES}$ would therefore increase with the coaptation height wherever the leaflets are in contact, regardless of the accuracy of the simulation. $\Omega_\mathrm{M}^\mathrm{ES}$, instead, is a complete leaflet surface, including the coaptation region, pressed against $\Omega^\mathrm{ES}$ over the portion of the valve visible in the images. Within the coaptation region, $\Omega_\mathrm{M}^\mathrm{ES}$ is constrained not by the images but by the morphing loads, and $\varepsilon_\mathrm{D}$ also reflects the assumptions of the morphing simulation.

\paragraph{Coaptation area.}
A nodal contact indicator $c_a$ was set to 1 at the nodes with positive contact pressure and to 0 elsewhere, and interpolated over each shell element with its bilinear shape functions $N_a$; the coapted region was taken as the part of the surface where the interpolated indicator is at least 0.5:
\begin{equation}\label{eq:S_AC}
A_\mathrm{C} = \sum_{e} \int_{\Omega_e} H\!\left( \sum_{a=1}^{4} N_a\, c_a - \tfrac{1}{2} \right) \mathrm{d}\Omega ,
\end{equation}
where $H$ is the Heaviside function and $\Omega_e$ is element $e$ in the configuration at ES. The integral was evaluated on sub-cells of about 0.05 mm. With this definition, the contact border runs between contacting and free nodes rather than along element edges, so $A_\mathrm{C}$ does not depend on element size.

\section{Sensitivity analysis}\label{supp:sens}
The factors and levels of the one-factor-at-a-time analysis (Section \secSensitivity{} of the main text) are summarized in Table \ref{tab:sensitivity}, in the order in which their results are reported.

\begin{table}[h]
\centering
\resizebox{\textwidth}{!}{%
\begin{tabular}{lll}
\toprule
\textbf{Factor} & \textbf{Baseline} & \textbf{Perturbed levels} \\
\midrule
Mesh size & 0.5 mm & 0.3--0.8 mm (0.3 mm as reference) \\
Segmentation & Reference observer (O1) & Two additional observers (O2, O3) \\
Chordal insertion density & Median density maps & $Q_1$ and $Q_3$ density maps \\
Leaflet material properties & Median set & Stiff and compliant sets \\
Leaflet thickness & Median thickness maps & $Q_1$ and $Q_3$ thickness maps \\
Chordal size and material properties & Median cross-sectional area & $Q_1$ and $Q_3$ of the cross-sectional area, with the Ogden parameters predicted at that area \\
\bottomrule
\end{tabular}
}
\caption{One-factor-at-a-time sensitivity analysis. Each factor was perturbed in TV3, TV6, and TV10, the three valves for which independent segmentations were available, while all other inputs were kept at their baseline values; the remaining seven valves were simulated at baseline only (Section \secSensitivity{} of the main text). $Q_1$ and $Q_3$: first and third quartile of the corresponding \textit{ex vivo} distribution. Stiff and compliant sets: representative draws of the population-predictive distribution whose equibiaxial response lies closest to the first and third quartile of the predicted stretch, respectively.}
\label{tab:sensitivity}
\end{table}

\paragraph{Mesh size.}
For each of the three valves, the leaflet surface was remeshed with reference element sizes from 0.3 to 0.8 mm, the 0.3 mm model being taken as the reference. To isolate the effect of discretization, CT insertions were transferred from the reference mesh to each variant by nearest-node projection, and CT origins and stress-free lengths were kept as in the reference model, so that the chordal apparatus was identical across variants. Each variant was simulated following the protocol of Section \ref{supp:numerics}, on the same type of cluster node (2 $\times$ Intel Xeon Gold 6148, 2.4 GHz), on 20 physical CPU cores and with the same parallelization settings, so that wall times are directly comparable. A wider tolerance was adopted for $S_\mathrm{I}$ (10\%) than for $\varepsilon_\mathrm{D}$ and $A_\mathrm{C}$ (5\%) because, unlike the two geometric metrics, it is evaluated element by element and is therefore directly affected by the interpolation order of the shell elements.

\paragraph{Inter-observer variability.}
The spread across observers of each output metric $x$ was summarized by its coefficient of variation over the three models of the same valve, as previously used to quantify segmentation-induced variability of image-based simulations \cite{voss2019match}:
\begin{equation}\label{eq:S_CV}
\mathrm{CV} = \frac{s}{\bar{x}} \times 100\%, \qquad
\bar{x} = \frac{1}{3}\sum_{k=1}^{3} x_k, \qquad
s = \sqrt{\frac{1}{2}\sum_{k=1}^{3} \left(x_k - \bar{x}\right)^2},
\end{equation}
where $x_k$ is the value obtained from the segmentation of observer $k$, i.e. the median for distributed quantities ($S_\mathrm{I}$, $\varepsilon_\mathrm{D}$) and the value itself for scalar ones ($A_\mathrm{C}$, leaflet area, number of CTs).

\paragraph{Chordal size and material properties.}
Cross-sectional area and Ogden parameters were always perturbed jointly to preserve the observed size-stiffness relationship and avoid unobserved combinations of area and stiffness \cite{haese2027chordae}. The perturbation was applied to all leaflet-by-radial-band cells simultaneously.

\clearpage
\section{Sensitivity analysis: detailed results}\label{supp:sensres}

\subsection{Mesh size}\label{supp:res_mesh}
\begin{figure}[htbp]
  \centering
  \includegraphics[width=1\textwidth]{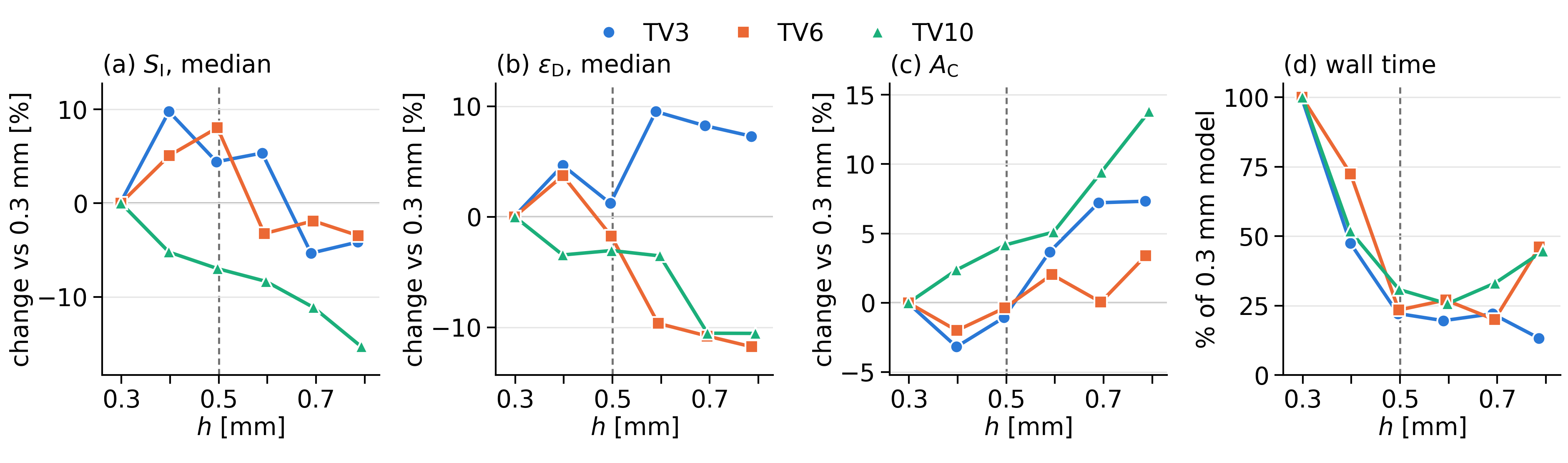}
  \caption{Mesh size sensitivity of TV3, TV6, and TV10, for reference element sizes $h$ from 0.3 (reference) to 0.8 mm; the dashed line marks the baseline size of 0.5 mm. Number of leaflet elements: 25552-35193 at 0.3 mm, 9204-12682 at 0.5 mm, 3654-5059 at 0.8 mm. \textbf{(a)}-\textbf{(c)} Change relative to the 0.3 mm model of the same valve of \textbf{(a)} the median of the maximum principal stress $S_\mathrm{I}$ (weighted by element area), \textbf{(b)} the median of the geometric error $\varepsilon_\mathrm{D}$ (over the nodes), and \textbf{(c)} the coaptation area $A_\mathrm{C}$. \textbf{(d)} Wall time on 20 CPU cores, as a percentage of that of the 0.3 mm model.}
  \label{fig_mesh}
\end{figure}

All 18 models (six element sizes for each of the three valves) reached the ES configuration with the leaflets in contact (Fig. \ref{fig_mesh}). The first metric to exceed its tolerance on coarser meshes was the geometric error at 0.6 mm, where the median of $\varepsilon_\mathrm{D}$ changed by $+10$\% in TV3 and $-10$\% in TV6 (at most 0.10 mm). At 0.5 mm, the closed configurations differed from those of the 0.3 mm models by less than 0.42--0.86 mm at 95\% of the nodes, and the total chordal force changed by 1--3\%. The central orifice of TV10 changed from 5.0 to 5.5 mm$^2$, whereas TV3 and TV6 closed without a residual orifice at every element size. The leaflet-wise medians of $S_\mathrm{I}$ were more sensitive than the whole-valve ones, with changes of up to $+16$\% (septal leaflet of TV3) and $+32$\% (posterior leaflet of TV6) at 0.5 mm. The closure simulations of the 0.5 mm models took 31.1--36.6 min of wall time on 20 CPU cores, against 117.8--140.2 min for the 0.3 mm models (22--31\%); coarser meshes brought little or no further saving (Fig. \ref{fig_mesh}d).

\subsection{Segmentation}\label{supp:res_seg}
The geometric error was insensitive to the observer. Per-model medians of $\varepsilon_\mathrm{D}$ ranged from 0.76 to 1.13 mm (CV 4-13\%). Each model was closer to the morphed configuration of its own observer than to those of the other two, which were 1.10-2.37 mm away (Fig. \ref{fig_operators}b).

The coaptation area was the output most affected by the observer (CV 20-37\%). $A_\mathrm{C}$ ranged from 683 to 981 mm$^2$ in TV3, from 804 to 1674 mm$^2$ in TV6, and from 874 to 1367 mm$^2$ in TV10; in TV6, the range across observers covered 72\% of the range across the ten baseline valves. In all three valves, the reference segmentations yielded the largest $A_\mathrm{C}$ (35-42\% of the leaflet area at ES, against 23-31\%), and the O2 segmentations, with the least leaflet tissue, yielded the smallest. In TV3, O3 segmented the largest leaflets (2515 vs. 1923-2260 mm$^2$ at ED) but also the widest annulus (projected annular area of the model at ES, Section \ref{supp:bio}: 1705 vs. 1122-1193 mm$^2$), and its $A_\mathrm{C}$ was intermediate. Accounting for the grouping of the models by valve, $A_\mathrm{C}$ correlated with the ratio of the leaflet area at ED to this projected annular area (repeated-measures correlation $r = 0.89$, 5 degrees of freedom, $p = 0.007$) and, less strongly, with the leaflet area alone ($r = 0.79$, $p = 0.03$), but not with the projected annular area ($r = -0.12$, $p = 0.79$). Three models closed with a central orifice larger than any baseline simulation (at most 11.8 mm$^2$, 0.7\% of the projected annular area): 47 mm$^2$ (3.8\%) in TV6 with O2, 62 mm$^2$ (3.2\%) in TV10 with O3, and 17 mm$^2$ (1.0\%) in TV3 with O3.

The median of $S_\mathrm{I}$ varied less (CV 11-23\%): from 11.1 to 18.0 kPa in TV3, from 11.5 to 14.0 kPa in TV6, and from 12.7 to 18.0 kPa in TV10. These ranges nonetheless cover much of the range across the ten baseline valves (7.6-16.9 kPa). The 95\textsuperscript{th} percentile varied irregularly (CV 9-48\%) and peaked in the two models in which one leaflet took the largest share of the leaflet area, that leaflet also having the highest leaflet-wise median of the valve: 115 kPa in TV6 with O3 (51-57 kPa in the other two models; anterior leaflet median 20.0 kPa), and 74 kPa in TV3 with O2 (42-51 kPa; septal leaflet median 23.1 kPa). Because the leaflet borders differed, leaflet-wise medians varied more than whole-valve ones, e.g., from 5.7 to 15.5 kPa for the septal leaflet of TV6.

\subsection{Insertion density}\label{supp:res_density}
The geometric error decreased monotonically with insertion density in all three valves. Per-model medians of $\varepsilon_\mathrm{D}$ were 1.03-1.53 mm with the $Q_1$ maps, 0.76-1.13 mm at baseline, and 0.56-0.95 mm with the $Q_3$ maps, i.e. 24-36\% above and 16-26\% below the baseline value of the same valve. With the $Q_1$ maps, the error concentrated in the leaflet bellies, mostly of the posterior leaflet in TV3, of all three leaflets in TV6, and of the anterior and posterior leaflets in TV10, where it exceeded 3 mm over large regions; with the $Q_3$ maps, it decreased mainly over the same regions (Fig. \ref{fig_density}). Denser insertions also brought the leaflets closer to $\Omega^\mathrm{ES}$, reducing $\varepsilon_\mathrm{ES}$ by 19-23\% from the baseline to the $Q_3$ maps (Table \ref{tab:S_position_sens}).

The leaflet stress followed the same trend. The median of $S_\mathrm{I}$ was 14.6-17.8 kPa with the $Q_1$ maps, 11.1-14.0 kPa at baseline, and 9.7-10.6 kPa with the $Q_3$ maps (18-40\% above and 13-24\% below baseline). Leaflet-wise medians decreased with density in eight of the nine leaflets; in the posterior leaflet of TV3, the median decreased from 15.8 to 10.0 kPa from $Q_1$ to baseline but rose to 10.8 kPa with the $Q_3$ maps.

The load carried by each CT decreased with the number of CTs. The median $F_\mathrm{CT}$ was 8.1-14.3 mN with the $Q_1$ maps, 5.2-7.7 mN at baseline, and 3.1-4.5 mN with the $Q_3$ maps, while the share of the transvalvular load carried by the chordal apparatus increased from 73-80\% to 87-100\%. The coaptation area changed little: by $-6.2$ to $-1.4$\% with the $Q_1$ maps and by $+2.6$ to $+7.4$\% with the $Q_3$ maps.

\subsection{Leaflet stiffness}\label{supp:res_stiff}
Leaflet stretch was the output most directly affected. The median of $\lambda_\mathrm{I}$ was 1.54-1.56 with the compliant set, 1.40-1.43 at baseline, and 1.29-1.33 with the stiff set, i.e., $+0.13$ to $+0.15$ and $-0.09$ to $-0.11$ relative to baseline, with similar changes on the three leaflets of each valve. 

The geometric error decreased monotonically with leaflet stiffness in all three valves. Per-model medians of $\varepsilon_\mathrm{D}$ were 1.03-1.44 mm with the compliant set, 0.76-1.13 mm at baseline, and 0.67-0.88 mm with the stiff set, i.e., 18-36\% above and 12-23\% below the baseline value of the same valve. With the compliant set, the closed leaflets billowed towards the atrium relative to baseline over all three leaflets, with a median signed distance between the two configurations of $+0.21$ to $+0.32$ mm; the stiff set displaced them towards the ventricle by a similar amount ($-0.26$ to $-0.34$ mm). Relative to $\Omega^\mathrm{ES}$, the stiff set also brought the leaflets closer to the image data, reducing $\varepsilon_\mathrm{ES}$ by 26-29\% (Table \ref{tab:S_position_sens}).

The coaptation area decreased with leaflet stiffness in all three valves, from 38-46\% of the leaflet area at ES with the compliant set to 33-37\% with the stiff set ($+14$ to $+29$\% and $-12$ to $-18$\% relative to baseline). TV3 and TV6 closed without a residual orifice at all three levels, whereas the central orifice of TV10 grew from 1.2 mm$^2$ with the compliant set to 5.5 mm$^2$ at baseline and 20.2 mm$^2$ (1.2\% of the projected annular area) with the stiff set.

The median of $S_\mathrm{I}$ was nearly insensitive to the compliant set ($-6$ to $-1$\%), whereas the stiff set raised it by 32\% in TV6 (from 14.0 to 18.5 kPa) and 31\% in TV10 (from 12.7 to 16.6 kPa), and left it unchanged in TV3 (10.9 vs. 11.1 kPa). The whole-valve medians masked a redistribution of stress among the leaflets. With the stiff set, the median $S_\mathrm{I}$ of TV6 and TV10 increased on the anterior ($+39$ to $+49$\%) and septal leaflets ($+56$ to $+81$\%) and decreased on the posterior leaflet ($-15$ to $-21$\%); with the compliant set, it increased on the posterior leaflet of all three valves ($+21$ to $+25$\%).

The chordal apparatus was less affected. The total chordal force changed by $-8$ to $+5$\% with no common sign across valves, and the force of each CT remained strongly correlated with its baseline value (Pearson $r = 0.87$-$0.97$). The share of the transvalvular load carried by the CTs increased with leaflet stiffness in TV3 (86, 92, and 99\%) and TV6 (84, 89, and 96\%), and remained at 85-86\% in TV10, where the stiff set instead shifted load to the free-margin CTs, whose total force increased from 0.67 to 1.20 N.

\subsection{Leaflet thickness}\label{supp:res_thick}
Relative to the reference (median) field, the area-weighted mean thickness at ED decreased by 17-19\% with the $Q_1$ field and increased by 22-26\% with the $Q_3$ field, the change being largest on the septal ($-25$ and $+29$\%) and smallest on the anterior leaflet ($\pm 11$\%).

Thickness acted mainly on the leaflet stress. The median of $S_\mathrm{I}$ increased by 7-25\% with the $Q_1$ field and decreased by 23-33\% with the $Q_3$ field, approximately in inverse proportion to thickness, whereas the membrane tension, i.e., $S_\mathrm{I}$ multiplied by the current thickness, changed little: the median element-wise ratio of tension to its baseline value was 0.95-1.01 in all six models. The leaflet kinematics was correspondingly insensitive: the median of $\lambda_\mathrm{I}$ changed by at most 0.03, that of $\varepsilon_\mathrm{D}$ by at most 0.08 mm, and $A_\mathrm{C}$ by at most 3\%, and the central orifice of TV10 changed from 5.5 to 9.0 and 5.8 mm$^2$ with the Q$_1$ and Q$_3$ fields.
The chordal apparatus was unaffected: the total chordal force changed by less than 3\%, and the force of each CT remained correlated with its baseline value with $r \geq 0.98$.

\subsection{Chordal size and material properties}\label{supp:res_ct}
In the nine leaflet-by-radial-band cells, the initial axial stiffness of a CT was 0.67-0.88 times the baseline value with the $Q_1$ set and 1.19-1.47 times with the $Q_3$ set: the smaller cross-sectional area of the $Q_1$ set outweighed its stiffer material. The recalibrated stress-free lengths differed from the baseline ones by $-0.03$ and $+0.02$\% in median, and by at most 0.3\%.

At ES, the CTs worked at small axial strains. The median strain was 0.09-0.17\% with the $Q_1$ set, 0.07-0.12\% at baseline, and 0.05-0.09\% with the $Q_3$ set, and the maximum strain did not exceed 1.8\%. The corresponding median elongations were 0.01-0.03 mm, on CTs with a median length of 22-28 mm. Accordingly, the leaflets were almost insensitive to chordal stiffness. The median of $\lambda_\mathrm{I}$ changed by at most 0.003. The closed configurations differed from baseline by at most 0.01 mm in median and, in five of the six models, by less than 0.4 mm between the 5th and 95th percentiles. The median of $\varepsilon_\mathrm{D}$ changed by $-3$ to $+2$\% (at most 0.03 mm) and $A_\mathrm{C}$ by $-0.1$ to $+1.7$\%. The central orifice of TV10 measured 5.6, 5.5, and 5.3 mm$^2$ with the $Q_1$, median, and $Q_3$ sets. The median of $S_\mathrm{I}$ changed by $-2$ to $+5$\%, with one exception: with the $Q_3$ set it decreased by 11\% in TV6 (from 14.0 to 12.4 kPa). This decrease was confined to the posterior leaflet, which closed with a locally different fold: the signed distance from the baseline configuration ranged from $-0.74$ to $+0.58$ mm between the 5th and 95th percentiles, and the median $S_\mathrm{I}$ of that leaflet decreased from 16.8 to 12.3 kPa.

The chordal load was also nearly unchanged. The total chordal force changed by less than 1\%, except in TV3 with the $Q_1$ set ($+5$\%). The force of each CT remained correlated with its baseline value with $r \geq 0.98$, and the share of the transvalvular load carried by the CTs changed by at most 0.7 percentage points.

The position of the leaflets relative to $\Omega^\mathrm{ES}$ was likewise unchanged (Table \ref{tab:S_position_sens}), although the total chordal cross-section changed by a factor comparable to that of the density maps.

\section{Data-only geometric error \texorpdfstring{$\varepsilon_\mathrm{ES}$}{epsilon-ES}}\label{supp:eps}

\paragraph{Methods.}
For each point of $\Omega^\mathrm{ES}$, the leaflet surface reconstructed from the ES segmentation (Section \ref{supp:geom}), $\varepsilon_\mathrm{ES}$ is the Euclidean distance to the closest point of the simulated leaflet surface at ES, triangulated from its shell elements. Distributed values are reported as median and interquartile range over the points of $\Omega^\mathrm{ES}$, without area weighting; the points are nearly uniformly spaced within each valve, and area-weighted medians differ from the unweighted ones by at most 0.06 mm. For the regional values, each point of $\Omega^\mathrm{ES}$ was attributed to the closest node and element of the simulated surface, and its distance from the annulus was measured from the nodes of the annular boundary of the simulated surface, as for $\varepsilon_\mathrm{D}$ below. $\varepsilon_\mathrm{ES}$ complements the geometric error $\varepsilon_\mathrm{D}$ (Section \secMetrics{} of the main text and Section \ref{supp:metrics}) in three respects. First, it is measured from the segmented surface to the simulated one: the coapting portions of the simulated leaflets, which extend into the RV beyond the region resolved by 3D-TEE and are represented by a single surface in $\Omega^\mathrm{ES}$, do not contribute to it, so that it is not inflated by the coaptation height. Second, its reference is the segmented surface itself, not the morphed configuration $\Omega_\mathrm{M}^\mathrm{ES}$; it therefore does not depend on the loads of the morphing simulation, and its reference is not the configuration on which the CT stress-free lengths were calibrated (Section \ref{supp:numerics}). Third, and conversely, it covers only the portion of the valve visible in the images, and does not assess the shape of the coapting portions of the leaflets.

\begin{figure}[htbp]
  \centering
\includegraphics[width=1\textwidth]{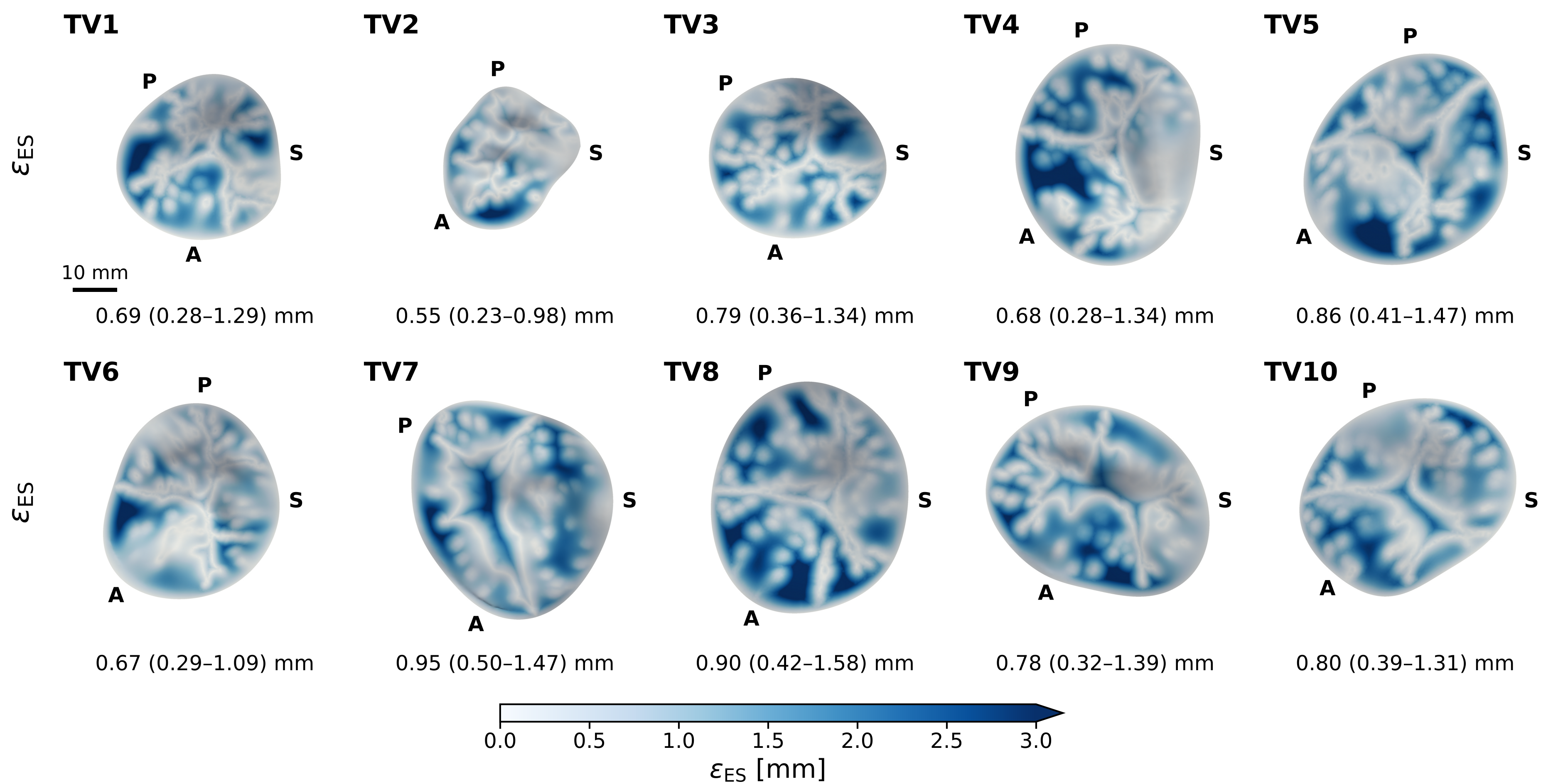}
  \caption{Data-only geometric error $\varepsilon_\mathrm{ES}$ of the ten baseline simulations: for each point of the segmented ES leaflet surface $\Omega^\mathrm{ES}$, distance to the closest point of the simulated leaflet surface at ES. Atrial view, septal leaflet on the right, all valves at the same scale as in Fig. \figCohortMaps{} of the main text; $\Omega^\mathrm{ES}$ covers only the portion of the leaflets resolved by 3D-TEE. Values below each map are median (IQR) over the points of $\Omega^\mathrm{ES}$. A: anterior, P: posterior, S: septal leaflet.}
  \label{figS_epsES}
\end{figure}

\paragraph{Results.}
In the ten baseline simulations, the per-valve medians of $\varepsilon_\mathrm{ES}$ ranged from 0.55 to 0.95 mm (cohort median 0.78 mm), with third quartiles of 0.98--1.58 mm (Fig. \ref{figS_epsES}), i.e. of the same order as $\varepsilon_\mathrm{D}$ (0.61--1.13 mm and 1.11-1.93 mm, respectively; Table \tabCohort{} of the main text). In the three valves of the sensitivity subset, the median $\varepsilon_\mathrm{ES}$ was 0.79, 0.67, and 0.80 mm (TV3, TV6, TV10), against an $\varepsilon_\mathrm{D}$ of 1.02, 0.76, and 1.13 mm. Measured on the image data alone, without the morphed reference and without the coaptation region, the error of the simulated closure is therefore of the same order as $\varepsilon_\mathrm{D}$. Its distribution also resembled that of $\varepsilon_\mathrm{D}$, whose median was highest in the band 3 to 6 mm from the annulus in all ten valves (0.70--1.53 mm, against 0.54--0.99 mm within 3 mm of the annulus and 0.61--1.14 mm farther from it), and exceeded the valve median within 2 mm of the free margin in six valves (0.56--2.17 mm). In all ten valves, the median $\varepsilon_\mathrm{ES}$ was highest in the radial band 3 to 6 mm from the annulus (0.67--1.33 mm, against 0.49--0.80 mm within 3 mm of the annulus and 0.50-0.93 mm farther from it). The error was lower along the coaptation lines, i.e. where the closest simulated node was in contact (nine valves), and around the CT insertions (nine valves), where the leaflets are held by the chordae, than in the leaflet bellies between insertions, where the simulated leaflets lay on the atrial side of $\Omega^\mathrm{ES}$ (Section \ref{supp:pos}).

\section{Position of the leaflets relative to \texorpdfstring{$\Omega^\mathrm{ES}$}{Omega-ES} and chordal insertion density}\label{supp:pos}

\paragraph{Methods.}
The position of the simulated leaflets relative to the image data was quantified outside the coaptation region, i.e. on the non-contact nodes: the nodes of the simulated configuration at ES with zero contact pressure, which is the same nodal indicator used to compute $A_\mathrm{C}$ (Section \ref{supp:metrics}). Contact nodes were excluded because, where the leaflets coapt, $\Omega^\mathrm{ES}$ represents the two coapting leaflets by a single surface. For each non-contact node $\mathbf{x}$, the signed distance from $\Omega^\mathrm{ES}$ is
\begin{equation}
d_\mathrm{s}(\mathbf{x}) = \operatorname{sign}\!\left[(\mathbf{x} - \mathbf{x}_\mathrm{c}) \cdot \mathbf{n}\right] \, \lVert \mathbf{x} - \mathbf{x}_\mathrm{c} \rVert ,
\end{equation}
where $\mathbf{x}_\mathrm{c}$ is the closest point of the triangulated $\Omega^\mathrm{ES}$ and $\mathbf{n}$ the normal of $\Omega^\mathrm{ES}$ at $\mathbf{x}_\mathrm{c}$, oriented, for each connected portion of $\Omega^\mathrm{ES}$, so that its area-weighted mean points towards the atrium. Positive values therefore indicate that the simulated leaflet lies on the atrial side of the segmented surface. For each simulation, we report the median and interquartile range of $d_\mathrm{s}$ and the fraction of non-contact nodes with $d_\mathrm{s} > 0$ (atrial fraction), over all non-contact nodes and over those 3 to 6 mm from the annulus, where $\varepsilon_\mathrm{D}$ was highest; the distance from the annulus is the Euclidean distance from the nodes of the annular boundary, as for the regional values of $\varepsilon_\mathrm{D}$ (Section \ref{supp:eps}). The same $\Omega^\mathrm{ES}$ was used as the reference for all simulations of a valve. $\varepsilon_\mathrm{ES}$ is defined as in Section \ref{supp:eps}.\\
The chordal insertion density at ED was computed as the number of CTs, $N_\mathrm{CT}$, divided by the leaflet area of the ED mesh, $A^\mathrm{ED}$, i.e. the sum of the areas of its shell elements; per leaflet, the CTs inserting on that leaflet were divided by its area at ED. Since every insertion carries its own CT (Section \ref{supp:insertion}), we also report the total chordal cross-section, $\Sigma A_0$, i.e. the sum of the cross-sectional areas assigned to all CTs (Table \ref{tab:material_params}).

\paragraph{Results.}
\textit{Baseline simulations.} Outside the coaptation region, the simulated leaflets lay on the atrial side of $\Omega^\mathrm{ES}$ in all ten valves (Table \ref{tab:S_position_baseline}): the median signed distance ranged from $+0.33$ to $+0.90$ mm (cohort median $+0.71$ mm), with 65--76\% of the non-contact nodes on the atrial side. The offset was largest in the band 3 to 6 mm from the annulus, $+0.66$ to $+1.72$ mm, with 73--97\% of the nodes on the atrial side. The nodes in contact were 22--45\% of the leaflet nodes. The median insertion density maps yielded 9.3--12.2 insertions/cm$^2$ of leaflet area at ED. The anterior leaflet had the lowest density in all ten valves (6.8--9.7 insertions/cm$^2$), against 11.9--13.0 and 8.9--14.1 insertions/cm$^2$ on the posterior and septal leaflets.\\
\textit{Sensitivity runs} (Table \ref{tab:S_position_sens}). From the $Q_1$ to the median and $Q_3$ density maps, the median signed distance decreased from $+1.16$--$1.25$ to $+0.51$--$0.85$ and $+0.42$--$0.62$ mm, and $\varepsilon_\mathrm{ES}$ from 1.08--1.25 to 0.67--0.80 and 0.52--0.65 mm, i.e. by 19--23\% from baseline to the $Q_3$ maps; the atrial fraction remained at 71--79\%. The $Q_1$, median, and $Q_3$ maps corresponded to 6.7--7.2, 10.9--12.0, and 15.9--16.8 insertions/cm$^2$, and $\Sigma A_0$ was 2.4--2.6 times larger with the $Q_3$ than with the $Q_1$ maps. From the compliant to the median and stiff leaflet sets, the median signed distance decreased from $+1.10$--$1.23$ to $+0.51$--$0.85$ and $+0.04$--$0.51$ mm, and $\varepsilon_\mathrm{ES}$ from 1.00--1.10 to 0.67--0.80 and 0.47--0.60 mm, i.e. by 26--29\% from baseline to the stiff set (and an increase of 26--64\% with the compliant set). In TV6, the stiff set brought the median signed distance close to zero ($+0.04$ mm, 53\% of the nodes on the atrial side, first quartile $-1.08$ mm), part of the leaflet having moved to the ventricular side of $\Omega^\mathrm{ES}$. With the chordal stiffness sets, the median signed distance changed by $-0.04$ to $+0.04$ mm relative to baseline and $\varepsilon_\mathrm{ES}$ by at most 0.014 mm, although $\Sigma A_0$ was 0.63--0.64 and 1.53--1.54 times the baseline value, a change comparable to that produced by the $Q_1$ and $Q_3$ density maps. In the band 3 to 6 mm from the annulus, 78--98\% of the non-contact nodes lay on the atrial side in every sensitivity run.

\begin{table}[p]
\centering
\resizebox{\textwidth}{!}{%
\begin{tabular}{lcccccccccccc}
\toprule
 & \multicolumn{2}{c}{Non-contact nodes} & \multicolumn{2}{c}{Non-contact, 3--6 mm} & & & & \multicolumn{4}{c}{Insertion density [1/cm$^2$]} & \\
\cmidrule(lr){2-3}\cmidrule(lr){4-5}\cmidrule(lr){9-12}
Valve & $d_\mathrm{s}$ [mm] & Atrial [\%] & $d_\mathrm{s}$ [mm] & Atrial [\%] & Contact [\%] & $N_\mathrm{CT}$ & $A^\mathrm{ED}$ [cm$^2$] & All & ANT & POST & SEPT & $\Sigma A_0$ [mm$^2$] \\
\midrule
TV1 & $+$0.71 (0.00, $+$1.44) & 74 & $+$1.32 ($+$0.57, $+$1.96) & 87 & 39 & 232 & 20.0 & 11.6 & 9.1 & 12.8 & 12.7 & 55.5 \\
TV2 & $+$0.33 ($-$0.26, $+$1.01) & 65 & $+$0.66 ($-$0.06, $+$1.64) & 73 & 34 & 127 & 11.4 & 11.1 & 8.8 & 12.1 & 12.7 & 28.6 \\
TV3 & $+$0.85 ($+$0.05, $+$1.51) & 76 & $+$1.37 ($+$0.70, $+$1.87) & 93 & 38 & 270 & 22.6 & 11.9 & 9.2 & 11.9 & 13.9 & 58.6 \\
TV4 & $+$0.72 ($+$0.02, $+$1.82) & 76 & $+$1.07 ($+$0.33, $+$2.39) & 88 & 23 & 250 & 22.7 & 11.0 & 8.7 & 11.9 & 13.5 & 58.8 \\
TV5 & $+$0.86 (0.00, $+$1.65) & 75 & $+$1.72 ($+$1.00, $+$2.40) & 97 & 27 & 257 & 24.3 & 10.6 & 8.6 & 12.0 & 12.8 & 63.9 \\
TV6 & $+$0.51 ($-$0.16, $+$1.08) & 72 & $+$0.99 ($+$0.54, $+$1.52) & 92 & 45 & 370 & 31.4 & 11.8 & 8.8 & 12.8 & 14.1 & 84.3 \\
TV7 & $+$0.65 ($-$0.92, $+$1.41) & 67 & $+$1.47 ($+$0.97, $+$2.24) & 97 & 22 & 295 & 24.2 & 12.2 & 9.7 & 13.0 & 12.4 & 64.0 \\
TV8 & $+$0.90 (0.00, $+$1.93) & 74 & $+$1.65 ($+$0.71, $+$2.65) & 87 & 35 & 344 & 34.0 & 10.1 & 6.8 & 12.0 & 12.6 & 84.8 \\
TV9 & $+$0.51 ($-$0.36, $+$1.38) & 68 & $+$1.49 ($+$0.78, $+$2.20) & 95 & 28 & 247 & 26.6 & 9.3 & 8.0 & 12.3 & 8.9 & 62.7 \\
TV10 & $+$0.74 ($-$0.09, $+$1.49) & 72 & $+$1.36 ($+$0.78, $+$2.02) & 93 & 40 & 320 & 29.4 & 10.9 & 8.6 & 12.7 & 12.7 & 76.5 \\
\midrule
Cohort median & $+$0.71 & 73 & $+$1.37 & 93 & 34 & 264 & 24.2 & 11.1 & 8.8 & 12.2 & 12.7 & 63.3 \\
\bottomrule
\end{tabular}
}
\caption{Position of the simulated leaflets relative to $\Omega^\mathrm{ES}$ and chordal insertion density, baseline simulations. $d_\mathrm{s}$: signed distance from $\Omega^\mathrm{ES}$, median (first quartile, third quartile), positive on the atrial side, over all non-contact nodes and over the non-contact nodes 3 to 6 mm from the annulus; Atrial: fraction of the same nodes with $d_\mathrm{s} > 0$; Contact: fraction of the leaflet nodes in contact at ES; $N_\mathrm{CT}$: number of CTs; $A^\mathrm{ED}$: leaflet area at ED; insertion density: $N_\mathrm{CT}$ per unit leaflet area at ED, for the whole valve (All) and for the anterior (ANT), posterior (POST), and septal (SEPT) leaflets; $\Sigma A_0$: total cross-sectional area of the CTs. Cohort median: median across valves of the per-valve values.}
\label{tab:S_position_baseline}
\end{table}

\begin{table}[p]
\centering
\resizebox{\textwidth}{!}{%
\begin{tabular}{llccccccccc}
\toprule
 & & & \multicolumn{2}{c}{Non-contact nodes} & \multicolumn{2}{c}{Non-contact, 3-6 mm} & & & & \\
\cmidrule(lr){4-5}\cmidrule(lr){6-7}
Valve & Model & $\varepsilon_\mathrm{D}$ [mm] & $d_\mathrm{s}$ [mm] & Atrial [\%] & $d_\mathrm{s}$ [mm] & Atrial [\%] & $\varepsilon_\mathrm{ES}$ [mm] & Contact [\%] & Density [1/cm$^2$] & $\Sigma A_0$ [mm$^2$] \\
\midrule
TV3 & Baseline & 1.02 & $+$0.85 & 76 & $+$1.37 & 93 & 0.79 (0.36--1.34) & 38 & 11.9 & 58.6 \\
 & Density $Q_1$ & 1.27 & $+$1.25 & 79 & $+$2.42 & 98 & 1.08 (0.48--1.98) & 37 & 7.2 & 35.1 \\
 & Density $Q_3$ & 0.82 & $+$0.61 & 76 & $+$0.91 & 90 & 0.61 (0.28--1.06) & 41 & 16.8 & 85.9 \\
 & Leaflet, compliant & 1.20 & $+$1.23 & 79 & $+$1.86 & 97 & 1.00 (0.47--1.68) & 45 & -- & -- \\
 & Leaflet, stiff & 0.79 & $+$0.51 & 69 & $+$0.92 & 86 & 0.58 (0.25--1.02) & 34 & -- & -- \\
 & Chordae $Q_1$ & 1.02 & $+$0.81 & 75 & $+$1.40 & 93 & 0.79 (0.35--1.35) & 38 & 11.9 & 37.8 \\
 & Chordae $Q_3$ & 1.04 & $+$0.85 & 76 & $+$1.39 & 94 & 0.79 (0.36--1.35) & 38 & 11.9 & 90.5 \\
\midrule
TV6 & Baseline & 0.76 & $+$0.51 & 72 & $+$0.99 & 92 & 0.67 (0.29--1.09) & 45 & 11.8 & 84.3 \\
 & Density $Q_1$ & 1.03 & $+$1.16 & 74 & $+$2.28 & 96 & 1.25 (0.51--2.04) & 45 & 7.1 & 49.6 \\
 & Density $Q_3$ & 0.56 & $+$0.42 & 71 & $+$0.88 & 90 & 0.52 (0.22--0.92) & 47 & 16.2 & 120.7 \\
 & Leaflet, compliant & 1.03 & $+$1.10 & 80 & $+$1.54 & 97 & 1.10 (0.59--1.65) & 50 & -- & -- \\
 & Leaflet, stiff & 0.67 & $+$0.04 & 53 & $+$0.52 & 78 & 0.47 (0.22--0.80) & 39 & -- & -- \\
 & Chordae $Q_1$ & 0.74 & $+$0.52 & 72 & $+$0.99 & 92 & 0.68 (0.30--1.10) & 45 & 11.8 & 53.8 \\
 & Chordae $Q_3$ & 0.75 & $+$0.55 & 73 & $+$1.01 & 93 & 0.68 (0.30--1.11) & 46 & 11.8 & 128.7 \\
\midrule
TV10 & Baseline & 1.13 & $+$0.74 & 72 & $+$1.36 & 93 & 0.80 (0.39--1.31) & 40 & 10.9 & 76.5 \\
 & Density $Q_1$ & 1.53 & $+$1.21 & 73 & $+$2.59 & 95 & 1.22 (0.53--2.02) & 36 & 6.7 & 46.3 \\
 & Density $Q_3$ & 0.95 & $+$0.62 & 72 & $+$1.03 & 93 & 0.65 (0.31--1.06) & 42 & 15.9 & 118.3 \\
 & Leaflet, compliant & 1.44 & $+$1.20 & 78 & $+$1.98 & 97 & 1.06 (0.50--1.64) & 43 & -- & -- \\
 & Leaflet, stiff & 0.88 & $+$0.35 & 63 & $+$0.77 & 84 & 0.60 (0.31--1.07) & 35 & -- & -- \\
 & Chordae $Q_1$ & 1.13 & $+$0.76 & 72 & $+$1.36 & 95 & 0.81 (0.40--1.32) & 40 & 10.9 & 48.1 \\
 & Chordae $Q_3$ & 1.11 & $+$0.77 & 73 & $+$1.35 & 93 & 0.81 (0.40--1.32) & 40 & 10.9 & 117.3 \\
\bottomrule
\end{tabular}
}
\caption{Position of the simulated leaflets relative to $\Omega^\mathrm{ES}$ in the sensitivity runs of TV3, TV6, and TV10: baseline, first- ($Q_1$) and third-quartile ($Q_3$) chordal insertion density maps (Table \tabDensity{} of the main text), compliant and stiff leaflet sets (Table \ref{tab:leaflet_params}; Table \tabLeafletStiff{} of the main text), and CT sets at the first and third quartiles of the cross-sectional area (Table \ref{tab:material_params}; Table \tabCtStiff{} of the main text). $\varepsilon_\mathrm{D}$: median geometric error; $d_\mathrm{s}$: median signed distance from $\Omega^\mathrm{ES}$, positive on the atrial side, over all non-contact nodes and over the non-contact nodes 3 to 6 mm from the annulus; Atrial: fraction of the same nodes with $d_\mathrm{s} > 0$; $\varepsilon_\mathrm{ES}$: median (IQR) over the points of $\Omega^\mathrm{ES}$; Contact: fraction of the leaflet nodes in contact at ES; Density: chordal insertion density at ED, whole valve; $\Sigma A_0$: total cross-sectional area of the CTs. The leaflet sets do not change the chordal apparatus (--: as at baseline); the CT sets change the cross-sectional area of each CT but not the number and position of the insertions.}
\label{tab:S_position_sens}
\end{table}
\clearpage
\section{Segmentation-level inter-observer agreement}\label{supp:seg}

\begin{figure}[htbp]
  \centering
\includegraphics[width=.8\textwidth]{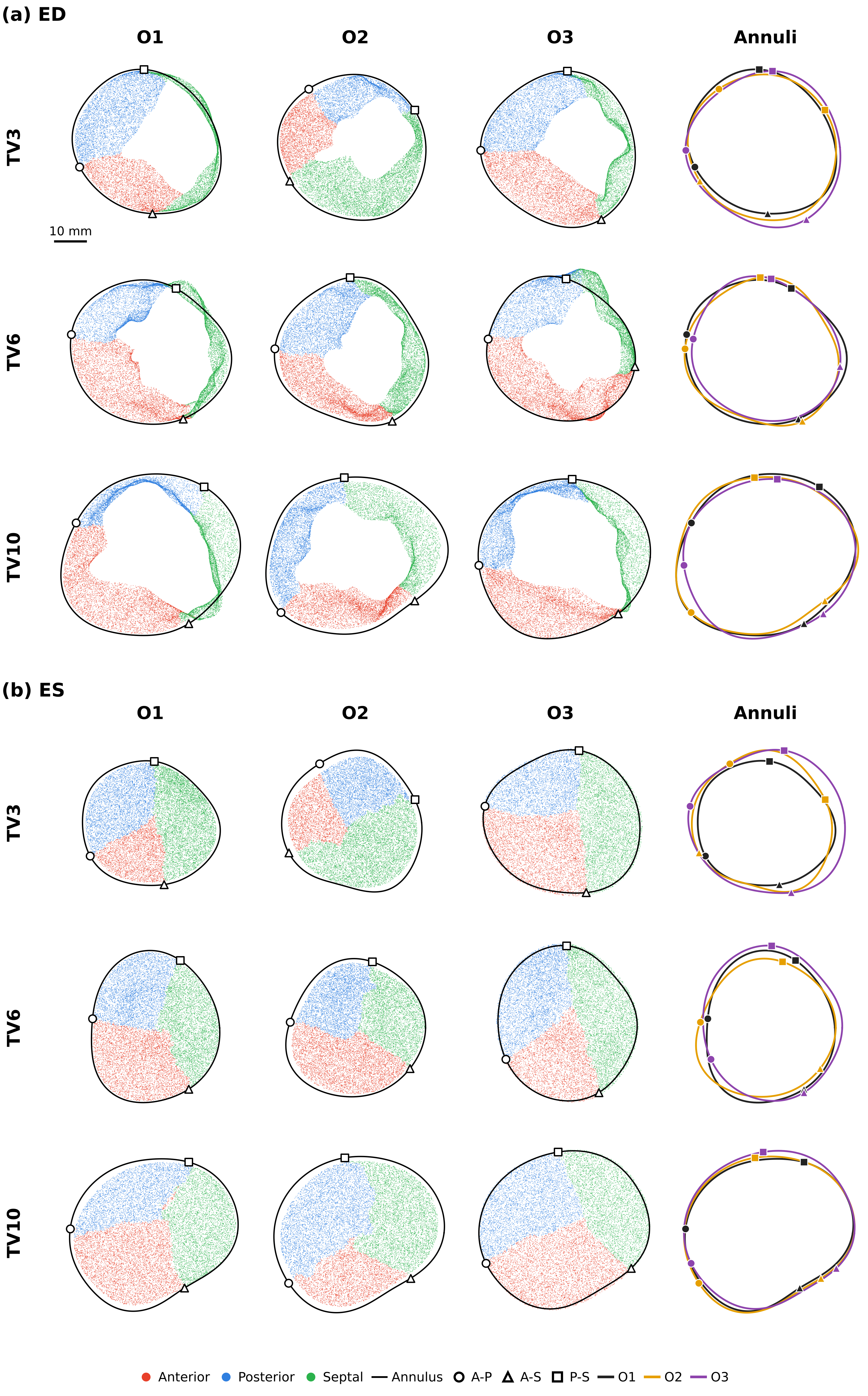}
\caption{Segmentations of TV3, TV6, and TV10 by the reference observer (O1) and the two additional observers (O2, O3), at ED (a) and ES (b). Atrial view in the common annular frame of each valve and frame: origin at the centroid of the three annuli, 0$^\circ$ (3 o'clock) at the middle of the septal annular arc of O1; all panels at the same scale. Columns O1--O3: reconstructed leaflet surface coloured by leaflet label, annular curve (Fourier fit of the annulus label, black), and commissures, located where the label of the closest leaflet changes along the annulus (A--P: anteroposterior, A--S: anteroseptal, P--S: posteroseptal). Right column: annular curves and commissures of the three observers.}
  \label{figS_seg}
\end{figure}

\paragraph{Methods.}
The three segmentations of each dataset (TV3, TV6, TV10) were compared at ED and ES. The label maps of the three observers share the voxel grid of the 3D-TEE volume, so all comparisons were performed in image space without registration. The leaflet surfaces reconstructed from each segmentation (Section \secGeometry{} of the main text and Section \ref{supp:geom}) were mapped back from the local reference frame of their observer to image space, and each surface point was assigned the label of the closest leaflet voxel of its own segmentation. For each pair of observers, five metrics were computed. The surface distance was quantified by the closest-point distance from 30\,000 area-weighted samples of one surface to the other, computed in both directions: the mean symmetric distance (MSD) is the average of the two directional means, and HD95 the 95\textsuperscript{th} percentile of the pooled distances. The label agreement, i.e. the fraction of the surface whose closest point on the other surface carries the same leaflet label, averaged over the two directions, quantified the agreement on the borders between adjacent leaflets. The volumetric overlap of the segmentations was quantified by the Dice similarity coefficient, $\mathrm{DSC} = 2|A \cap B| / (|A| + |B|)$, where $A$ and $B$ are the voxel sets of the two label maps, computed with the three leaflet labels merged and the annulus label excluded, and, for reference, for each leaflet label separately. Since the leaflet labels are only a few voxels thick, the DSC is sensitive to differences in the thickness of the tracing \cite{herz2021tvleaflets}; the surface distance was therefore taken as the primary measure of agreement. For the annulus distance, the annulus voxels of each segmentation were fitted, as a function of the angle about their centroid, with Fourier series of $x$, $y$, and $z$ (six harmonics), rejecting outliers beyond three robust standard deviations; the curve was resampled at 360 equally spaced points, and the mean symmetric closest-point distance between the two curves was computed (fourth-order fits, as in the main pipeline, changed it by at most 0.05 mm). Finally, the commissures of each observer were located along its annular curve, where the label of the closest leaflet voxel changes (circular mode filter over $\pm 7^\circ$), and the distance between corresponding commissures of the two observers was computed; at ED, these commissures were within 1--2.5 mm of those located by the reconstruction pipeline (Section \ref{supp:geom}). For each observer, the annulus was described by the perimeter of the annular curve, its area projected on the best-fit plane, and its height, i.e. the range of the distance of the curve from that plane, and the leaflets by the area of the leaflet surface. The septal-lateral and anteroposterior diameters reported in Table \tabOperators{} of the main text are the chords of the annular curve, projected on the common annular plane defined below, through its centroid, along the direction of the middle of the septal annular arc of O1 and perpendicular to it. The spread across observers was summarized by the coefficient of variation (CV), defined as in Eq.~(\ref{eq:S_CV}), i.e. the sample standard deviation of the three values over their mean. For display (Fig. \ref{figS_seg}), a common annular frame was defined for each valve and frame, with origin at the centroid of the annulus voxels of the three observers, axis normal to their best-fit plane, atrial view, and 0$^\circ$ (3 o'clock) at the middle of the septal annular arc of the reference observer (O1).

\paragraph{Results.}
Between pairs of observers, the MSD was 0.93--1.77 mm at ED and 0.54--1.11 mm at ES, and HD95 was 2.9--6.8 mm and 1.3--4.3 mm, respectively (Table \ref{tab:S_seg_pairs}). The DSC of the merged leaflet labels was 0.49--0.63 at ED and 0.64--0.73 at ES, lower than the inter-observer values reported for the merged leaflets of the TV on 3D transthoracic echocardiography (median 0.81) \cite{herz2021tvleaflets} and of the mitral valve on 3D-TEE (0.75--0.83) \cite{munafo2024mvpipeline}, both obtained by observers of the same group following a common labelling protocol. In the present study, the observers belonged to independent laboratories, and they also differed in the thickness of the traced leaflets. In TV3 at ES, for instance, the leaflet labels of O1 and O3 occupied 4956 and 8946 mm$^3$, which alone limits the DSC to $2 \times 4956 / (4956 + 8946) = 0.71$ even if one label contained the other entirely; the DSC of this pair was 0.64. Per leaflet, the DSC also reflects the placement of the borders between leaflets and ranged from 0.04 to 0.67, the lowest values occurring in the TV3 pairs involving O2. At ES, the CVs of the annular perimeter, projected annular area, and leaflet area were 2--7\%, 3--13\%, and 12--24\% (Table \ref{tab:S_seg_operators}), larger than at ED because O3 consistently traced the widest annulus and segmented the leaflets further toward it: its annular perimeter was the longest in all three valves (146--159 mm, against 127--157 mm), and its leaflet surface the largest (2028--2037 mm$^2$, against 1362--1684 mm$^2$). The annular height varied more (CV 16--60\%), being determined by a few millimeters of out-of-plane excursion. Most of the disagreement concerned the borders between adjacent leaflets (Fig. \ref{figS_seg}, Table \ref{tab:S_seg_pairs}), corresponding commissures of two observers being 3.1 to 35.1 mm apart at ED. In TV3, all three commissures of O2 were rotated with respect to those of O1 and O3, by 21--35 mm along the annulus at ED, while the leaflet order was preserved, so that only 43--44\% of the leaflet surface had the same label in the pairs involving O2, against 91\% between O1 and O3. In TV6, the anteroseptal commissure of O3 was 20 mm away from those of O1 and O2 at ED, so that the anterior leaflet of O3 extended over 48\% of the length of the annular curve, against 33--35\%. In TV10, the commissures of O1 and O2 were 9.5--27.5 mm apart at ED. Excluding the TV3 pairs involving O2, the label agreement ranged from 64\% to 91\% at ED and from 74\% to 89\% at ES.

Since the borders were not harmonized (Section \ref{Imseg}), the share of the ED leaflet area taken by each leaflet varied: the septal leaflet took 54\% of it in the O2 model of TV3 (33-37\% in the other two), and the anterior leaflet 47\% in the O3 model of TV6 (31-37\%). The annulus was segmented more consistently than the leaflets. At ED, the CV across observers was 1-4\% for the septolateral and anteroposterior diameters, 0-3\% for the perimeter, and 3-5\% for the projected area, compared to 10-19\% for the leaflet area (Table \ref{tab:operators}; Tables \ref{tab:S_seg_pairs} and \ref{tab:S_seg_operators}). O2 segmented the smallest leaflets in all three valves, 15-29\% below the mean of O1 and O3, on annuli within 4\% of theirs in perimeter and projected area. As a result, the leaflet area of the models at ED differed by up to 32\% within a valve, and the number of CTs differed by up to 35\%. At ES, the spread of the annular dimensions was larger (CV up to 7\% for the diameters and perimeter, and 13\% for the projected area) because O3 segmented the widest annulus in all three valves. The annular curves of O1 and O2, on the other hand, were closer at ES than at ED (0.8-2.7 vs. 1.6-4.6 mm). Across all observer pairs, the annular curves were 0.8-3.7 mm apart at ES.

\begin{table}[htbp]
\centering
\resizebox{\textwidth}{!}{%
\begin{tabular}{llccccccccccccc}
\toprule
 & & \multicolumn{2}{c}{MSD [mm]} & \multicolumn{2}{c}{HD95 [mm]} & \multicolumn{2}{c}{DSC [-]} & \multicolumn{2}{c}{Same label [\%]} & \multicolumn{2}{c}{Annulus distance [mm]} & \multicolumn{3}{c}{Commissure distance, ED [mm]} \\
\cmidrule(lr){3-4}\cmidrule(lr){5-6}\cmidrule(lr){7-8}\cmidrule(lr){9-10}\cmidrule(lr){11-12}\cmidrule(lr){13-15}
Valve & Observers & ED & ES & ED & ES & ED & ES & ED & ES & ED & ES & A--P & A--S & P--S \\
\midrule
TV3 & O1--O2 & 1.56 & 0.78 & 6.4 & 2.4 & 0.52 & 0.70 & 44 & 44 & 4.6 & 2.6 & 24.8 & 23.0 & 24.4 \\
 & O1--O3 & 1.28 & 0.94 & 4.0 & 3.5 & 0.50 & 0.64 & 91 & 77 & 3.6 & 3.4 & 6.7 & 12.8 & 4.2 \\
 & O2--O3 & 1.11 & 0.94 & 3.5 & 4.1 & 0.54 & 0.65 & 43 & 53 & 2.9 & 3.1 & 21.4 & 35.1 & 20.7 \\
\midrule
TV6 & O1--O2 & 1.74 & 0.55 & 5.0 & 1.3 & 0.49 & 0.73 & 89 & 89 & 2.3 & 1.8 & 4.4 & 3.1 & 10.5 \\
 & O1--O3 & 1.22 & 0.85 & 4.3 & 3.2 & 0.59 & 0.68 & 86 & 85 & 1.9 & 3.4 & 3.6 & 20.4 & 6.8 \\
 & O2--O3 & 1.66 & 0.93 & 5.6 & 4.3 & 0.50 & 0.67 & 76 & 83 & 1.8 & 3.7 & 4.9 & 20.5 & 4.5 \\
\midrule
TV10 & O1--O2 & 1.54 & 0.69 & 6.3 & 1.8 & 0.59 & 0.70 & 64 & 74 & 1.6 & 0.8 & 27.5 & 9.5 & 20.0 \\
 & O1--O3 & 0.93 & 1.11 & 3.0 & 3.0 & 0.63 & 0.65 & 78 & 78 & 2.7 & 2.1 & 13.8 & 7.3 & 13.2 \\
 & O2--O3 & 1.77 & 0.54 & 6.8 & 2.1 & 0.57 & 0.72 & 86 & 82 & 2.7 & 1.7 & 16.7 & 5.6 & 7.0 \\
\bottomrule
\end{tabular}
}
\caption{Segmentation-level agreement between observers. MSD: mean symmetric distance between the leaflet surfaces of two observers; HD95: 95\textsuperscript{th} percentile of the same distances; DSC: Dice similarity coefficient of the label maps, with the three leaflet labels merged; Same label: fraction of the leaflet surface whose closest point on the other surface carries the same leaflet label; Annulus distance: mean symmetric closest-point distance between the two annular curves; Commissure distance: distance between corresponding commissures at ED (A--P: anteroposterior, A--S: anteroseptal, P--S: posteroseptal), located along the annular curve of each observer where the label of the closest leaflet changes.}
\label{tab:S_seg_pairs}
\end{table}

\begin{table}[htbp]
\centering
\resizebox{\textwidth}{!}{%
\begin{tabular}{llcccccccc}
\toprule
 & & \multicolumn{4}{c}{ED} & \multicolumn{4}{c}{ES} \\
\cmidrule(lr){3-6}\cmidrule(lr){7-10}
Valve & Observer & $P_\mathrm{A}$ [mm] & $A_\mathrm{A}$ [mm$^2$] & $h_\mathrm{A}$ [mm] & $A_\mathrm{L}$ [mm$^2$] & $P_\mathrm{A}$ [mm] & $A_\mathrm{A}$ [mm$^2$] & $h_\mathrm{A}$ [mm] & $A_\mathrm{L}$ [mm$^2$] \\
\midrule
TV3 & O1 & 143 & 1564 & 4.9 & 2262 & 127 & 1242 & 4.2 & 1362 \\
 & O2 & 143 & 1568 & 5.6 & 1926 & 137 & 1406 & 6.9 & 1362 \\
 & O3 & 151 & 1704 & 7.1 & 2520 & 146 & 1624 & 5.4 & 2028 \\
 & CV [\%] & 3 & 5 & 19 & 13 & 7 & 13 & 25 & 24 \\
\midrule
TV6 & O1 & 148 & 1656 & 6.2 & 3145 & 137 & 1433 & 5.6 & 1635 \\
 & O2 & 149 & 1593 & 8.6 & 2136 & 134 & 1373 & 6.2 & 1442 \\
 & O3 & 143 & 1524 & 7.9 & 2849 & 147 & 1547 & 11.0 & 2034 \\
 & CV [\%] & 2 & 4 & 16 & 19 & 5 & 6 & 39 & 18 \\
\midrule
TV10 & O1 & 164 & 2091 & 2.2 & 2940 & 154 & 1792 & 6.4 & 1684 \\
 & O2 & 164 & 2035 & 5.4 & 2411 & 157 & 1863 & 6.4 & 1650 \\
 & O3 & 163 & 1988 & 8.8 & 2762 & 159 & 1898 & 9.5 & 2037 \\
 & CV [\%] & 0 & 3 & 60 & 10 & 2 & 3 & 24 & 12 \\
\bottomrule
\end{tabular}
}
\caption{Annulus and leaflet geometry segmented by each observer. $P_\mathrm{A}$: perimeter of the annular curve; $A_\mathrm{A}$: area of the annulus projected on its best-fit plane; $h_\mathrm{A}$: annular height, i.e. range of the distance of the annular curve from that plane; $A_\mathrm{L}$: area of the segmented leaflet surface. CV: coefficient of variation across the three observers.}
\label{tab:S_seg_operators}
\end{table}
\clearpage
\section{Baseline biomechanics of the ten valves}\label{supp:bio}

\paragraph{Methods.}
\textit{Leaflet stretch.} The deformation from $\Omega^\mathrm{ED}$, the stress-free configuration, to the configuration at ES was evaluated at the centroid of each S4R element from the mid-surface nodal coordinates. With $\mathbf{G}_\alpha$ and $\mathbf{g}_\alpha$ ($\alpha = 1,2$) the covariant tangent vectors of the element at ED and ES, obtained from its bilinear shape functions, and $G_{\alpha\beta} = \mathbf{G}_\alpha \cdot \mathbf{G}_\beta$ and $C_{\alpha\beta} = \mathbf{g}_\alpha \cdot \mathbf{g}_\beta$, the in-plane principal stretches $\lambda_\mathrm{I} \geq \lambda_\mathrm{II}$ are the square roots of the eigenvalues of $G^{-1}C$, and the area stretch is $J_\mathrm{s} = \lambda_\mathrm{I}\lambda_\mathrm{II} = \sqrt{\det C / \det G}$. The stretch along a unit vector $\mathbf{M} = M^\alpha \mathbf{G}_\alpha$ of the ED tangent plane is $\lambda_\mathbf{M} = (M^\alpha C_{\alpha\beta} M^\beta)^{1/2}$. The circumferential direction is the projection on the ED tangent plane of the azimuthal direction about the valve axis, i.e. the axis of the local reference frame through the centroid of the annulus at ED, and the radial direction is the ED tangent direction orthogonal to it, running from the annulus to the free margin; they define the circumferential and radial stretches $\lambda_\mathrm{C}$ and $\lambda_\mathrm{R}$. Statistics of the stretches are weighted by the element area at ES, as those of $S_\mathrm{I}$.\\
\textit{Regional stress.} The median and the 95\textsuperscript{th} percentile (P95) of $S_\mathrm{I}$, weighted by element area, were computed in the band 3 to 6 mm from the annulus and in the region farther than 6 mm from it, with the distance measured from the element center at ES to the nodes of the annular boundary. Since $S_\mathrm{I}$ peaks locally at the CT insertions, both statistics were also computed after excluding the elements that contain a node of a CT distributing coupling (Section \ref{supp:insertion}).\\
\textit{Forces.} The force transmitted by each CT is the magnitude of the reaction force at its origin at ES. The force on each papillary body is the magnitude of the vector resultant of the reaction forces of its CTs; the bodies are named after the commissure beneath which they were placed (anteroposterior, A--P; anteroseptal, A--S; posteroseptal, P--S). The pressure resultant is $|\mathbf{F}_p|$, with $\mathbf{F}_p = -p \sum_e \mathbf{a}_e$, where $\mathbf{a}_e$ is the area vector of element $e$ at ES and $p = 22$ mmHg; the CT load share is defined as in Table \tabCohort{} of the main text. The projected annular area is the area enclosed by the annular boundary of the simulated leaflets at ES, projected on the plane normal to the valve axis.

\paragraph{Results.}
The leaflet deformation from ED to ES was predominantly radial (Table \ref{tab:S_biomech}a). The per-valve medians of $\lambda_\mathrm{I}$ were 1.37--1.43, those of $\lambda_\mathrm{R}$ 1.26--1.38, and those of $\lambda_\mathrm{C}$ 0.99--1.12; $\lambda_\mathrm{R}$ exceeded $\lambda_\mathrm{C}$ over 61--84\% of the leaflet area, and the minor principal stretch was below unity (per-valve medians of $\lambda_\mathrm{II}$ 0.89--0.97). The median area strain $J_\mathrm{s} - 1$ was 26--36\% (cohort median 34\%), and 20--42\% on the individual leaflets, with no leaflet systematically more stretched than the others. The median $S_\mathrm{I}$ was 9.9--20.0 kPa in the band 3 to 6 mm from the annulus and 5.9--19.0 kPa farther from it, and P95 was 25--61 and 29--63 kPa, respectively (Table \ref{tab:S_biomech}b). 

The elements coupled to CT insertions covered 27--33\% of the leaflet area; excluding them changed the regional medians to 9.2--20.9 and 6.8--18.9 kPa and P95 to 26--62 and 28--59 kPa. The total chordal force was 1.79--5.92 N (median 4.56 N), against a pressure resultant of 2.19--5.80 N, and the force on each papillary body was 0.53--2.42 N (median 1.37 N over the 30 papillary bodies; Table \ref{tab:S_biomech}c). The ten models included 127 to 370 CTs (9.3-12.2 insertions/cm$^2$ of leaflet area at ED), 18-21\% of which belonged to the free margin. The force per CT, $F_\mathrm{CT}$, was strongly skewed: per-valve medians ranged from 5.2 to 13.8 mN, and maxima from 56 to 153 mN. The corresponding nominal stress, $F_\mathrm{CT}/A$, had per-valve medians of 24-58 kPa. The chordal load depended mainly on the radial position of the insertion: in all ten valves, the median $F_\mathrm{CT}$ decreased from the near-annulus band (22-56 mN) to the belly (5.5-16.5 mN) and to the free-margin band (2.2-8.3 mN). Marginal CTs carried 8-17\% of the total chordal force, less than their share of the CT count, and their median force was lower than that of non-marginal CTs in eight of ten valves (3.3-13.8 vs. 4.5-15.4 mN). No leaflet systematically carried more load per CT than the others. Valves with more CTs transferred a larger share of the load to the chordal apparatus (Spearman $\rho = 0.78$, $p = 0.008$), independently of valve size: the leaflet area at ED correlated with the number of CTs ($\rho = 0.81$) but not with the load share ($\rho = 0.43$, $p = 0.21$).

\begin{table}[htbp]
\centering
\small
\textbf{(a) Leaflet stretch from ED to ES}\\[3pt]
\resizebox{\textwidth}{!}{%
\begin{tabular}{lcccccccc}
\toprule
 & & & & \multicolumn{4}{c}{$J_\mathrm{s} - 1$ [\%]} & \\
\cmidrule(lr){5-8}
Valve & $\lambda_\mathrm{I}$ [-] & $\lambda_\mathrm{R}$ [-] & $\lambda_\mathrm{C}$ [-] & All & ANT & POST & SEPT & $\lambda_\mathrm{R} > \lambda_\mathrm{C}$ [\%] \\
\midrule
TV1 & 1.37 (1.29--1.44) & 1.28 (1.12--1.38) & 1.03 (0.93--1.16) & 27 & 33 & 20 & 26 & 73 \\
TV2 & 1.39 (1.33--1.43) & 1.32 (1.21--1.40) & 1.04 (0.96--1.15) & 33 & 33 & 31 & 35 & 81 \\
TV3 & 1.40 (1.33--1.46) & 1.31 (1.17--1.40) & 1.03 (0.91--1.17) & 26 & 35 & 25 & 21 & 76 \\
TV4 & 1.40 (1.34--1.46) & 1.27 (1.15--1.37) & 1.11 (1.01--1.23) & 36 & 34 & 38 & 36 & 70 \\
TV5 & 1.41 (1.35--1.45) & 1.26 (1.13--1.39) & 1.12 (0.98--1.30) & 36 & 35 & 33 & 39 & 61 \\
TV6 & 1.43 (1.36--1.48) & 1.35 (1.24--1.43) & 0.99 (0.90--1.11) & 29 & 32 & 34 & 24 & 84 \\
TV7 & 1.41 (1.35--1.45) & 1.34 (1.21--1.42) & 1.05 (0.94--1.20) & 36 & 40 & 32 & 42 & 77 \\
TV8 & 1.42 (1.36--1.46) & 1.35 (1.25--1.42) & 1.03 (0.92--1.17) & 35 & 32 & 37 & 37 & 82 \\
TV9 & 1.43 (1.38--1.47) & 1.38 (1.26--1.44) & 1.03 (0.94--1.15) & 34 & 35 & 29 & 34 & 83 \\
TV10 & 1.41 (1.34--1.46) & 1.35 (1.23--1.42) & 1.02 (0.93--1.14) & 32 & 27 & 39 & 32 & 83 \\
\midrule
Cohort median & 1.41 & 1.33 & 1.03 & 34 & 33 & 32 & 34 & 79 \\
\bottomrule
\end{tabular}
}\\[5pt]
\textbf{(b) Regional maximum principal stress $S_\mathrm{I}$ [kPa]}\\[3pt]
\resizebox{.8\textwidth}{!}{%
\begin{tabular}{lccccccccc}
\toprule
 & \multicolumn{4}{c}{All elements} & \multicolumn{4}{c}{Excluding CT-coupled elements} & \\
\cmidrule(lr){2-5}\cmidrule(lr){6-9}
 & \multicolumn{2}{c}{3--6 mm} & \multicolumn{2}{c}{$>$ 6 mm} & \multicolumn{2}{c}{3--6 mm} & \multicolumn{2}{c}{$>$ 6 mm} & CT-coupled \\
\cmidrule(lr){2-3}\cmidrule(lr){4-5}\cmidrule(lr){6-7}\cmidrule(lr){8-9}
Valve & Median & P95 & Median & P95 & Median & P95 & Median & P95 & area [\%] \\
\midrule
TV1 & 10.3 & 32 & 5.9 & 29 & 10.2 & 35 & 6.8 & 30 & 33 \\
TV2 & 11.0 & 25 & 9.7 & 31 & 11.3 & 26 & 9.7 & 28 & 32 \\
TV3 & 13.1 & 40 & 9.9 & 38 & 12.5 & 39 & 9.1 & 36 & 33 \\
TV4 & 9.9 & 31 & 14.3 & 53 & 9.2 & 31 & 14.5 & 44 & 31 \\
TV5 & 13.3 & 31 & 16.1 & 53 & 13.0 & 28 & 16.6 & 49 & 30 \\
TV6 & 20.0 & 61 & 12.8 & 50 & 20.9 & 62 & 14.4 & 50 & 32 \\
TV7 & 11.6 & 43 & 14.6 & 48 & 12.0 & 45 & 14.7 & 46 & 33 \\
TV8 & 12.7 & 36 & 15.9 & 43 & 13.6 & 34 & 15.9 & 41 & 30 \\
TV9 & 12.0 & 39 & 19.0 & 47 & 11.8 & 38 & 18.9 & 44 & 27 \\
TV10 & 10.3 & 33 & 14.5 & 63 & 10.3 & 34 & 16.2 & 59 & 31 \\
\midrule
Cohort median & 11.8 & 34 & 14.4 & 48 & 11.9 & 34 & 14.6 & 44 & 32 \\
\bottomrule
\end{tabular}
}\\[5pt]
\textbf{(c) Chordal and pressure forces at ES}\\[3pt]
\resizebox{.8\textwidth}{!}{%
\begin{tabular}{lccccccc}
\toprule
 & & \multicolumn{3}{c}{Papillary body [N]} & & & \\
\cmidrule(lr){3-
5}
Valve & $\Sigma F_\mathrm{CT}$ [N] & A--P & A--S & P--S & $|\mathbf{F}_p|$ [N] & CT load [\%] & $A_\mathrm{A}^\mathrm{ES}$ [mm$^2$] \\
\midrule
TV1 & 3.15 & 0.86 & 0.94 & 1.25 & 3.35 & 83 & 1082 \\
TV2 & 1.79 & 0.54 & 0.68 & 0.53 & 2.19 & 74 & 706 \\
TV3 & 3.83 & 1.01 & 1.35 & 1.31 & 3.56 & 92 & 1122 \\
TV4 & 4.29 & 1.16 & 1.53 & 1.49 & 4.87 & 79 & 1639 \\
TV5 & 5.04 & 1.70 & 1.82 & 1.32 & 5.30 & 82 & 1714 \\
TV6 & 4.57 & 1.38 & 1.62 & 1.34 & 4.37 & 89 & 1356 \\
TV7 & 5.13 & 1.04 & 1.43 & 2.42 & 5.18 & 85 & 1718 \\
TV8 & 5.92 & 1.82 & 1.76 & 2.04 & 5.80 & 87 & 1837 \\
TV9 & 4.57 & 1.43 & 2.08 & 0.87 & 5.04 & 79 & 1674 \\
TV10 & 4.55 & 1.55 & 1.33 & 1.53 & 4.85 & 85 & 1622 \\
\midrule
Cohort median & 4.56 & 1.27 & 1.48 & 1.33 & 4.86 & 84 & 1630 \\
\bottomrule
\end{tabular}
}
\caption{Baseline biomechanics of the ten valves at ES. (a) Major principal stretch $\lambda_\mathrm{I}$, radial stretch $\lambda_\mathrm{R}$, and circumferential stretch $\lambda_\mathrm{C}$ from ED to ES, median (IQR) weighted by element area; median area strain $J_\mathrm{s} - 1$ over the whole valve (All) and for the anterior (ANT), posterior (POST), and septal (SEPT) leaflets; fraction of the leaflet area with $\lambda_\mathrm{R} > \lambda_\mathrm{C}$. (b) Median and 95\textsuperscript{th} percentile (P95) of $S_\mathrm{I}$, weighted by element area, in the band 3 to 6 mm from the annulus and farther than 6 mm from it, over all elements and excluding the elements coupled to CT insertions; CT-coupled area: fraction of the leaflet area covered by those elements. (c) $\Sigma F_\mathrm{CT}$: total chordal force; papillary body: magnitude of the resultant force of the CTs of each papillary body, named after the commissure beneath which it lies (A--P: anteroposterior, A--S: anteroseptal, P--S: posteroseptal); $|\mathbf{F}_p|$: resultant of the transvalvular pressure on the leaflets; CT load: share of the transvalvular load carried by the chordal apparatus (Table \tabCohort{} of the main text); $A_\mathrm{A}^\mathrm{ES}$: projected annular area of the model at ES. Cohort median: median across valves of the per-valve values.}
\label{tab:S_biomech}
\end{table}

\end{document}